\documentclass[%
 reprint,
 aps,amssymb,amsmath,
 prfluids,longbibliography,
onecolumn,11pt
]{revtex4-2}

\usepackage{graphicx}
\usepackage{bm}
\usepackage[mathlines]{lineno}
\usepackage{lineno}
\usepackage{soul}
\usepackage{graphicx}
\usepackage{xcolor}
\usepackage{natbib}
\usepackage{multirow}
\usepackage[normalem]{ulem}
\usepackage{comment}

\newcommand{\half}{{\textstyle \frac12}}

\newcommand{\rmd}{\mathrm{d}}
\newcommand{\rme}{\mathrm{e}}
\newcommand{\rmi}{\mathrm{i}}
\newcommand{\eps}{\epsilon}
\newcommand{\bV}{\mathbf{V}}
\newcommand{\bU}{\mathbf{U}}

\newcommand{\bu}{\mathbf{u}}
\newcommand{\bx}{\mathbf{x}}
\newcommand{\bk}{\mathbf{k}}
\newcommand{\p}{\partial}
\newcommand{\order}[1]{\mathcal{O}(#1)}

\newcommand{\kx}{\mathbf{k}\cdot\bx}

\newcommand{\mR}{\mathcal{R}}
\newcommand{\mo}{\mathcal{O}}
\newcommand{\hzeta}{\hat{\zeta}}
\newcommand{\hw}{\hat{w}}
\newcommand{\hu}{\hat{u}}
\newcommand{\hv}{\hat{v}}
\newcommand{\hbu}{\hat{\bu}}

\newcommand{\hp}{\hat{P}}

\newcommand{\cV}{\mathcal{V}}

\newcommand{\hmN}{\hat{\mathcal{N}}}
\newcommand{\hmF}{\hat{\mathcal{F}}}

\newcommand{\fst}{^{(1)}}
\newcommand{\snd}{^{(2)}}

 \newcommand{\sdl}[1]{}

\newcommand{\tzt}{\tilde{\zeta}}
\newcommand{\ttht}{\tilde{\theta}}
\newcommand{\ta}{{a}}
\newcommand{\tpsi}{{\psi}}
\newcommand{\hA}{{\hat{A}}}

\newcommand{\be}{\begin{equation}}
\newcommand{\ee}{\end{equation}}
\newcommand{\tz}{\tilde{\zeta}}
\newcommand{\tzc}{\tz_c}

\makeatletter
\newcommand{\specialnumber}[1]{%
\def\tagform@##1{\maketag@@@{(\ignorespaces##1\unskip\@@italiccorr#1)}}}
\newcommand{\specialeqref}[2]{\begingroup
\def\tagform@##1{\maketag@@@{(\ignorespaces##1\unskip\@@italiccorr#2)}}%
\eqref{#1}\endgroup}
\begin{document}

\preprint{Submitted to Physical Review Fluids}

\title{
  Statistics of weakly nonlinear waves on currents with strong vertical shear
}

\author{Zibo Zheng$^1$}
\email{zibo.zheng@ntnu.no}
\author{Yan Li$^{1,2,}$\footnote{Corresponding author: yan.li@uib.no}}
%
\author{Simen {\AA.} Ellingsen$^1$}
\email{simen.a.ellingsen@ntnu.no}
\affiliation{%
 $^1$Department of Energy and Process Engineering, Norwegian University of Science and Technology, N-7491 Trondheim, Norway,
 \\
 $^2$Department of Mathematics, University of Bergen, N-5020 Bergen, Norway.
}%

\date{December 12, 2022}

\begin{abstract}
We investigate how the presence of a vertically sheared current affects wave statistics, including the probability of rogue waves, and apply it to a real-world case using measured spectral and shear current data from the Mouth of the Columbia River.
A theory for weakly nonlinear waves valid to second order in wave steepness is derived, and used to analyze statistical properties of surface waves; the theory extends the classic theory by Longuet-Higgins [{\it J.~Fluid Mech.} {\bf 12}, 3 (1962)] to allow for an arbitrary depth-dependent background flow, $U(z)$, with $U$ the horizontal velocity along the main direction of wave propagation and $z$ the vertical axis. Numerical statistics are collected from a large number of realisations of random, irregular sea-states following a JONSWAP spectrum, on linear and exponential model currents of varying strengths. A number of statistical quantities are presented and compared to a range of theoretical expressions from the literature; in particular the distribution of wave surface elevation, surface maxima, and crest height; the exceedance probability including the probability of rogue waves; the maximum crest height among $N_s$ waves, and the skewness of the surface elevation distribution. 
We find that compared to no-shear conditions, opposing vertical shear ($U'(z)>0$) leads to increased wave height and increased skewness of the nonlinear-wave elevation distribution, while a following shear ($U'(z)<0$) has opposite effects. 
With the wave spectrum and 
velocity profile
measured in the Columbia River estuary by Zippel \& Thomson [{\it J.~Geophys. Res: Oceans} {\bf 122}, 3311 (2017)] our second--order theory predicts that the probability of rogue waves is significantly reduced and enhanced during ebb and flood, respectively, adding support to the notion that shear currents need to be accounted for in wave modelling and prediction. 
\end{abstract}

\maketitle

\section{Introduction}

Waves in the ocean are almost invariably affected by interaction with their surroundings, ambient currents in particular. While large-scale ocean currents may be approximately depth-independent, this is often not the case for smaller scale currents such as those driven by wind shear, or currents in the near-shore environment including river deltas and tidal currents.
Of particular interests is the role of these environmental factors on the occurrence probability of extremely large waves 
\citep{kharif08,Cavaleri18,dudley19}, 
known also as rogue, giant, or freak waves, defined as waves whose amplitude far exceeds that of their surrounding wave field. To this end, many formation mechanisms of rogue waves have been proposed, including (but not limited to)
dispersive focusing of linear waves \cite{kharif08}, nonlinear effects such as the modulational instability  \cite{benjamin67} and quartet resonances \citep{janssen03} as well as 
refraction by currents \citep{white98} and bathymetry \citep{janssen09,gao21},  
nonlinear interaction between surface waves and depth transitions \citep{trulsen20,li21}. 
In this paper, our main attention is paid to the effect of a background depth-varying current on the statistics of weakly non-linear waves, rogue wave events in particular.

In order to obtain a proper statistical description of rogue wave events, a theory for second-order interaction of waves in a random sea has been widely used in both analytical \citep{longuet62,Longuet-Higgins63,Tayfun80,Tayfun83,Tayfun86,dalzell99,Forristall00,Arena02} and numerical studies \citep{Toffoli07,Toffoli08}. In contrast to linear waves in a random sea for which the wave elevation can be represented as a Gaussian random process \citep{longuet52}, second-order nonlinear waves can lead to considerable deviations from Gaussian wave statistics due to the steepened crests and flattened troughs caused by second-order (bound) waves. To describe the altered statistics, analytical models for wave crest and elevation distributions have been proposed for deep-water random waves, see, e.g., \cite{Tayfun80,Tayfun86,Forristall00}. 
These generally agree well with both laboratory and field measurements for narrowband and broadband wave fields (see, e.g., \cite{Forristall00,petrova06,Toffoli08,fedele09,Fedele19}) with moderate steepness.
In more nonlinear sea states discrepancies arise from third and higher order nonlinear effects, e.g., the well-known Benjamin-Feir instability \cite{benjamin67} 
and the resonant wave quartets \citep{janssen03}.  Hence, a second-order theory such as the one we present herein, is limited to the cases where higher-order corrections are comparatively small.

Many studies have suggested several different ways by which the probability of rogue waves is increased in the presence of currents with horizontal, but not vertical, spatial variation (c.f.\ \cite{shrira14,shrira14b}). A current whose magnitude and direction varies slowly in space relative to the rapidly varying wave phase has mostly been considered as a (local) Doppler shift on the wave dispersion relation and as a medium of refraction in the conservation of wave action \citep{peregrine76,white98}. Due to this, White \& Fornberg \cite{white98} attribute the enhanced probability of larger wave events in currents to the local refraction by currents. Many varieties of the third-order nonlinear Schr{\"o}dinger equations have been developed for slowly (horizontally) varying currents, see, e.g., \cite{stocker99,Curtis18,Hjelmervik09}. An opposing current has been found to lead to strengthened modulational instability \citep{janssen09,onorato11}  and Shrira \& Slunyaev \cite{shrira14b} found that trapped waves by a jet current can also lead to an enhanced formation probability of rogue waves 
while \citet{Hjelmervik09} found that a wave impinging on an opposing jet has increased significant wave height, but decreased kurtosis, and \emph{vice versa}.

The aforementioned works have focused on a current whose velocity profile does not have significant gradients in the vertical direction. Among the studies of waves in a horizontally uniform and depth varying current, a majority have examined waves propagating along or against currents which vary linearly with depth, which in two dimensions permits the use of a velocity potential \cite{Ellingsen16}, considerably simplifying the analytical treatment \cite{Dalrymple74,thomas12,touboul16,liao17,Curtis18,hsu18}.
The assumption of a linearly varying current also results in significant simplification of the continuity and Euler momentum equations in three dimensions, based on which a second-order theory for three-dimensional waves was developed by 
\citet{Akselsen19}. 
A uniform vorticity plays a significant role in both the sideband instability and modulational growth rate for weakly nonlinear unidirectional Stokes waves \cite{baumstein98, thomas12}. A positive vorticity, which corresponds to a following current --- i.e. $U(z)>0$ and  $U'(z)<0$ with $U(z)$ the current oriented along the wave propagation direction, $z$ the vertical
coordinate,
and a prime denotes the derivative 
--- can remove the modulational instability altogether, demonstrated experimentally by \citet{steer20} 
and \citet{pizzo23}
(the definition of positive/negative shear in ref.~\cite{steer20} is different from ours due to a different choice of the coordinate system). 
\citet{francius17} have extended \cite{thomas12} to two-dimensional Stokes waves where new quartet and quintet instabilities have been discovered arising from the presence of a uniform vorticity, while \citet{abrashkin17} derived a nonlinear Schr\"{o}dinger equation for arbitrary, weak vertical shear in a Lagrangian framework,
generalized in ref.~\cite{pizzo23}.

Realistic natural currents have non-zero curvature in the depth direction which leads to additional effects on wave properties. A number of works, e.g., \cite{voronovich76,quinn17,Banihashemi17,li19,banihashemi19}, have demonstrated the importance of the depth-varying curvature of a current profile in the wave action equation. Effects of the curvature are wavenumber- and depth-dependent, leading to considerable deviations of the direction and speed of the propagation of wave energy from the cases where the curvature has been neglected \citep{banihashemi19}. Experimental studies, e.g. \cite{Cummins94,Waseda15,Smeltzer19}, have confirmed the importance of curvature in wave modelling. Cummins \& Swan \cite{Cummins94} carried out an experimental study of irregular waves propagating in an arbitrarily depth varying current and the wave spectra measured showed significant differences from those in a uniform and magnitude-equivalent current.
It was concluded by Waseda {\it et al.} \cite{Waseda15} from experiments that the variability of 
the ambient current affected the third-order resonant interaction of wave quartets more than its mean profile did. In field observations, ocean currents are found to have considerable effect on the significant wave height \citep{Ardhuin17}, estimation of Stokes drift and particle trajectories \citep{Ardhuin09}, and the dissipation of waves through breaking \citep{Zippel17}.

The objective of the paper is twofold. Firstly, we present a new framework to allow for the interaction of weakly nonlinear surface gravity waves and a vertically sheared current, generalising the work of Longuet-Higgins \cite{longuet62}. 
Secondly, we implement the new theory numerically to study how a current profile's shear and curvature affect wave statistics, e.g., wave crest distribution and skewness of the surface elevation of random waves. 

We highlight that the new framework presented in this paper does not rely on 
assumptions of weak vertical shear (such as \citet{Stewart74,skop87,kirby89,zakharov90}) 
or weak curvature (or `near-potentiality', e.g., \citet{shrira93} and \citet{Ellingsen17}). 
Although these simplifying assumption may be applicable to most realistic situations in the open ocean, their validity should not be taken for granted, and must be properly ascertained \cite{Ellingsen17}. 
Indeed the  shear of a current can be strong in oceanic and coastal waters. For example, a wind-driven shear current in the top few centimetres can have very strong shear (e.g. \cite{laxague17,laxague18}) and the surface current typically takes values $\sim 3\%$ of the wind speed \cite{wu83}. Estuarine tidal flow has been found to be very strongly sheared, for instance the Mouth of the Columbia River which we use as example herein \cite{kilcher10,Zippel17}. 
We therefore choose to use the numerical Direct Integration Method (DIM) proposed by \citet{li19} to calculate the linear wave surface and velocity fields, being equally applicable to any horizontally-uniform depth-dependent current profile regardless of its magnitude, shear, and curvature. As detailed in \citet{li19}, the computational cost of the DIM is comparable to that using analytical approximations which involve integration over the water column  \cite{Stewart74, skop87,kirby89,Ellingsen17}, and unlike the aforementioned approximations, it provides an error estimate at little extra cost. The computer code used to generate the results presented in this paper is included as supplementary material online.

This paper is laid out as follows. A second-order theory based on a perturbation expansion, the Direct Integration Method for linear waves \citep{li19}, and  double Fourier integrals for the second-order bound waves is presented in \S\ref{sec:theory}.  Using the assumption of narrow-banded waves the shear current-modified wave statistics (e.g., skewness and the exceedance probability of wave crest) are derived in \S\ref{sec:nbapprox}.  With the numerical implementation of the theory detailed in \S \ref{sec:num}, weakly nonlinear waves in a random sea are examined in \S\ref{sec:result}, for which the linear wave amplitude and phase used for random wave realisations are assumed to follow a Rayleigh distribution and a uniform distribution, respectively, following \citet{Tucker84}.  


\section{Theoretical description and methodology} 
\label{sec:theory}
\subsection{Problem statement} \label{sub:problem}
We consider three-dimensional surface gravity waves atop a background flow in deep water. Incompressible and inviscid fluids are assumed and the surface tension has been neglected for simplicity.  The background flow propagates in the horizontal plane and varies with depth (i.e. vertically sheared). 
Its 3-dimensional velocity vector is described by $\bU_3^*(z^*) = (\bU^*(z^*), 0)$, with $\bU^*$ the velocity vector in the horizontal plane,  $z^*$ the upward axis, and a vanishing vertical component. Dimensional variables are marked with an asterisk. 
A Cartesian coordinate system is chosen and the still water surface in the absence of waves and flow is located at $z^*=0$. The surface elevation due to the background flow in the absence of surface waves is described by $z^*=\eta^*$, which is assumed known and whose spatial and temporal variations are comparably negligible to the wave perturbed fields. Neglecting the influence of surface waves on the background flow field, the system of surface waves in a background flow can be described by the continuity and Euler momentum equations as follows (see, e.g., \cite{peregrine76})
\begin{linenomath}
\begin{align} 
     \nabla_3^* \cdot \bV_3^* =~&  0,
     \\
     \p_{t^*} \bV_3^* + (\bV_3^*\cdot\nabla_3^*) \bU_3^* +(\bU_3^*\cdot\nabla_3^*) \bV_3^*+  \nabla_3^* \left(P^*/\rho+gz^*\right) 
              = ~& -(\bV_3^*\cdot \nabla_3^*)\bV_3^*,
\end{align}
\end{linenomath}
for $-\infty<z^*<\zeta^*+\eta^*$. 
Here $\nabla_3^* = (\nabla^*,\p_{z^*})$ denotes the gradient operator in three dimensions and $\nabla^* =(\p_{x^*},\p_{y^*})$ the gradient in the horizontal plane; $\bV_3^*= (\bu^*,w^*)$ denotes the velocity field  due to surface waves in the presence of the background flow, with $\bu^*$ and $w^*$ the velocity vector in the horizontal plane and vertical component, respectively, $\bx^*$ the position vector in the horizontal plane, and $t^*$ is time; $P^*$ denotes the total pressure; $\rho$ and $g$ denote the fluid density and gravitational acceleration, respectively; $\zeta^*(\bx^*,t^*)$ denotes the surface elevation due to additional surface waves in the presence of the background flow, $\bU_3^*$.

We choose the characteristic length $L_c^*$ and velocity $u_c^*$ to nondimensionalize the variables. In all cases we consider in \S\ref{sec:num}, a wave frequency spectrum $S^*(\omega^*)$ is assumed which has a clear peak at a frequency $\omega_p^*$. Therefore, we form the characteristic length, $  L_c^*=g/\omega_p^{*2}$, and, characteristic velocity, $u_c^*=g/\omega_p^*$ using $g$ and $\omega_p^*$ for convenience while our specific choice does not affect the generality of the theory derived in \S\ref{sec:theory} and \ref{sec:nbapprox}. Explicitly,
\begin{equation}
  (x^*, y^*, z^*) = (x, y, z)L_c^*; ~~ t^*=\frac{L_c^*}{u_c^*}t;~~ \cV^*= u_c^* \cV;
\end{equation}
Here, $\cV$ represents any velocity component, and 
we define the wave--induced nondimensional pressure as
\be
  P= (P^*+\rho gz^*)/(\rho u_c^{*2}). 
\ee

The dimensionless continuity and Euler momentum equations become
\begin{linenomath}
\begin{align}
\nabla_3\cdot \bV_3=~&  0;\label{eq:gvn_contin}\\
\p_t \bV_3 + (\bV_3\cdot\nabla_3) \bU_3 +(\bU_3\cdot\nabla_3) \bV_3+  \nabla_3 P 
              = ~& -(\bV_3\cdot \nabla_3)\bV_3,\label{eq:gvn:euler}
\end{align}
\end{linenomath}
for $-\infty<z<\zeta+\eta$.

The governing equations \eqref{eq:gvn_contin} and \eqref{eq:gvn:euler} should be solved subject to the dynamic and kinematic boundary conditions at the surface, respectively, 
\begin{equation}\label{eq:boundary_cond}
    P-
     (\zeta+\eta) = 0~~\text{and}~
    w = ~ \p_{t}\zeta + (\bu+\bU)\cdot\nabla\zeta~~
    \text{for}~~z=\zeta+\eta, 
\end{equation}
and the deepwater seabed condition
\begin{equation} \label{eq:bc_sb} 
    (\bu,w)=~ 0~ \text{for}~~z\to -\infty.
\end{equation}

\subsection{Perturbation expansion and linear wave fields}
We seek the solution for unknown velocity ($\bV$) and elevation $(\zeta)$  of the boundary value problem described by \eqref{eq:gvn_contin} -- \eqref{eq:bc_sb} in a form of power series in wave steepness denoted by $\eps$; i.e. a so-called Stokes expansion. To leading order, they are given by
\begin{equation} \label{eq:perturb}
     [\zeta,\bu,w,P] = \eps [\zeta\fst,\bu\fst,w\fst,P\fst] +\eps^2 [\zeta\snd,\bu\snd,w\snd,P\snd], 
\end{equation}
where the terms are kept up to second order in wave steepness and the superscript `($j$)' denotes the $j$-th order in wave steepness. Inserting the perturbed solutions \eqref{eq:perturb} into the boundary value problem described by \eqref{eq:gvn_contin} -- \eqref{eq:bc_sb} and collecting the terms at the same order lead to the various boundary value problems at different orders in wave steepness. 
In the special case of linearly varying current, an explicit solution is available. We provide the expression, adapted from the solution by \citet{Akselsen19}, in appendix \ref{app:linshear}.

Linear surface elevation due to irregular surface waves can be described by 
\begin{equation} \label{eq:zeta_1st}
    \zeta\fst(\bx,t) =  \mR\left[
                        \dfrac{1}{4\pi^2}
                       \int
                       |\hat{\zeta}(\bk)|\rme^{\rmi \psi(\bk,\bx,t)} \rmd \bk
                       \right], 
\end{equation}
where $\mR$ denotes the real part, $\bk$ denotes a wavenumber vector in the horizontal plane,  $\hat{\zeta}(\bk)$ denotes the linear wave elevation transformed in the Fourier $\bk$ plane, $\psi(\bk,\bx,t) = \bk\cdot\bx -\omega(\bk)t + \theta(\bk)
$ denotes the rapidly varying phase with $\theta(\bk)$ the initial phase (angle) of the complex elevation $\hat{\zeta}(\bk)$ at the origin, $\omega(\bk)$ denotes the angular frequency of wave $\bk$. Integration is over the whole $\bk$ plane. Without the detailed derivations, this paper employs the Direct Integration Method (DIM) developed by \citet{li19}, which provides a shear-modified dispersion relation $\omega=\omega(\bk)$. The dispersion relation is solved numerically together with the linear wave fields $\bu\fst$, $w\fst$, and $P\fst$.

The linear velocity and pressure in the physical plane can be obtained through an inverse Fourier transform as follows
\begin{linenomath}
\begin{align}\label{eq:linear_sol}
    \left[
    \begin{array}{c}
        \bu\fst(\bx,z,t) \\
        w\fst(\bx,z,t)\\
        P\fst(\bx,z,t)
    \end{array}
    \right]
    =
    \mR\left\{
    \dfrac{1}{4\pi^2}
    \int
    \left[
        \begin{array}{c}
        \hat{\bu}\fst(\bk,z) \\
        \hat{w}\fst(\bk,z)\\
        \hat{P}\fst(\bk,z)
    \end{array}
    \right]\rme^{\rmi\psi(\bk,\bx,t)}
                          \rmd \bk
                          \right\}.
\end{align}
\end{linenomath}

Arbitrary linear wave fields can then be constructed by adding monochromatic components together, in the manner of Fourier transformation.
We will not consider changes in mean water level herein and set $\eta=0$ henceforth.

%

\subsection{Second-order equations of motions} \label{sec:second_ord}
Inserting the solution for unknown velocity ($\bV$) and surface elevation ($\zeta$) in a form of power series given by \eqref{eq:perturb} into the boundary value problem described by \eqref{eq:gvn_contin}--\eqref{eq:bc_sb}, collecting the terms at second order in wave steepness, and eliminating the horizontal velocity ($\bu\snd$) and pressure ($P\snd$) at second order leads to the following equations
\begin{linenomath}
\begin{subequations}
\begin{align}
    (\p_t+\bU\cdot\nabla) \nabla_3^2 w\snd - \bU''\cdot\nabla w\snd = ~& \mathcal{N}\snd(\bx,z,t),
    \label{eq:w2nd_xyzt}
\end{align}
for $-\infty<z<\zeta$, 
\begin{align}
     (\p_{t}+\bU\cdot\nabla)^2\p_zw\snd - \bU'\cdot(\p_t+\bU\cdot\nabla)\nabla w\snd - 
    \nabla^2 w\snd =~&  \mathcal{F}\snd(\bx,z,t)~
    \text{for}~z=0, 
    \label{eq:w2nd_bcsf}
    \\
    w\snd = ~& 0~\text{for}~z\to-\infty,
    \label{eq:w2nd_bcsb}
\end{align}
\end{subequations}
\end{linenomath}
where  $\bU''=\p_{zz}\bU$,  the forcing terms, $ \mathcal{N}\snd$ and $\mathcal{F}\snd$, on the right hand side of \eqref{eq:w2nd_xyzt} and \eqref{eq:w2nd_bcsf} are functions of linear wave fields and are given by
\begin{subequations}
\begin{align}
    \mathcal{N}\snd =~& \nabla\cdot\left[ (\bV\fst\cdot\nabla_3)\bu\fst \right]'
    - \nabla^2\left[ (\bV\fst\cdot\nabla_3)w\fst \right], 
    \\
    \mathcal{F}\snd =~& -\nabla^2(\bu\fst\cdot\nabla\zeta\fst)
    - [\nabla^2(\partial_t+\bU\cdot\nabla) {P\fst}'-\nabla^2 w{\fst}']\zeta
    - \zeta\fst\nabla^2(\bU'\cdot\nabla)P\fst
    \notag \\
    &+ (\partial_t+\bU\cdot\nabla)\nabla\cdot[(\bV\fst\cdot\nabla_3)\bu\fst]
    , 
\end{align}
\end{subequations}
with notation $(\cdots)^\prime\equiv \partial_z(\cdots)$.
Inserting the linear solution from \eqref{eq:linear_sol}, the forcing term is then 
\begin{subequations} \label{eq:ifft_Ns}
\begin{align}
    \mathcal{N}\snd =~& \mR
    \left[
    \dfrac{1}{16\pi^4}
      \iint
        \hmN\snd(\bk_1,\bk_2, \bx,z,t)
    \rmd \bk_1\rmd\bk_2 
    \right],
    \\
    \mathcal{F}\snd =~& 
    \mR
    \left[
    \dfrac{1}{16\pi^4}
      \iint
        \hmF\snd(\bk_1,\bk_2,\bx,z,t)
    \rmd \bk_1\rmd\bk_2 
    \right],
    %
\end{align}
where $\bk_1$ and $\bk_2$ denote the wave vector of two different linear wave trains;  the forcing terms in the Fourier space are decomposed into the two types of second--order wave interactions as
(see, e.g., \cite{hasselmann62,longuet62})
\begin{align}
    \hmN\snd=~&  
    \hmN\snd_{+}(\bk_1,\bk_2,z)\rme^{\rmi(\psi_1+\psi_2)}
    +\hmN\snd_{-}(\bk_1,\bk_2,z)\rme^{\rmi(\psi_1-\psi_2)},
    \label{eq:hmn_def}
    \\
    \hmF\snd =~& 
    \hmF\snd_{+}(\bk_1,\bk_2,z)\rme^{\rmi(\psi_1+\psi_2)}
    +\hmF\snd_{-}(\bk_1,\bk_2,z)\rme^{\rmi(\psi_1-\psi_2)},
    \label{eq:hmf_def}
\end{align}
\end{subequations}
where the subscripts `+' or `-' denote the components for the superharmonics and subharmonics, respectively;  the wave phases are denoted with shorthand: $\psi_j=\psi(\bk_j,\bx,t)$
; and the lengthy expressions of $\hmN_\pm$ and $\hmF_\pm$ are given in Appendix \ref{app:forcing}.

With the linear velocity fields solved for by using the DIM \citep{li19}, the second-order equations \eqref{eq:w2nd_xyzt}-- \eqref{eq:w2nd_bcsb} for the vertical velocity $w\snd$ can be solved numerically in Fourier space. Due to the interaction of different wave components and the main harmonic components of the forcing terms (i.e. $ \mathcal{N}\snd$ and $\mathcal{F}\snd$) in the Fourier plane, the second-order vertical velocity 
%
\begin{equation}
    w\snd(\bx,z,t) =
    \mR\left[
    \dfrac{1}{16\pi^4} 
     \iint \hat{w}\snd(\bk_1,\bk_2,\bx,z,t) \rmd \bk_1\rmd \bk_2
     \right]. 
\end{equation}
We can also decompose $\hw\snd$ in terms corresponding to the two types of second--order wave interactions as
\begin{equation}
    \hat{w}\snd(\bk_1,\bk_2,z,\bx,t) =  
    \hat{w}\snd_{+}(\bk_1,\bk_2,z)\rme^{\rmi(\psi_1+\psi_2)}
    + \hat{w}\snd_{-}(\bk_1,\bk_2,z)\rme^{\rmi(\psi_1-\psi_2)},
    \label{eq:hwsnd_spm}
\end{equation}
Each component on the right hand side of \eqref{eq:hwsnd_spm} for $\hat{w}\snd$ can be solved for numerically from the boundary value problem as follows
\begin{linenomath}
\begin{subequations}\label{eq:hw_2nd}
\begin{align}
    {\hat{w}^{(2)\prime\prime}_\pm} - \left(|\bk_\pm|^2 +\dfrac{\bk_\pm\cdot\bU''}{\bk_\pm\cdot\bU-\omega_\pm}\right) \hat{w}\snd_\pm
    =~&  \dfrac{\hmN_\pm\snd}{\bk_\pm\cdot\bU-\omega_\pm},
\end{align}
for $-\infty<z<0$, where $\bk_\pm=\bk_1\pm\bk_2$, $\omega_\pm =\omega(\bk_1)\pm\omega(\bk_2)$, and boundary conditions
\begin{align}\label{eq:cbc_2nd}
     -(\bk_\pm\cdot\bU -\omega_\pm)^2 \p_z\hat{w}_\pm\snd +\big[
     \bk_\pm\cdot\bU'(\bk_\pm\cdot\bU-\omega_\pm)
     +
     |\bk_\pm|^2 \big]
     \hat{w}_\pm\snd = ~& 
     \hmF\snd_\pm(\bk_\pm,z)
     ~\text{for}~z=\eta,
     \\
     \hat{w}_\pm\snd =~&  0~\text{for}~z\to -\infty.
\end{align}
\end{subequations}
\end{linenomath}
 In our problem setting the waves obtained from the second-order boundary value problem \specialeqref{eq:hw_2nd}{a,b,c} are bound since they do not satisfy the linear dispersion relation and can only propagate together with their linear free contents. 
Moreover, with the linear free waves obtained,  the second-order ordinary equation \specialeqref{eq:hw_2nd}{a} with two boundary conditions \specialeqref{eq:hw_2nd}{b,c} can be solved for numerically with a finite difference method where a central Euler approximation to the second-order derivative, ${\hat{w}_\pm}^{(2)\prime\prime}$, was used in this paper.  Especially for directionally spread irregular waves in a random sea, we remark that the numerical estimation of double Fourier integrals in a form as \specialeqref{eq:ifft_Ns}{a,b} is computationally expensive for statistical analysis. Nevertheless, the framework developed here can be easily reformulated such that a pseudo-spectral method for the second-order interaction of waves in a vertically sheared current can be used,  following papers, e.g., \cite{dommermuth87} and \cite{west87} for a high-order spectral method and \cite{li21b} for a semianalytical approach. In doing so, it allows for reducing the computational operations of $\mathcal{O}(N_g^2)$ to $\mathcal{O}(N_g\mathrm{In}N_g)$, with $N_g$ the total number of discrete points chosen for the grid of a computational domain. 

The second-order wave surface elevation $\zeta\snd$ can be obtained from the following kinematic boundary condition
\begin{align}
\label{eq:bc4zeta2nd}
     (\p_{t}+\bU\cdot\nabla)\zeta\snd = w\snd +\zeta\fst {w^{(1)\prime}}- \dfrac{1}{2} \bU'\cdot\nabla\big(\zeta\fst\big)^2 -\bu\fst\cdot\nabla\zeta\fst, 
\end{align}
which leads to the surface elevation $\zeta\snd$ given by
\begin{subequations}\label{eq:zeta_snd}
\begin{align}
     \zeta\snd(\bx,t)=~& \mR\left[ \dfrac{1}{16\pi^2}
     \iint\hzeta\snd(\bk_1,\bk_2;\bx,t)\rmd \bk_1\rmd\bk_2\right]~
     \text{with}~
     \\
    \hzeta\snd=~& \hzeta\snd_+(\bk_1,\bk_2)\rme^{\rmi(\psi_1+\psi_2)} + \hzeta\snd_-(\bk_1,\bk_2)\rme^{\rmi(\psi_1-\psi_2)},
\end{align}
\end{subequations}
where the elevation $\hzeta\snd_\pm$ is obtained from \eqref{eq:bc4zeta2nd} in the Fourier plane through substituting the vertical velocity $w\snd$ and the linear wave fields $\bu\fst$ and $\zeta\fst$. It's noteworthy that for $\bk_1=\bk_2$ the super-harmonics ($\hzeta\snd_+$) reduce to the well-known second-order Stokes waves. The sub-harmonics ($\hzeta\snd_-$) become a constant, which refers to a mean water level and is ignored in our experiment.

\subsection{Notation in the frequency domain}
\label{sec:freqdom}

The theory in \S\ref{sec:theory} so far was formulated in reciprocal horizontal ($\bk$) space. Often it is more convenient in practice to use a frequency domain formulation, for instance when working with power spectra, from time series from wave buoys, say. 
In the presence of a vertically sheared current the dispersion relation $\omega=\omega(\bk)$ is anisotropic in any reference system, i.e., $\omega$ is always a function of the direction of $\bk$, not only its modulus. This introduces subtleties in interpreting nondirectional wave frequency data in the presence of a sheared current as wavelength cannot be inferred from frequency alone. We herein work in two dimensions, i.e., waves propagating with known direction either along or against the current, thus eschewing this potential complication.

The linear and quadratic-order elevations are denoted
\begin{linenomath}
\begin{subequations}\label{zeta12_f}
\begin{align}
    \zeta\fst(\bx,t) = ~& \mR\left(
    \int a(\omega) \rme^{\rmi\psi} \rmd\omega \right), 
    \\
    \zeta\snd(\bx,t) = ~& 
    \mR\left\{
    \iint 
    a_1a_2 \left[\hA^+_{12}\rme^{\rmi(\psi_1+\psi_2)} + \hA^-_{12}\rme^{\rmi(\psi_1-\psi_2)} \right]
    \rmd\omega_1\rmd\omega_2
    \right\}.
\end{align}
where $\ta(\omega)$ denotes the linear (real) amplitude of a wave with frequency $\omega$ and complex phase 
$\tpsi(\omega)=\kx-\omega t+\theta(\omega)$, where we solve the dispersion relation $\omega=\omega(\bk)$ for the wave vector with a given frequency using the DIM method as noted.
The following notations are used: $\ta_n=\ta(\omega_n)$,  $\tpsi_n=\tpsi(\omega_n)$, $\hA^\pm_{12} = \hA^\pm(\omega_1,\omega_2)$ with 
\begin{align}
    \hA^\pm(\omega_1,\omega_2) = \dfrac{|\hzeta\snd_\pm(\omega_1,\omega_2)|}{a_1a_2},
\end{align}
\end{subequations}
\end{linenomath}
where $\hzeta\snd_\pm$ was given by \specialeqref{eq:zeta_snd}{b} with the difference that it is expressed here in the frequency domain instead.

\section{Waves of a narrow bandwidth}
\label{sec:nbapprox}
In this section we present the skewness and probability density function of the surface displacement and wave crests in the special case where the bandwidth of the wave spectrum is narrow. 
We now use the frequency-domain formulation of \S\ref{sec:freqdom}. Consider an ensemble of waves described in the form \eqref{zeta12_f} where the amplitude $a(\omega)$ becomes an independent random  variable denoted by $\tilde{a}(\omega)$ which follows a Rayleigh distribution based on a spectrum $S(\omega)$ and where the phase $\theta$ becomes another independent random variable, $\ttht$, which is uniformly distributed in the range $[0, 2\pi\rangle$. Therefore, $\zeta(\bx,t)\to \tzt(\tilde{a}(\omega),\tilde{\theta}(\omega))$. 
The $j$-th spectral moment $m_j$ is defined as
\begin{linenomath}
\begin{align}\label{eq:mj}
    m_j = \int\omega^jS(\omega)\rmd\omega; ~~j\in \{0,1,2,...\}.
\end{align}
\end{linenomath}
Assuming zero mean water level as before, the standard deviation, $\sigma$, and skewness, $\lambda_3$, of the surface elevation are 
\begin{equation}\label{eq:Defs_sigma_skew}
    \sigma = \sqrt{\langle \tzt^2 \rangle}
    ~~\text{and}~~
    \lambda_3 =
    \langle \tzt^3 \rangle/\sigma^3,
    \specialnumber{a,b}
\end{equation}
where $\langle...\rangle$ denotes the expectation value of random variables. Assuming the energy spectrum $S(\omega)$ to have a narrow bandwidth ($\nu=\sqrt{1-{m_2^2}/{(m_0m_4)}}\ll 1$), 
we follow the detailed derivations of \citet{fedele09} using the elevations \specialeqref{zeta12_f}{a,b}, and obtain to $\mo(\eps)$
\begin{equation}\label{eq:sigma_skew}
    \sigma^2 = m_0
    ~\text{and}~
    \lambda_3 = 6\sigma \hA^+_{mm},
    \specialnumber{a,b}
\end{equation}
where $\hA^+_{mm} = \hA(\omega_m,\omega_m)$ denotes the second-order superharmonic amplitude of the spectral mean wave, with $\omega_m$ the spectral mean frequency given by
\begin{linenomath}
\begin{align} \label{eq:omg_m}
    \omega_m= m_1/m_0. 
\end{align}
\end{linenomath}
The skewness given by \specialeqref{eq:sigma_skew}{b} agrees with \citet{fedele09}, \citet{Srokosz86} and \citet{li21} for waves in the absence of a shear current, which is clear when noting that the superharmonic amplitude $\hA^+_{mm}$ can be written as ${k_{m}}/{2}\equiv \omega_m^2/(2g)$ in the case for second-order deepwater Stokes waves (see, e.g., \cite{longuet62}).  
It is different from \citet{fedele09} to the extent that it does not account for the effect of bandwidth as it is not so straightforward due to a shear current. Nevertheless, it allows us to take into account the effect of a shear current to some extent. Especially, if all linear waves follow the same power energy spectrum with a narrow bandwidth, i.e., $m_j$ are identical for all cases, the spectral mean given by \eqref{eq:omg_m} is identical regardless of a shear current. A shear current affects the skewness given by \specialeqref{eq:sigma_skew}{b} through the second-order superharmonic amplitude of the spectral mean wave, compared with the cases in the absence. 

Following \citet{Longuet-Higgins63}, we obtain that the normalized surface displacements follow the distribution
\begin{equation}\label{pdf_zeta}
  p_\zeta(
  \tzt
  )=\dfrac{1}{\sqrt{2\pi}}\mathrm{e}^{-\tzt^2/2}
    \left [1+\dfrac{\lambda_3}{6}
  \tzt
  (
  \tzt
  ^2-3)\right ].
\end{equation}
For linear waves, where $\lambda_3=0$, expression \eqref{pdf_zeta} becomes a Gaussian distribution. Different from \citet{Longuet-Higgins63}, the probability density function given by \eqref{pdf_zeta} can account for the effect of a shear current due to that the skewness $\lambda_3$ is modified according to
\specialeqref{eq:sigma_skew}{b} which considers the effect of a shear current. 

Similarly, following \citet{Forristall00}, the `exceedance probability', i.e., the probability that a randomly chosen wave crest $X_c$ exceeds the value $\tzt_c$, is found as
\begin{linenomath}
\begin{equation}\label{eq:epd}
P(X_c>\tilde{\zeta_c})=
\exp\left[-\dfrac{1}{8(\hA_{mm}^{+}\sigma)^2}\left(\sqrt{1 + \dfrac{16\tzt_c}{H_s}\hA_{mm}^+\sigma} -1 \right)^2
\right],
\end{equation}
\end{linenomath}
where $H_s$ is the significant wave height. The exceedance probability given by \eqref{eq:epd} agrees with (2.12) by \citet{li21} with the same chosen notations  whereas the main difference lies in that 
the effect of a shear current enters here via 
the superharmonic amplitude of the spectral mean wave, $\hA_{mm}^+$. In the limit of infinitesimal wave, i.e., $m_0\to0^+$, the exceedance probability of wave crest becomes
\begin{linenomath}
\begin{align}\label{eq:rayepd}
    P(X_c>\tilde{\zeta_c})= \exp\left(-8\dfrac{\tzt_c^2}{H_s^2} \right),
\end{align} 
\end{linenomath}
which is the Rayleigh distribution as expected. For second-order deepwater Stokes waves in the absence of a shear current which admits $\hA_{mm}^+ = k_m/2\equiv \omega_m^2/(2g)$, the exceedance probability given by \eqref{eq:epd} is identical to eq.(4) in \citet{Forristall00}.   We will refer repeatedly to \eqref{pdf_zeta} and \eqref{eq:epd} in section \ref{sec:maxcrest}.

\section{Numerical setup}\label{sec:num}
In our simulations, 
we generate two-dimensional (long-crested or uni-directional) waves from realistic spectra. 
Doing so implies that the possible triad resonant interactions in three dimensions  considered in previous papers, e.g., \cite{Craik68,zakharov90,Akselsen19} are assumed negligible in the simulations.
We choose the characteristic velocity, $u_c^*= g/\omega_p^*$, as defined in \S\ref{sub:problem}. 
Here, $\omega_p^*$ is the peak frequency of the spectrum; although $\omega_p=1$ by definition, we find it instructive to retain it in some equations below.

We begin by defining the terms following and opposing shear for two-dimensional flow, i.e., where all waves propagate parallel or antiparallel to the mean current. We will assume that waves travel along the positive $x$ axis. We then define
\begin{itemize}
    \item[] $\bullet$ ~~Following shear: $U'(z)<0$; ~~~~~~~~~~~~
      $\bullet$ ~~Opposing shear: $U'(z)>0$.
\end{itemize}
Following (opposing) shear corresponds to the situation where the flow increases (decreases) in the direction of propagation with increasing depth. 

Note carefully the distinction between following (opposing) shear and following (opposing) current. When seen in an Earth--fixed reference system, currents in nature are often strongest near the surface and decrease to zero at larger depths, such as in the Columbia River Mouth current we regard in section \ref{sec:C_R}. 
In such a case a ``following surface current'' $U(z)>0$ would correspond to opposing shear and {\it vice versa}. For clarity of comparison between cases we shall work in a surface-following frame and, therefore, assume $U(0)= 0$, in which case following shear implies positive $U(z)$ for a monotonically varying $U$. Doing so allows us to focus only on the effects due to the profile shear and curvature of a current.

\subsection{Realisation of random seas states for linear waves}
\label{sec:randomSeaState}

We follow Tayfun \cite{Tayfun80} and Tucker {\it et al.} \cite{Tucker84} for the realisation of random sea states, which assumes Rayleigh distributed amplitude of linear waves and uniformly distributed wave phases in the range of $[0,2\pi\rangle$. 
The energy spectrum we choose for computation is JONSWAP spectrum \cite{hasselmann73} with a peak enhancement (or peakedness) parameter of $\gamma =3.3$ and moderately narrow bandwidth\cite{Dysthe05,Socquet-Juglard05}, which is shown in figure \ref{fig:spec_Uz}(a).

The JONSWAP spectrum is given by (recall that $\omega_p=1$) 
\newcommand{\talG}{\tilde{\alpha}_G}
\newcommand{\talJ}{\tilde{\alpha}_J}
\begin{equation}\label{eq:jonswap}
    S_J(\omega)=
    \dfrac{\talJ}{
    \omega^5}\exp{\left[-1.25\omega^{-4}\right]}\gamma^{b(\omega)},
\end{equation}
where the peak enhancement factor $\gamma$ appears with an exponent 
\begin{equation}
 b(\omega)=\exp{\left[-\dfrac{(\omega-1)^2}{2\sigma_J^2}\right]},
\end{equation}
and
\begin{equation}
\sigma_J= 
\begin{cases}
 0.07,& \omega \leq 1 \\
 0.09,& \omega > 1. 
\end{cases}
\end{equation}

The parameter  $\Tilde{\alpha}_J$ is chosen such that 
the JONSWAP spectrum is fixed for all numerical cases, i.e., independent of a current profile. 
The frequency is truncated at $0.01\omega_p$ and $2.6\omega_p$. The bandwidth parameter is defined as 
\begin{equation}\label{eq:nu}
    \nu=\sqrt{1-\frac{m_2^2}{m_0m_4}}
\end{equation}
and here $\nu=0.5284$. For another widely used bandwidth parameter $\nu_{L}=\sqrt{m_0m_2/m_1^2-1}$ proposed by \citet{Longuet75}, the value becomes $0.2689$.
We choose bulk steepness 
$\epsilon=\half H_s=0.14$ in all cases.
As noted, the  peak frequency ($\omega_p=1$), significant wave height ($H_s$), and the moments ($m_j$) of the JONSWAP spectrum are fixed for all cases, regardless of the profile of a shear current. However, the spectrum peak wavenumber $k_p\equiv k(\omega_p)=k(1)\neq 1$ in the presence of a current, since the linear dispersion relation $k(\omega)$ depends on $U(z)$,
as explained in \S\ref{sec:theory} and \S\ref{sec:nbapprox}.  

Once the input spectrum is determined, the amplitudes $a_i$ of a total of $N_s$ linear elementary waves are generated with 
a prescribed significant wave height%
, with
\begin{equation}
    \sum_{i=1}^{N_s} \dfrac{\tilde{a}_i^2}{2}=\int_\omega S(\omega) \rmd \omega
    ~\text{and}~
    \zeta\fst(x,t)=\sum_{i=1}^{N_s} \tilde{a}_i\cos(k_ix-\omega_it+\ttht_i),
\end{equation}
where the energy spectrum is discretised with unequal frequency intervals and an identical area of $N_s$ energy bins (i.e., constant $S(\omega_i)\rmd \omega_i$). For a train of random waves, we assume the amplitude $\tilde{a}_i$ follows a Rayleigh  distribution and the phase $\ttht$  a uniform distribution in the range $[0,2\pi\rangle$ similar to \S\ref{sec:nbapprox} and \citet{Tayfun86}. The wave numbers $k_i$ are found numerically from $\omega_i$ using the DIM algorithm as described. We especially computed the temporal evolution of the linear surface elevation at $x=0$ and then, the second-order correction of the wave surface are calculated from \specialeqref{eq:zeta_snd}{a} and \specialeqref{eq:zeta_snd}{b}.

We also make a flow diagram of numerical implementations, which is shown in Appendix~\ref{sec:flow_diag}. In our simulations, $128$ elementary waves are generated from the relevant input wave spectra and ran from $0\leq$t$\leq5638$. 2000 realizations were simulated to assure that the skewness of the wave surface elevation was converged.

\subsection{Current profiles and cases considered}

\begin{figure}
\includegraphics[width=\textwidth]{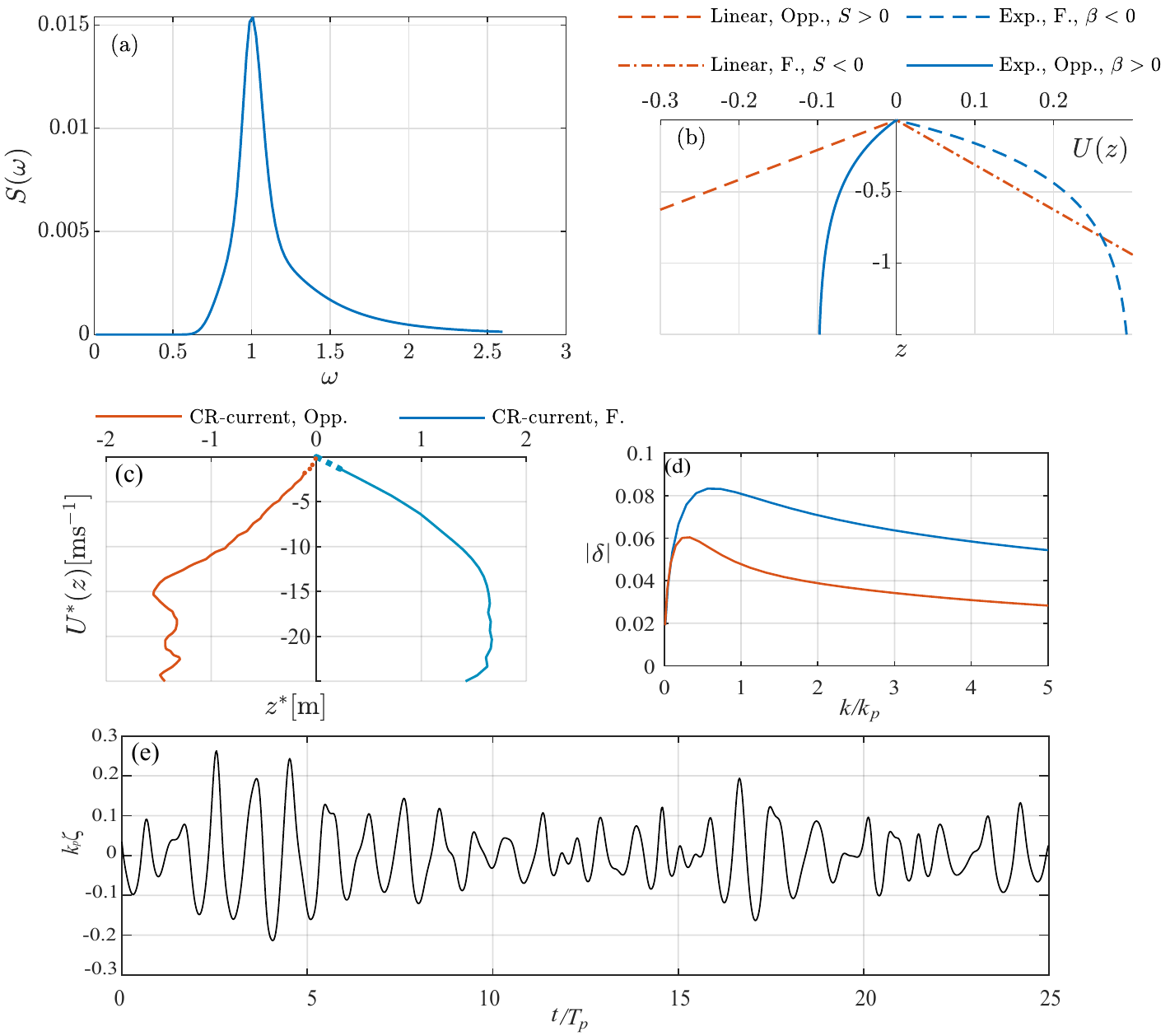}
      \caption{(a): JONSWAP power energy spectrum of linear waves with nondimensional peak frequency 
      $\omega_p=1$
      and bulk steepness $\epsilon = 0.14$; (b) examples of linear and exponential (`Exp.') shear profiles where both opposing (`Opp.') and following (`F.') shear are shown; (c) two tidal current profiles 
      from ref.\ 
      \citet{Zippel17} mearured at the mouth of Columbia river (`CR'), during ebb tide (following shear, `F.'), and flood (mostly opposing shear, `Opp.'), respectively.
      Note that in an Earth-fixed coordinate system (see Fig.~3 of \cite{Zippel17}) these correspond to opposing and following surface currents, respectively.
      Dashed lines are extrapolations from $z = 1.35$\,m to the surface%
      ; (d) wave--averaged shear 
      $|\delta(k)|$ 
      for the two profiles in panel c; 
      (e) extract of the time series of wave surface elevation for illustration, here without current.
      }
      \centering
      \label{fig:spec_Uz}
\end{figure}

We consider three different current profiles with different parameters, which are typical of the open ocean, including an exponential profile, a linearly sheared current, and one that was measured at the mouth of Columbia River from Zippel \& Thomson \cite{Zippel17}, as shown in figure \ref{fig:spec_Uz}(b) and (c). 

\newcommand{\lin}{_\textrm{lin}}

\subsubsection{Model profiles}

The exponential and linear profile of shear current are parameterized as
\begin{equation}\label{eq:bU_def}
    \bU_{\text{exp}}(z) = \beta [\exp(\alpha z)-1]\mathbf{e}_x
    ~\text{and}~
    \bU\lin = S z\mathbf{e}_x,
    \specialnumber{a,b}
\end{equation}
respectively, where $\mathbf{e}_x$ is a unit vector along the positive $x$ axis, 
the subscripts `exp' and `L' denote the exponential and linear profile, respectively, $\alpha$ ($\alpha>0$), $\beta$, and $S$ are dimensionless parameters that define the magnitude and shear strength of a current profile relative to the peak wave parameters. 
Note that we choose a reference system following the free surface so that 
 $\bU(0)=0$. 
This eschews arbitrary Doppler shift terms which would clutter the formalism, reduces the number of free parameters, and makes results from different profiles immediately comparable.
The choice also emphasizes that it is the shear $U'(z)$ and curvature $U''(z)$ which cause statistics to be altered, not the strength of the current itself. 
The surface shear is  obtained from \eqref{eq:bU_def}
\begin{equation}
   \bU'_{\text{exp}}(0)  = \alpha\beta \mathbf{e}_x
    ~\text{and}~
    \bU'\lin(0) = S\mathbf{e}_x,
    \specialnumber{a,b}
\end{equation}
which denote the profile shear of an exponential and linearly sheared current at still water surface, respectively.

Recall that following (opposing) shear correspond to $U'(z)<0$ ($>0$). %
We wish our model current to have strong, but not unreasonable vertical shear. To determine how strongly the current shear affects the dispersion of a wave of wave number $k^*$ or frequency $\omega^*$ (whichever is known), the proper parameter to consider is the wave--weighted depth-averaged shear \cite{Ellingsen17}, respectively
\begin{equation}\label{eq:delta}
    \delta =
    \frac{1}{ c_0^*} \int_{-\infty}^0 U^{*\prime}(z^*)\rme^{2k^*z^*} \rmd z^* = 
    \sqrt{k}\int_{-\infty}^0 U'(z)\rme^{2kz}\rmd z
\end{equation}
nondimensionlized as explained in Section \ref{sub:problem},
and $c_0^*=\sqrt{g/k^*}$.
Inserting $U'(z)=\alpha\beta\exp(2\alpha z)$ gives 
\begin{equation}
    |\delta| = \frac{|\alpha\beta|\sqrt{k}}{\alpha+2k},
\end{equation}
whose maximum value is found at $k=\alpha/2$ and in either case,
$    |\delta|_\text{max} = |\alpha\beta|/\sqrt{8} $. 
In the following sections we use $\alpha=2.5$ and $|\beta|\leq 0.3$ giving $|\delta|_\text{max}\lesssim 0.17$.

\subsubsection{Profile from the Mouth of Columbia River} \label{sec:CRcurrent}

The profiles of tidal currents in the Mouth of the Columbia River have been used as a test-case in a wide array of studies of wave-shear current interactions (e.g.\ \cite{Zippel17,li19,Banihashemi17,Dong12,elias12,maxwell19,campana15,lund18}) due to the availability of high quality current profile measurements \cite{Zippel17,kilcher10} and strong vertical shear. Herein we use the profiles measured by \citet{Zippel17} using an acoustic Doppler current profiler (ADCP) mounted on a drifter. 
The currents were measured between $1.35$\,m and $25$\,m depth, but we require profiles ranging all the way to the undisturbed surface level. What the profile might look like in the top $1.35$\,m is not obvious; the shear strength can drop sharply closer to the surface \cite{Kudryavtsev08}, but could also increase all the way to the top centimetres \cite{laxague18}. We use a polynomial extrapolation as shown in figure \ref{fig:spec_Uz}c; we show in appendix \ref{sec:3profile} that two other common approaches produce no discernable difference in the resulting skewness. 
The current profiles reported in \citet{Zippel17} and shown in figure Fig.~\ref{fig:skku_zip}a
are fitted with a 7th order polynomial to the surface. 
The wave-averaged dimensionless shear $\delta$ of Eq.~\eqref{eq:delta} for the two profiles in Fig.~\ref{fig:spec_Uz}c are seen in Fig.~\ref{fig:spec_Uz}d, peaking near $0.095$ for the following current. 

Note that the currents taken from \citet{Zippel17} are not extreme for the location --- the shear current used in e.g.~\citet{li19b} taken from the measurements during the RISE project \cite{kilcher10} peaks at a value $\delta\approx 0.19$, more than our strongest exponential model current. For comparison with the results of \citet{Zippel17} for ebb and flow respectively, we choose the more conservative profiles in the latter.

We remark that \citet{zakharov90} proposed a set of analytical theory for second-order wave-shear current problem with the assumptions $U'<0$ and $U_\text{max}/c\ll 1$. Here, $U_\text{max}$ and $c$ refer to the maximum velocity of shear current and phase velocity of surface wave, respectively. From Fig.\ref{fig:spec_Uz}c the parameter $U_\text{max}/c$ of Columbia River current for peak wave could reach $0.2$. 
Hence, the theory by \citet{zakharov90} is not expected to be quantitatively accurate for the Columbia River current cases considered herein.

\section{Results}\label{sec:result}

We present second order statistical quantities for waves on model shear currents, generalising a number of classical results.
The example for time series of wave surface elevation is shown in Fig. \ref{fig:spec_Uz}(e). All the statistical quantities are based on very long time series.

\subsection{The distribution of wave surface elevation}
\label{sec:elevation}
%

\begin{figure}[htb]
\includegraphics[width=0.95\linewidth]{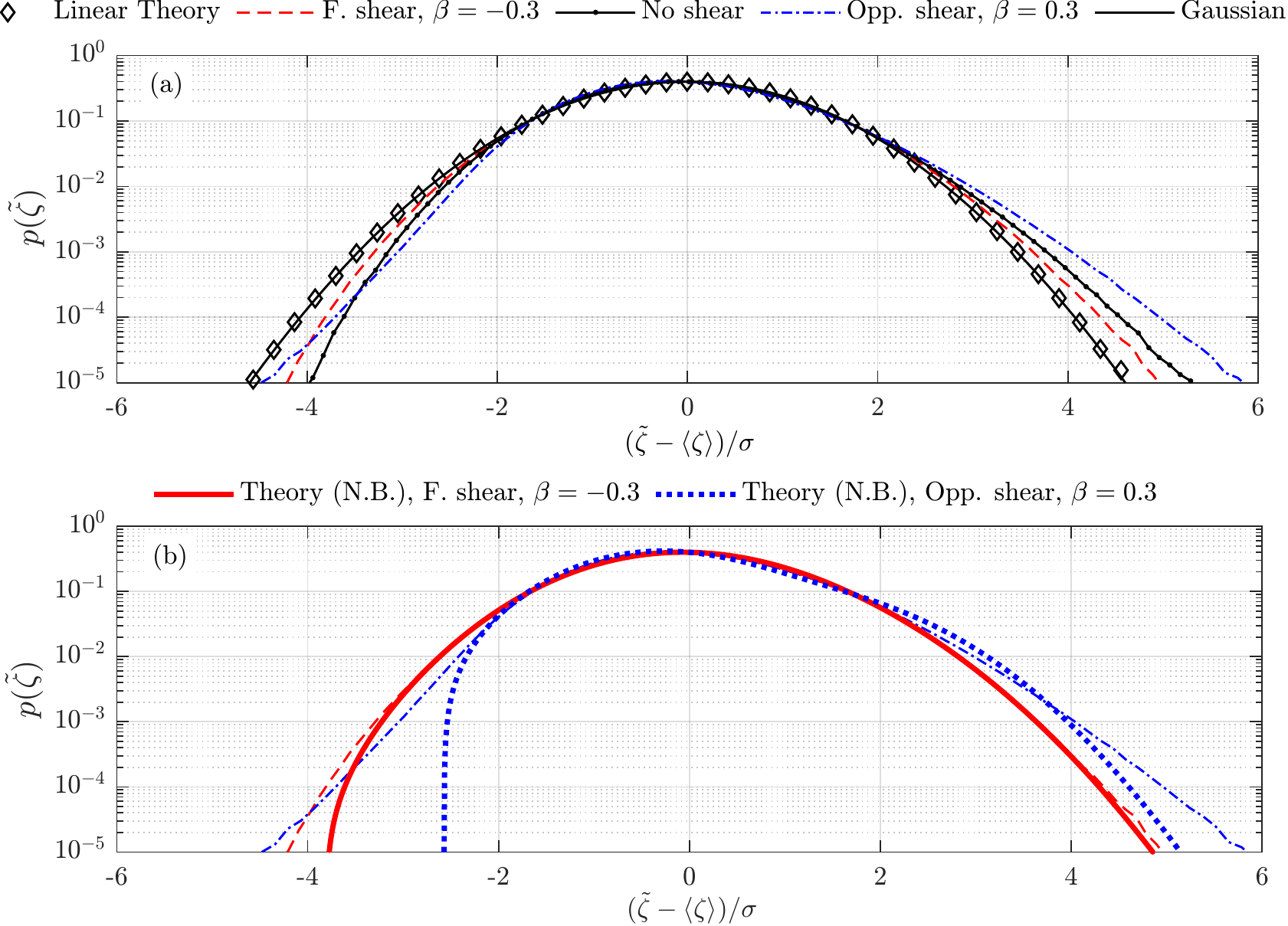}
\caption{Probability density function (PDF) of wave surface elevation 
for a moderately narrowband Gaussian input spectrum assuming the
exponential  current profile \protect\specialeqref{eq:bU_def}{a}
with $\beta$ the magnitude of the shear at a still water surface. 
Numerical results for $\beta=-0.3$ (following shear, `F. shear') and $\beta=0.3$ (opposing shear: `Opp. shear') are compared to
(a) the linear prediction and the case without current, and (b) the narrow-band (N.B.) theory based on \eqref{pdf_zeta}.
}
\label{fig:pdf_surf}
\end{figure}

In this section we examine the effects of sub-surface shear on the distribution of 
surface elevation 
to second order in steepness. We compare the case of no current to cases with following and opposing shear.
We also show comparisons of the same case with shear between the broadband and narrow-band theory presented in \S\ref{sec:theory} and \S\ref{sec:nbapprox}, respectively. %
A moderately narrowband spectrum is considered, with the linear wave field amplitudes chosen from a Gaussian distribution with zero mean and variance $\sigma^2$.

Fig.\ref{fig:pdf_surf} plots the numerically calculated PDFs of wave surface elevation in the presence of a model current (equation \eqref{eq:bU_def}a) varying exponentially with depth, comparing our numerical results based on the broad-band theory presented in \S\ref{sec:theory}, together with different theoretical predictions: a Gaussian distribution, and theoretical predictions based on a narrow-band assumption presented in \S\ref{sec:nbapprox}. We firstly discuss the results shown by Fig.(\ref{fig:pdf_surf}a). When both second-order corrections and shear are omitted, the numerically calculated PDF (diamond symbols) should coincide with the Gaussian input distribution (zero mean, variance $\sigma^2$) which indeed it does,
as expected. 
The probability of amplitudes greater than about two standard deviations from the mean are decreased for negative values (deep troughs) and increased for positive (high crests), conforming with the known properties of second-order Stokes waves: the wave crests get higher and wave troughs get flatter.

The presence of opposing shear $U'(z)>0$ enhances the wave crests and flatten the wave troughs compared to no current, while following shear current has the opposite effects. %
 The effect on second--order statistics from the shear is considerable
 in the range of larger wave crests ($>2\sigma$)  but modest for wave troughs (negative elevation) in this case. 
 
A comparison of the probability density function of surface elevation for the cases in the presence of shear is shown in Fig.(\ref{fig:pdf_surf}b) comparing the numerical results based on the full theory of \S\ref{sec:theory} and the narrow-band approximation in \S \ref{sec:nbapprox}. It is seen that the narrow-band assumption agrees with the broad-band theory up to three and two standard deviations for the cases with following (`F. shear') and opposing shear (`Opp. shear'), respectively; for following shear the approximation would be good enough for most practical purposes, except extreme statistics.  
The narrow-band approximation underestimates the probability of the most extreme events in both cases, but to very varying degrees as the figure shows.

\subsection{The distribution of wave maxima and crest height}
\label{sec:maxcrest}

The crest height is conventionally defined as the highest surface elevation reached inside discrete time intervals. Within each time interval, the surface elevation is above the mean--surface level, $\zeta>0$, i.e., delimited by consecutive zero crossings $\zeta(t)=0$ so that $\zeta'(t)>0$ ($<0$) at the beginning (end). This contrasts, in general, with a {\it surface elevation maxima}  $\zeta_m$, which is any point where $\zeta'(t)=0$ and $\zeta''(t)<0$. 
Surface elevation maxima can be negative for a broad-band spectrum, whereas 
for a sufficiently narrow
spectrum, the two are positive and coincide: every maximum is also a wave crest.

As discussed by \citet[][Chapter 2]{goda10}, when the spectrum is not narrow there is no universal and unique definition of wave height in a time series. The most common definition based on zero-crossings described above is theoretically somewhat unsatisfactory in a broadband setting; a more theoretically coherent method proposed by \citet{janssen03,janssen14} based on the envelope of $\zeta$ is also in use \cite{barbiarol19}. For theoretical derivations the envelope procedure becomes more cumbersome for weakly non-linear waves, requiring expressions for third and fourth statistical moments, needed to adequately describe a generic wave distribution. 
In the following we use the customary definition using zero-crossing, as described above, bearing in mind that the identification of individual waves, and hence its distribution of maxima, will carry some dependence on the spectral shape which vanishes in the narrow-band limit.

For a narrow frequency spectrum according to linear theory, the dimensionless wave crest heights $\tilde{\zeta_c}$, normalised by significant wave height $H_s$, is distributed according to the Rayleigh probability function as given by \eqref{eq:rayepd}.
It is difficult, however, to determine theoretically the probability distribution of crest heights if the waves have a broad frequency spectrum. Hence, \citet{Cartwright56} made a compromise by calculating the distribution of surface elevation maxima denoted by $\zeta_m$,  
adapting the theory of Rice \cite{Rice45} from in electrical signal processing
to an ocean waves setting.  
Their result based on linear theory for a broadband spectrum is
\begin{linenomath}
\begin{align}\label{eq:cart}
p(\xi)=\dfrac{1}{\sqrt{2\pi}}\nu\exp\left (-\dfrac{\xi^2}{2\nu^2}\right )+\dfrac{\xi\sqrt{1-\nu^2}}{2}& \exp\left(-\dfrac{1}{2}\xi^2\right)\left[1 + \text{erf}\left (\dfrac{\xi\sqrt{1-\nu^2}}{\sqrt{2}\nu}\right)\right],
\end{align}
\end{linenomath}
where $\xi={\zeta_m}/{\sigma}$ denotes the 
normalised 
maxima,
the bandwidth parameter $\nu$ is defined in \eqref{eq:nu},  
$m_j$ is the $j$-th moment of the energy spectrum given by \eqref{eq:mj}, 
and $\mathrm{erf}$ is the error function. 

\begin{figure}
\centering
  \includegraphics[width=0.9\textwidth]{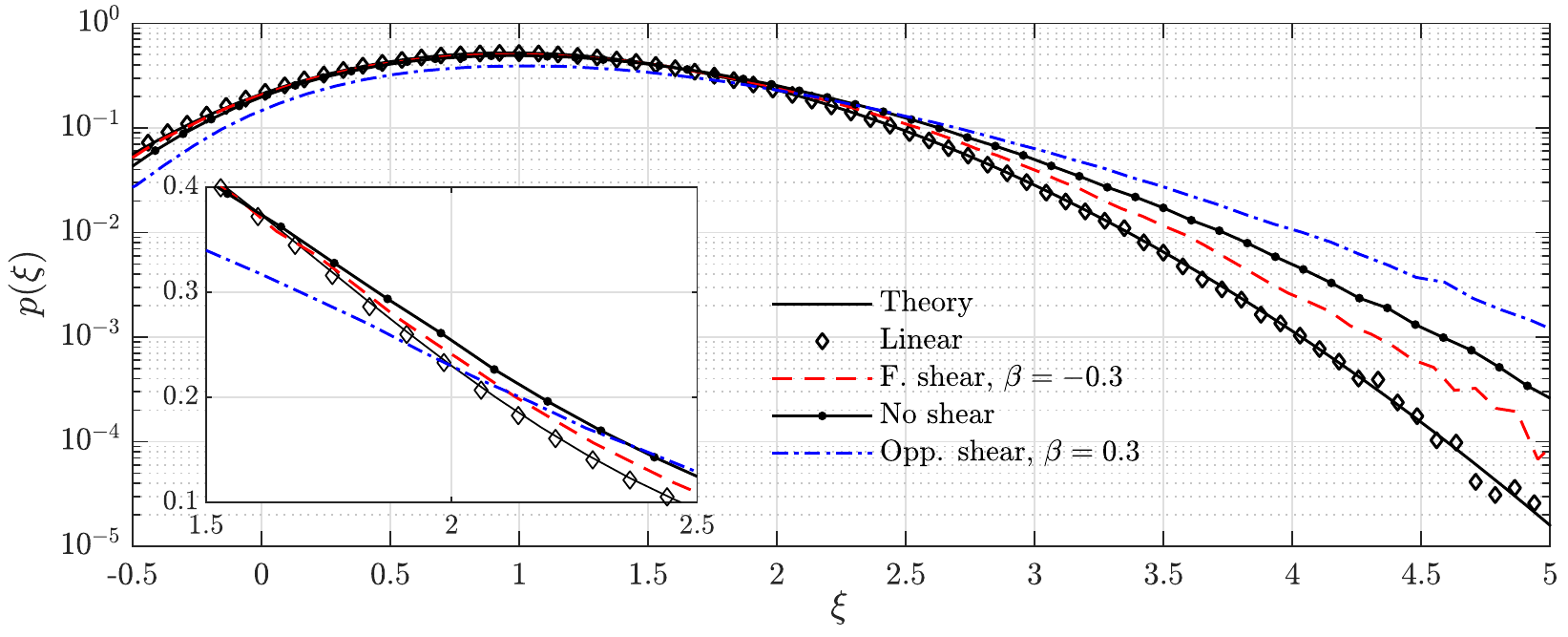}
  \caption{ Probability density function of the dimensionless maxima ($\xi=\zeta_m/\sigma$) of the wave elevation. The theoretical estimates (`Theory') are based on \eqref{eq:cart} and the other cases shown are the same as Fig.\ref{fig:pdf_surf}{a}. 
  }
  \label{fig:pdf_maxima}
\end{figure}

Fig. \ref{fig:pdf_maxima} shows the PDF of the 
surface elevation maxima for linear and nonlinear results. We also plot the theoretical estimates with \eqref{eq:cart}, which is given by solid line in the figure. When nonlinear effects and shear are both omitted, the numerically calculated PDF (diamond symbols)
should coincide with equation \eqref{eq:cart}, which indeed it does as the figure shows.
The second-order results show increased probability of large wave maxima in all cases. Notice that negative-valued surface maxima occurs for a broadband spectrum, corresponding to nonzero $p(\xi)$ for $\xi<0$. The probability of a negative maxima increases monotonically with bandwidth parameter $\nu$. 

The most prominent nonlinear effect in Fig.~\ref{fig:pdf_maxima} is for opposing shear, where probability for large maxima above approximately two standard deviations is enhanced in our simulation, whereas maxima below this threshold are made less probable. 
The current with following shear has the opposite influence. This phenomenon is consistent with the PDF of wave surface elevation studied in \S \ref{sec:elevation}.

\begin{figure}
\includegraphics[width=\linewidth]{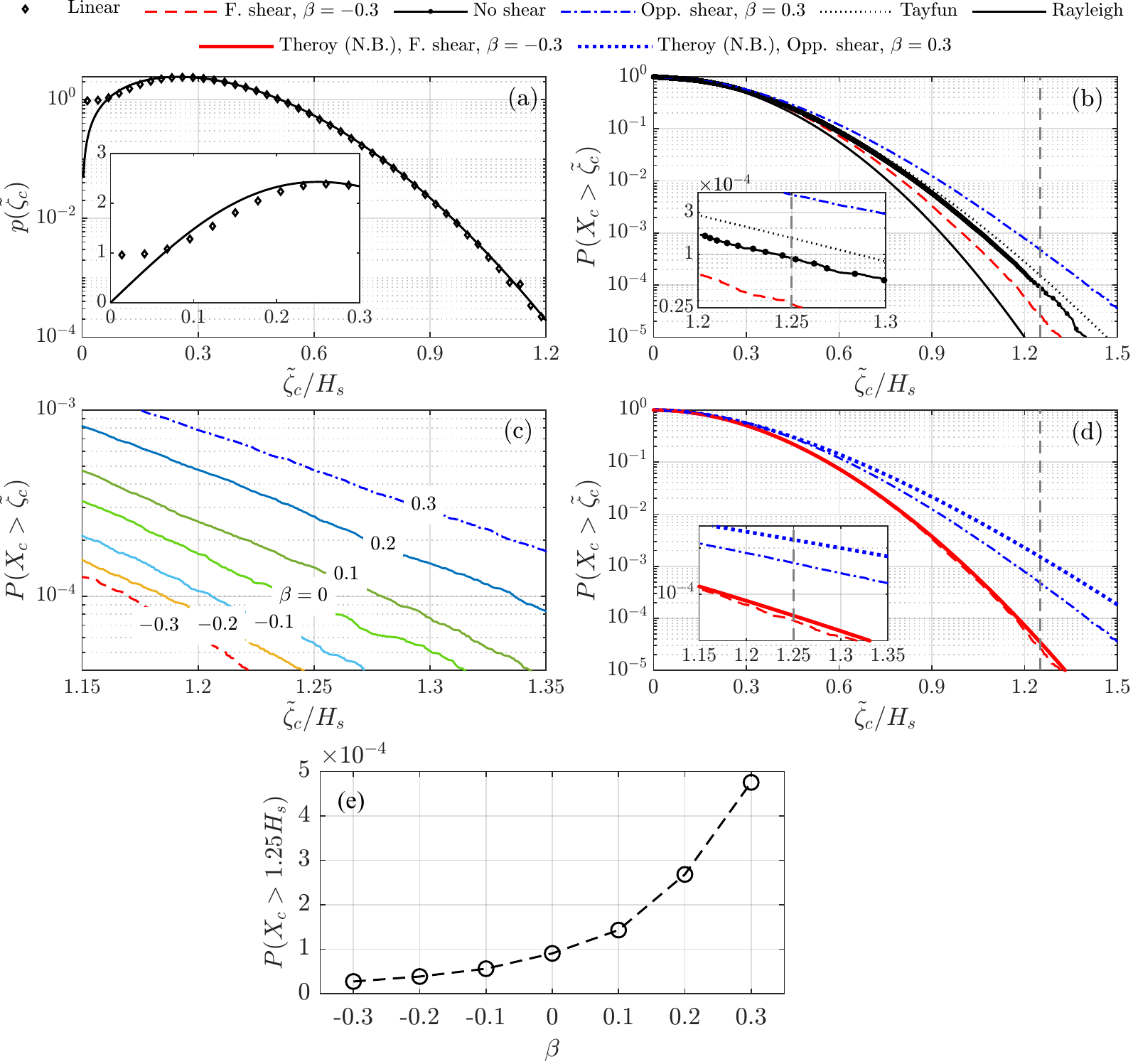} \\
\caption{
Numerically calculated probability density function (panel (a))
and exceedance probability (panels (b,c,d)) 
for wave crests. An exponential shear profile, Eq.~\protect\specialeqref{eq:bU_def}{a}, was assumed. 
(a) Linear waves based on numerical simulations and the Rayleigh probability density function;  (b,c) nonlinear wave fields for varying shear strength; (d) the broad-band and narrow-band results for cases with shear based on the theory in \S\ref{sec:theory} and \S\ref{sec:nbapprox}, respectively. We used \eqref{eq:epd} with $\beta=0$ for the Tayfun distribution.
(e) Occurrence probability of rogue wave for all the exponential shear cases in panel c.
}
\label{fig:epd}
\end{figure}

There exists a few commonly used expressions for crest height distribution obtained by empirical fitting, theoretical
considerations or parameterization \citep{Haring77,Kriebel94, Huang86, Forristall00, Kriebel91, fedele09, Prevosto00}. One example we use in this section is the distribution derived by Tayfun \cite{Tayfun83} for a narrow-band spectrum, which corresponds to our narrow-band equation \eqref{eq:epd} in the limiting case of no current, i.e., $k_m^*\to k_{m0}^*=\omega_m^{*2}/g$ (shear-free dispersion relation in nondimensional units).
To the best of our knowledge, theoretical expressions for wave crest distribution with a broad-band frequency spectrum have not been reported. 

Fig.\ \ref{fig:epd} shows the numerical PDF and exceedance probability of the scaled crest height compared to the Rayleigh and Tayfun distributions. Notice in Fig.\ (\ref{fig:epd}a) that for very low crests $\tzc\lesssim 0.1 H_s$ the probability density of wave crest height deviates noticeably from the Rayleigh curve, consistent with Fig.\ \ref{fig:pdf_maxima}. 
The reason is that finite bandwidth allows negative maxima (hence a finite probability density at zero crest height), whereas the narrow-band Rayleigh distribution only allows positive maxima. The physical significance of this difference is perhaps not so high being primarily a result of the definition of a crest, referring somewhat arbitrarily to the mean water level. The tail of our numerical results without shear still agrees well with those produced by the Rayleigh distribution \citep{fedele09}, perhaps surprising in light of the linear theory for broadband waves due to \citet{Cartwright56}. 
This can be explained by noting that in the context of their theory our spectrum is still relatively narrow, since the bandwidth parameter 
$\nu\approx 0.53$ as defined in Eq.~\eqref{eq:nu} is considerably smaller than unity.

It can be observed in Fig.\ \ref{fig:epd}b and \ref{fig:epd}c that, when nonlinear second--order corrections are accounted for, the tail of the simulated curve for the case with no shear clearly exceeds the Rayleigh distribution values, yet remain lower than the Tayfun distribution curve. This observation was also made by Fedele \& Tayfun \cite{fedele09} who considered broadband waves without current; They showed that in that case the Tayfun distribution is an upper bound for the wave crest distribution to second order in steepness.

With the additional presence of a shear current and broader spectrum, crest distributions can clearly exceed that of Tayfun. The numerical results show substantial differences between the three currents considered,
consistent with the general trend observed before: opposing shear makes high crests more probable and \emph{vice versa}. 
The gray dashed vertical line in Fig.\ \ref{fig:epd} refers to the conventional criterion for rogue waves, which is $\tilde{\zeta_c}/H_s=1.25$ \citep{Dysthe08}. Compared with the no-shear current case, the opposing shear current leads to 
significant
enhancement in the occurrence probability of rogue wave, as shown in Fig. \ref{fig:epd}e. The presence of following shear current has the opposite influence. The exceedance probability increases monotonously as a function of the shear strength $\beta$, which is shown in Figure (\ref{fig:epd}b,c).

We note in passing, however, that whereas the probability of \emph{unusually high} (rogue) waves is decreased on following shear, the significant wave height itself will often be increased. A typical situation where this occurs is when the shear current, measured in a land-fixed reference system, has its greatest velocity at the surface. In this case the current itself is opposing in an earth-fixed frame of reference, so waves generated elsewhere will steepen as they encounter the current. Thus the expectation in many real scenarios would be that following shear makes for rougher seas overall, whereas with opposing shear, while calmer on the whole, have an increased probability of \emph{surprisingly} high crests. This point was discussed in depth by Hjelmervik \& Trulsen \cite{Hjelmervik09}.

Fig.~\ref{fig:epd}d compares the exceedance probability of wave crest between the narrow-band predictions and numerical results for the cases with a shear current, the former of which are obtained by using \eqref{eq:epd}. We observe
that the narrow-band assumption leads to a small and large overestimate of the occurrence probability of wave crest for the case with a following and opposing shear current, respectively. The differences for the following current are nearly negligible, as being consistent with Fig.\ref{fig:pdf_surf}{b}, but are much more pronounced for the opposing shear case. Fig. \ref{fig:epd}d suggests aligned conclusion with  \citet{fedele09} in which it is stated that the narrow-band assumption would produce an upper bound of the exceedance probability of wave crest as aforementioned.
Since the effect of current shear on waves
 depend on both the shift in wavelength as reflected from the linear dispersion relation as well as the amplitude of the second-order superharmonic bound waves,
the overall effect of current on waves of a broad-band spectrum 
will in general differ in a non-trivial way from that only on the 
 amplitude of the spectral mean wave, $\hat{A}_{mm}^+$. As a result, the assumption of narrow bandwidth seems to lead to larger overestimate for opposing shear compared to the case of a following shear.

\subsection{The distribution of maximum wave crest}
Consider next the distribution of the height of the highest wave crest among a randomly chosen sequence of $N$ consecutive waves, where a `wave' in this context is a time interval wherein the surface elevation contains one maximum and one minimum.
A long time ago \citet{longuet52} derived an expression for maximum wave crest distribution based on linear waves with a narrow band frequency spectrum. 
\citet{Cartwright56} extended the theory to allow for a broadband spectrum, still in the linear wave regime. More recently, the Gumbel distribution was used to solve this problem up to second order \citep{Krogstad04, Prevosto00, Socquet-Juglard05}; for a linear narrow-band process, the expressions in these references are the same. In this section we use the expression from \citet{Cartwright56} for comparison:
\begin{equation}\label{eq:LHmax}
\dfrac{\zeta_{\max}}{\sigma}=\sqrt{2\ln \left  [(1-\nu^2)^{\frac{1}{2}}N \right ]} + \gamma_E/\sqrt{2\ln \left [(1-\nu^2)^{\frac{1}{2}}N \right ]},
\end{equation}
where $\zeta_{\max}$ is the maximum crest height from a continuous wave train, 
$\gamma_E\approx 0.5772$ is Euler's constant.

\begin{figure}
\centering
\includegraphics[width=0.95\textwidth]{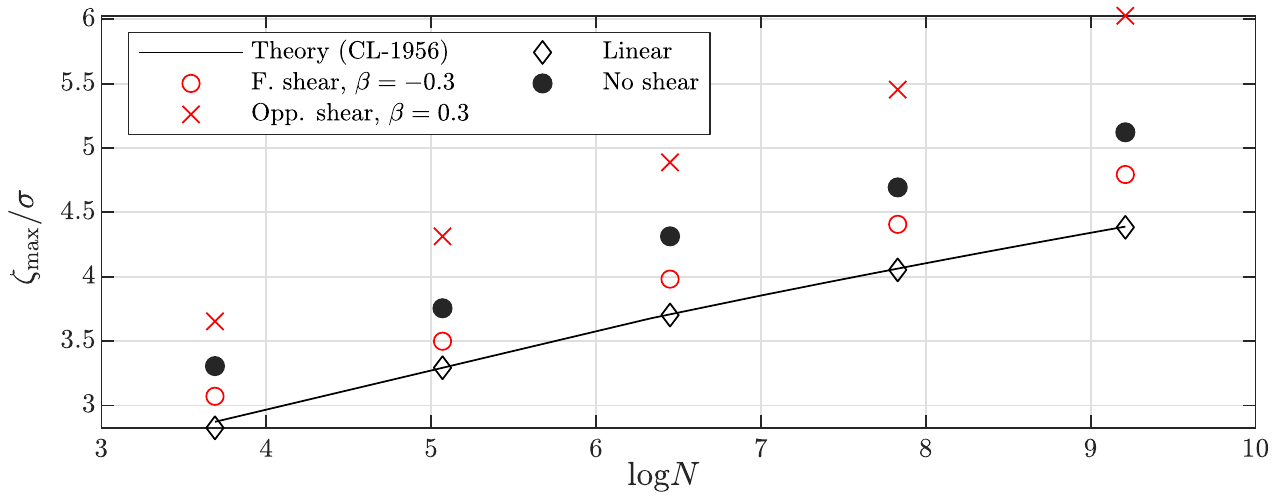}
\caption{ The average of crest height of scenes containing the largest $N$ waves. In the figure, the theoretical predictions (the black solid line) are based on \eqref{eq:LHmax}  for linear waves.
}
\label{fig:max_crest}
\end{figure}

Fig.\ \ref{fig:max_crest} gives the comparison of largest crest height between our numerical results and equation \eqref{eq:LHmax}. 
Each point is obtained as follows: a time series containing $2\times 10^6$ waves is divided into $160$ segments. From each segment a sequence of $N$ consecutive waves is chosen randomly from which the highest crest is found, then the average is taken over all the highest crests and plotted in the figure.
Fig. \ref{fig:max_crest}a shows that, once again, our simulated results of linear wave fields fit well with the theoretical solution.  

Compared with linear results, second order correction makes a considerable contribution to largest crest heights. The largest crest heights rise by around 10\% to 20\%. A similar phenomenon was observed by \citet{Socquet-Juglard05}, who used a narrow-band frequency spectrum and found the largest crest heights of nonlinear wave field increased by about 20\% compared with linear wave fields. Moreover, it is clear that the additional presence of sub-surface shear also has notable influence on largest crest heights. The opposing and following shear current increase or decrease the largest crest heights by about 18\% or 8\%, respectively for the case with $\beta = 0.3$ and $\beta=-0.3$, compared with the case with no shear current. Note that the comment at the end of the previous section still applies: the current will often change a free wave surface in such a way that in absolute terms, the crest heights are actually increased by opposing shear, which is a following current in the earth-fixed frame of reference, and \emph{vice versa}.

%
\subsection{Skewness}
In this section, we discuss the influence of a shear current on skewness, which is a measure of the lack of symmetry. Unlike skewness, kurtosis is not expected to be well approximated by second-order theory, and therefore not included in this paper.

Skewness of second-order waves can be expressed as a function of wave steepness, which is given by equation \eqref{eq:sigma_skew} in the limiting case of a narrow-band wave spectrum. The skewness should generally depend on both the bandwidth parameter ($\nu$) and spectrum shape, as has been shown by \citet{Srokosz86}. 

We consider two types of shear currents, as given in equations \specialeqref{eq:bU_def}{a,b}. From the point of view of the waves, which can ``feel'' the current only down to about half a wavelength's depth, the significant difference is that a linear current has the same shear at all depths, affecting the wave dispersion for all wavelengths, whereas the exponential profile is felt strongly by the short waves with $k\gtrsim \alpha k_{p,0}$ and hardly at all for long waves $k\ll \alpha k_{p,0}$.

\begin{figure}
\includegraphics[width=0.85\linewidth]{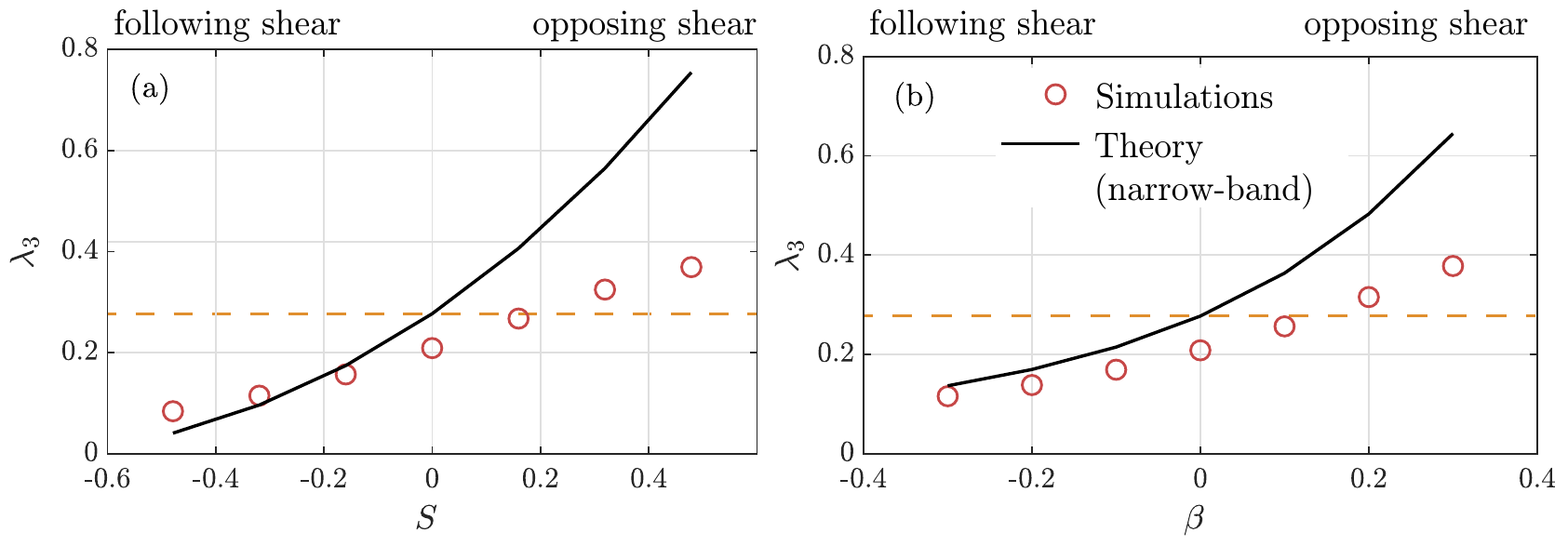}
\caption{
Skewness of the wave surface elevation for the cases with a linear shear current (a) and exponential shear current (b). The narrow-band theoretical predictions in  solid black lines are based on \protect\specialeqref{eq:sigma_skew}{b}. The dashed line is the no-shear case, for reference.
}
\label{fig:sk}
\end{figure}

Fig. \ref{fig:sk}a and  \ref{fig:sk}b show the skewness of linear and exponential shear current cases, respectively, calculated according to its definition given by \specialeqref{eq:Defs_sigma_skew}{b}. 
The theoretical 
narrow--band
predictions in solid blue lines are based on \specialeqref{eq:sigma_skew}{b} with the assumption of narrow-band  waves  in both the absence (i.e. $S=0$  and $\beta=0$ in Fig.\ \ref{fig:sk}a and  \ref{fig:sk}b, respectively) and presence of a shear current. For both 
linear and exponential current
cases the skewness increases monotonically with $S$ and $\beta$, respectively. 
In the range of shear strengths examined in Fig. \ref{fig:sk}, the skewness always remains positive. The strongest shear current enhances the skewness by about 86\% compared with the cases in the absence of a shear current. The narrow-band assumption for the cases with an exponential shear current always leads to an overestimate of the skewness, compared with the numerical simulations due to the theory in \S\ref{sec:theory} applicable to arbitrary bandwidth. In contrast, it may lead to underestimated values for the linear, following current cases in the regime where $S\leq -0.2$. The inaccuracy induced by the narrow-band assumption is obvious, which may arise from that the JONSWAP spectrum chosen is not very narrow and that the strong profile shear can lead to a considerable change in the wavelength of all waves prescribed on the JONSWAP spectrum.

\subsection{The Mouth of the Columbia River}\label{sec:C_R}
\begin{figure}
\includegraphics[width=0.8\linewidth]{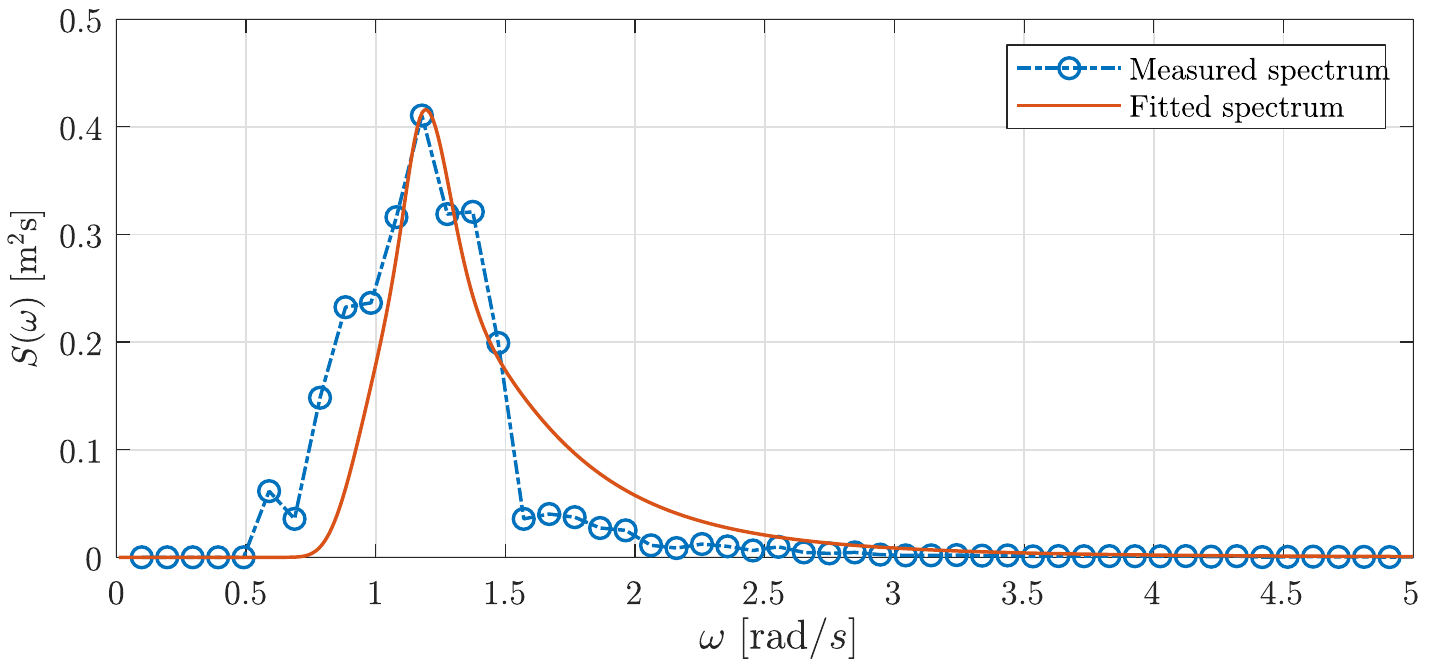}
\caption{Power energy spectrum for the Columbia River wave data.
}
\label{fig:CR_spectrum}
\end{figure}

As a real--life example we consider the real measured data described in Section \ref{sec:CRcurrent} to demonstrate and quantify the significant misprediction of wave statistics that would result from neglecting the current's vertical shear. The currents considered, adapted from figure 3 of Zippel \& Thomson \cite{Zippel17} are shown in Fig.~\ref{fig:skku_zip}a, using the same color coding as in said figure. The surface current was subtracted and the profiles extended to the surface as explained in section \ref{sec:CRcurrent}. As input wave spectrum we fit a JONSWAP spectrum
with bandwidth parameter $\nu=0.6618$
to a representative example among the manywave spectra measured by \citet{Zippel17}, 
shown in figure \ref{fig:CR_spectrum}. The fit is not excellent, but sufficient to provide a representative example. 

Figure \ref{fig:spec_Uz}d shows the weak-shear parameter $\delta(\omega)$ when $\omega$ is the given parameter; we argue in appendix \ref{app:SJ} that the appropriate value in this case is $\delta_\omega(\omega) = 2\delta(\omega^2/g)$ where $\delta(k)$ is defined in \eqref{eq:delta}.

\subsubsection{Skewness}

\begin{figure}
\includegraphics[width=\linewidth]{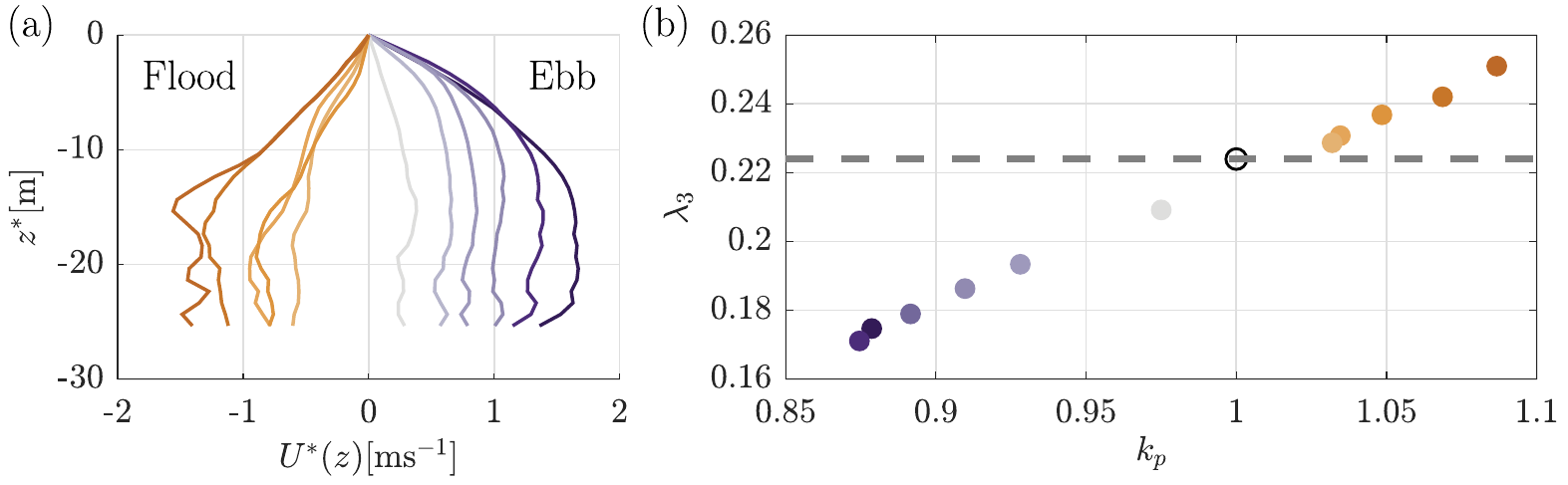}
\caption{Skewness of wave surface elevation with Columbia River current and wave spectrum data
(a) Considered current profiles, reproduced with kind permission from figure 3 of \cite{Zippel17} with the same colour coding, shifted to the surface level and with surface current subtracted. (b) Numerically obtained skewness for the measured wave spectrum of ref.~\cite{Zippel17} on the currents in panel (a), with corresponding color coding; the ascissa is the shear-shifted peak wave number with $k_p=1$ corresponding to zero shear (open circle).
}
\label{fig:skku_zip}
\end{figure}

The skewness of simulated results with Columbia River current data are given in Fig. \ref{fig:skku_zip}b, where $k_{p}$ is the dimensionless peak wavenumber which depends on the shear current as aforementioned. We chose to use $k_p$
as a representation of the shear strength as it expresses the amount by which the shear changes the wavelength of the wave with peak frequency. 

Failure to take into account the presence of shear causes overprediction of skewness by $\approx 24\%$ or underprediction by $\approx 13\%$ during ebb and flood, respectively, as is shown in Fig. \ref{fig:skku_zip}. 
Absolute numbers provided by a second-order theory like ours carry significant uncertainty, particularly when the spectrum is not narrow, but show a clear and consistent trend. Held together with Zippel \& Thomson's conclusion that wave steepness can be mispridicted by $\pm 20\%$ in these waters in the same conditions if shear is not accounted for \cite{Zippel17}, there is compelling evidence that shear can be highly significant to the estimation of wave statistics from measured spectra.

\subsubsection{Rogue wave probability}\label{sec:CR}

\begin{figure}
\centering
\includegraphics[width=0.95\textwidth]{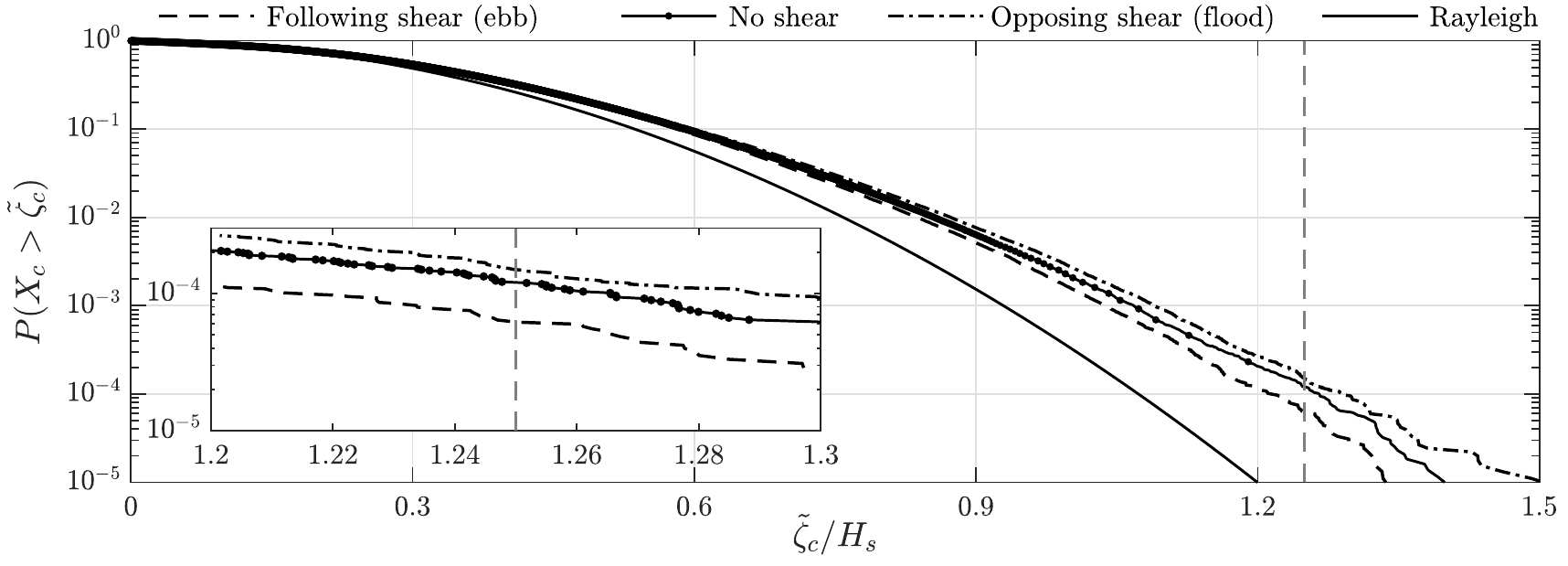}
\caption{Exceedance probability of simulated results with the current measured by Zippel \& Thomson \cite{Zippel17} in the Columbia River (CR) shown in Fig.~\ref{fig:spec_Uz}c, equal to the strongest currents in either direction in Fig.~\ref{fig:skku_zip}a. The profiles of the following and opposing CR-current are shown in Figure \ref{fig:spec_Uz}c. 
}
\label{fig:epdZippel}
\end{figure}

We also carried out simulations with data from Columbia River (CR) using both the wave spectrum and shear profiles measured in this location by \citet{Zippel17}.
As usual, rogue wave probability is defined as the probability of crests exceeding $1.25 H_s$.

As observed for the model currents in Figure \ref{fig:epd}, opposing shear enhances the crest heights of large waves while following shear weakens them, leading to increased and decreased exceedance probability, respectively. The rogue wave probability on opposing shear (i.e., a following surface current during ebb) is increased by $36\%$ while on following shear (opposing surface current, during flow) it is decreased by 45\%; from  $1.12\times 10^{-4}$ to $6.20\times 10^{-5}$ and $1.52\times 10^{-4}$, respectively.
Given that our theory is second order only, these numbers are not quantitatively accurate, but show clearly that shear currents must be accounted for in prediction and modelling of extreme waves.

Note carefully that the rogue wave probability is the probability of \emph{surprisingly} high waves, as discussed by \citet{Hjelmervik09}. Although rogue waves are more than twice as probable on the wave--following flow current than the wave--opposing ebb current, the significant wave height itself is typically much greater in the former case (more than twice as high in the conditions measured in \cite{Zippel17}, for instance), making for rougher conditions overall. The effect of shear is to reduce the prevalence of very large waves during ebb, a beneficial effect with respect to sealoads and maritime safety.


\section{Conclusions}
In this paper, we develop the second-order (deterministic) theory using perturbation expansion, which is extended from \citet{longuet62} to allow for a depth-dependent background flow whose profile shear can be strong. The new theory can be used to investigate the wave-current interaction and applicable to waves of an arbitrary bandwidth. The linear wave field is solved with the DIM method proposed by Li \& Ellingsen \cite{li19}. We derived a boundary value problem for the second-order waves, which can be solved numerically. With the additional assumption of narrow-band waves, a second-order accurate statistical model is derived for the skewness, probability density function of surface elevation, and the probability distribution of wave crest, which have accounted for the presence of a depth-dependent background flow.

We carried out numerical simulations for the analysis of wave statistics and examined effects of a shear current. We used a JONSWAP spectrum and several different shear currents as input to generate linear random waves. The second-order waves are solved for numerically based our newly derived theory. The measured wave spectrum and currents from Columbia River by Zippel \& Thomson \cite{Zippel17} were also used in our simulations.

For linear wave fields the probability distribution of wave surface elevation and wave maxima and average maximum wave crest all satisfy theoretical expressions well as expected. The nonlinear wave fields show similar properties compared with well-known second-order Stokes waves. The wave crests are higher and troughs are flatter than linear wave fields. As a result, the positive tails of the probability density function for wave surface elevation and wave maxima from nonlinear wave fields are longer than linear wave fields while the negative tails of surface elevation are shorter. Also, the largest wave crests in nonlinear wave fields are substantially greater. We found that the opposing shear currents can strengthen such `nonlinear properties' while the following shear currents can weaken them.

 We also found that the additional assumption of narrow-band waves  leads to in general negligible and pronounced differences for the following- and opposing-shear case, respectively, when comparing the second-order statistical model  with the more general deterministic theory which is applicable to waves with an arbitrary bandwidth.

\label{sec:conclus}

\begin{acknowledgments}
Z.B.\ Zheng acknowledges the support from China Scholarship Council through project 201906060137. Y.~Li is supported by
the Research Council of Norway (RCN) through the FRIPRO mobility project 287389. S.{\AA}.\ Ellingsen is supported by the European Research Council Consolidator Grant no.\ 101045299 (\emph{WaTurSheD}), and the RCN grant 325114 (\emph{iMod}). We thank Dr.\ Seth Zippel and Professor Jim Thomson for the use of the data collected from the Data Assimilation and Remote Sensing for Littoral Applications (DARLA) project and the Rivers and Inlets (RIVET) program (see, e.g., \cite{Zippel17} for details). 
The computer code (MATLAB) used to generate our data is included as supplementary material.
We thank the anonymous referees for their valuable suggestions and comments which have improved the quality of the paper. 
\end{acknowledgments}

\appendix
\section{Flow diagram of numerical implementations}
\label{sec:flow_diag}

\begin{figure*}
\centering
\includegraphics[width=0.9\columnwidth]{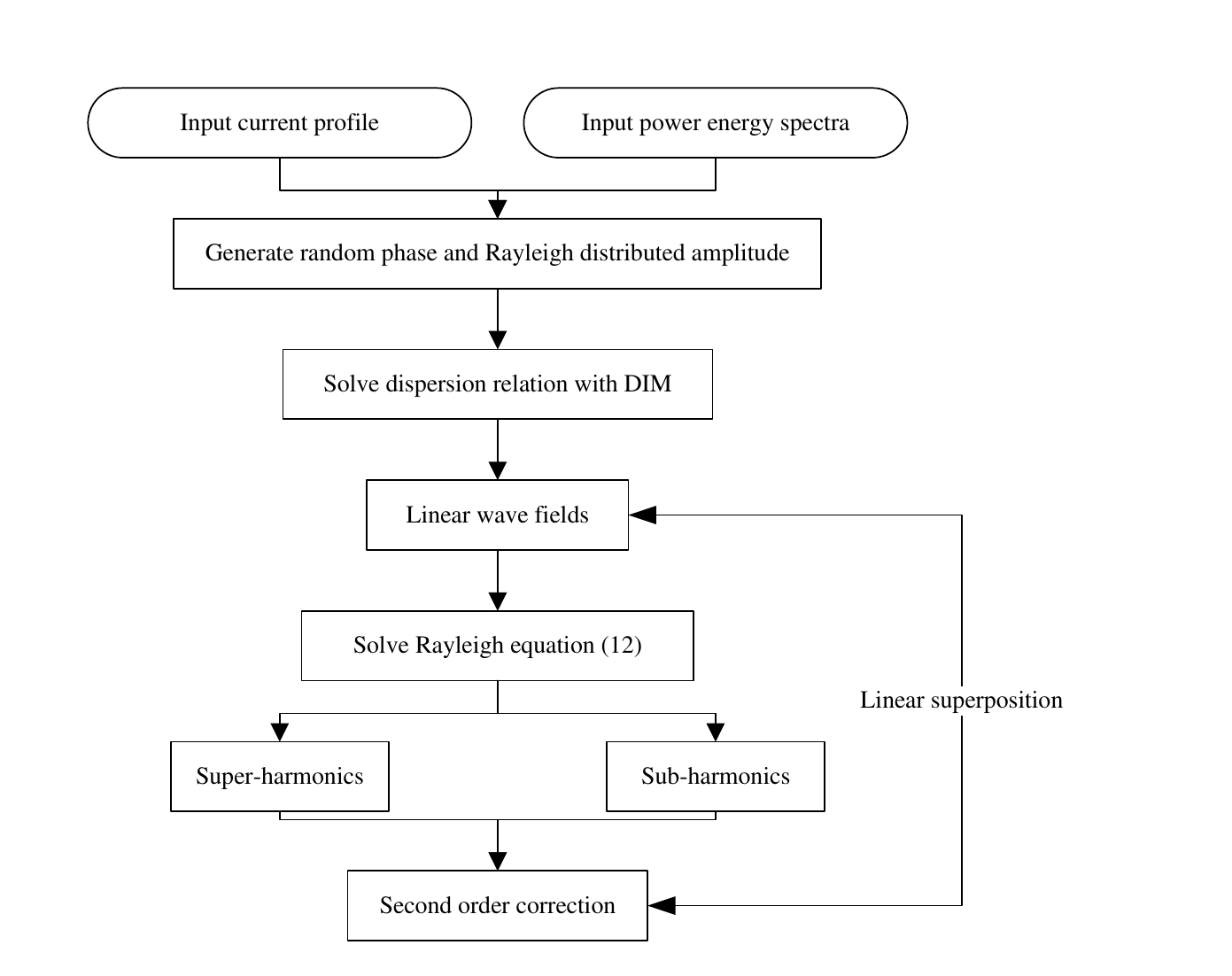}
\caption{Numerical procedures of the simulation.}
\label{fig:flowDiagram}
\end{figure*}

A flow diagram of the numerical implementation used to generate statistics is shown in Figure \ \ref{fig:flowDiagram}.

\section{The forcing terms of the Rayleigh equation}
\label{app:forcing}

With the linear wave fields given by \specialeqref{eq:linear_sol}{a,b,c}, the nonlinear forcing terms in \eqref{eq:hmn_def} are expressed as
\begin{linenomath}
\begin{subequations}\label{appeq:NF}
\begin{align}
    \hmN\snd_\pm=&[\bk_\pm\cdot \partial_z \mathbf{N}_{h,\pm} + k_\pm^2 N_{Rz,+}]\cos\psi_\pm,\\
    \hmF\snd_\pm=&[k_\pm^2 N_{F1,\pm}-N_{F2,\pm}+N_{F3,\pm+}-N_{F4,\pm}-(\bU\cdot \bk_\pm-\omega_\pm)\bk_\pm\cdot N_{h,+}]\sin\psi_\pm,
\end{align}
\end{subequations}
\end{linenomath}
with $\psi_\pm = \psi_1\pm\psi_2$, $\mathbf{N}_{h,i} = [N_{Rx,i}, N_{Ry,i}]$,
\begin{linenomath}
\begin{subequations}
\begin{align}
    \left[
    \begin{array}{ccc}
         N_{Rx,\pm} \\
         N_{Ry,\pm}\\
         N_{Rz,\pm}
    \end{array}
    \right] 
    =
    \frac12
    \left[
    \begin{array}{ccc}
     - (k_{1x}\hu_{1}\fst\hu_{2}\fst\pm k_{2x}\hu_{2}\fst\hu_{1}\fst + 
              k_{1y}\hu_{1}\fst\hv_{2}\fst\pm   k_{2y}\hu_{2}\fst\hv_{1}\fst
              \mp\hu_{1}^{(1)\prime} \hw_{2}\fst -\hu_{2}^{(1)\prime}\hw_{1}\fst
              ) 
     \\
     - (k_{1x}\hv_{1}\fst\hu_{2}\fst\pm k_{2x}\hv_{2}\fst\hu_{1}\fst + 
              k_{1y}\hv_{1}\fst\hv_{2}\fst\pm   k_{2y}\hv_{2}\fst\hv_{1}\fst
              \mp\hv^{(1)\prime}_{1} \hw_{2}\fst -\hv^{(1)\prime}_{2}\hw_{1}\fst
              )
     \\ 
           k_{x1}\hw_{1}\fst\hu_{2}\fst + k_{x2}\hw_{2}\fst\hu_{1}\fst
           +k_{y1}\hw_{1}\fst\hv_{2}\fst + k_{y2}\hw_{2}\fst\hv_{1}\fst    
           \mp \hw_{1}^{(1)\prime}\hw_{2}\fst \mp \hw_{1}\fst \hw_{2}^{(1)\prime}
    \end{array}
    \right]
\end{align}
\end{subequations}
\end{linenomath}
and
\begin{linenomath}
\begin{subequations}
\begin{align}
    N_{F1\pm}=&-\half(
    k_{1x}\hu\fst_{2}\hzeta\fst_{1} + k_{1y}\hv\fst_{2}\hzeta\fst_{1} \pm 
    k_{2x}\hu\fst_{1}\hzeta\fst_{2} \pm k_{2y}\hv\fst_{1}\hzeta\fst_{2}
    )\\
    N_{F2\pm}=&\half(
    \bk_1^2(\bk_1\cdot\bU-\omega_1)\hzeta\fst_2{\hp_{1}^{(1)\prime}} \pm \bk_2^2(\bk_2\cdot\bU-\omega_2)\hzeta\fst_1{\hp_{2}^{(1)\prime}})
    \\
    N_{F3\pm}=&-\half(
    \bk_1^2\hzeta\fst_2{\hw_{1}^{(1)\prime}} \pm \bk_2^2\hzeta\fst_1{\hw_{2}^{(1)\prime}}
    )\\
     N_{F4\pm}=&\half(
    \bk_1^2\bk_1\cdot\bU'\hp_{1}\fst\hzeta\fst_2 \pm \bk_2^2\bk_2\cdot\bU'\hp_{2}\fst\hzeta\fst_1
    )
\end{align}
\end{subequations}
\end{linenomath}
where $\bk_1=[k_{1x}, k_{1y}]$ and $\bk_2=[k_{2x}, k_{2y}]$

\section{Analytical solution for linearly sheared current}
\label{app:linshear}
We assume the shear profile is given by $\bU=(S_0z,0)$. The linear solution can be easily solved, which is expressed as \citep{Ellingsen16,Akselsen19}
\begin{linenomath}
\begin{subequations}
\begin{align}
    \hw\fst(\bk,z) =& \hw\fst_0(\bk)\rme^{kz}\\
    \hbu\fst(\bk,z) =& \rmi\dfrac{k^2\bU'+[(\bU\cdot\bk-\omega)k-k_xS_0]\bk}{(\bU\cdot\bk-\omega)k^2} \hw\fst_0\rme^{kz}\\
    \hp\fst(\bk,z) = & -\rmi\dfrac{(\bU\cdot\bk-\omega)k-k_xS_0}{k^2} \hw\fst_0\rme^{kz}\\
    \hw\fst_0(\bk)=&-\rmi \hzeta\fst(\bk)\omega
\end{align}
\end{subequations}
\end{linenomath}
where $\bk=(k_x,k_y)$, $k=\sqrt{k_x^2+k_y^2}$ and the subscript `0' denotes the evaluation at a undisturbed surface $z=0$. The dispersion relation for linear waves in a linearly sheared current is given by \citep{Ellingsen16,Akselsen19}
\begin{equation}
    \omega=-\dfrac{S_0k_x}{2k}\pm\sqrt{ k + \dfrac{S_0^2k_x^2}{4k^2}  },
\end{equation}
where '$+$' and '$-$' denotes the waves propagating 'downstream' and 'upstream' relative to the current, respectively.

Substituting the linear solution into the forcing terms of second-order equations (\ref{eq:hw_2nd}), we obtain an inhomogeneous boundary value problem for the second-order vertical velocity $w\snd$. The general solution to this boundary value problem in the Fourier space should admit the form 
\begin{equation} \label{eq:w2n}
    \hw\snd_\pm(\bk_1,\bk_2,z)= B_{1\pm}(\bk_1,\bk_2)\rme^{k_\pm z}  + \hw_{cross}(\bk_1,\bk_2,z),
\end{equation}
where the deepwater boundary condition was used,  the first term on the right hand side of the equation is due to the forcing at a still water surface and the homogeneous Rayleigh equation,  and $\hw_{cross}$is a particular solution of the inhomogeneous Rayleigh equation given by \citep{Akselsen19}
\begin{linenomath}
\begin{align}
    \hw_{cross}(\bk_1,\bk_2,z)= &-\dfrac{i}{2k_\pm} \dfrac{\hw\fst_{0,1}\hw\fst_{0,2}}{k_{\pm x}S_0} \dfrac{k_{1x}k_{2y}-k_{1y}k_{2x}}{k_1k_2} 
    \rme^{(k_1 + k_2)z}
    \sum_{i,j=1}^3
    \left[\dfrac{\pm b_{ij}}{(\xi_i-z)^{j-1}}
    \right.\notag\\
    &
    \left.\times \tilde{E}_j[k_\pm(\xi_i-z)]
    \right],
\end{align}
\end{linenomath}
with $\hw\fst_{0,j}=\hw\fst_{0}(\bk_j)$ for $j=1$ and $j=2$,
\begin{linenomath}
\begin{subequations}\label{eq:bij}
\begin{align}
    b_{ij}=~& \sum_{m=j}^3\dfrac{-a_{im}}{(\xi_i-\xi_3)^{m-j+1}},~~~~i=1,2;~b_{31}=-b_{11}-b_{21};~ b_{32}=b_{33}=0,\\
    \xi_1=~& \dfrac{\omega_1}{k_{1x}S_0},~~~~\xi_2=\dfrac{\omega_2}{k_{2x}S_0},~~~~\xi_3=\dfrac{\omega_\pm } {k_{x}S_0},\\
    \tilde{E}_j(\mu)=~& \rme^\mu \mu^{j-1}\int_\mu^\infty\dfrac{\rme^{-\tau}}{\tau^j}\rmd\tau. 
\end{align}
\end{subequations}
\end{linenomath}
Assuming $\xi_1\neq\xi_2$, the coefficients in (\ref{eq:bij}) are expressed as
\begin{linenomath}
\begin{subequations}
\begin{align}
    a_{i1}=& (-1)^i\left[ k_1k_2-\bk_1\cdot\bk_2 - \dfrac{k_1+k_2}{\xi_1-\xi_2} \dfrac{k_{1x}k_{2y}-k_{1y}k_{2x}}{k_1k_2} \tan\theta_m
    \right] \tan\theta_i\\
    a_{i2}=& (-1)^i\dfrac{1}{k_i} \left[ k_1k_2-\bk_1\cdot\bk_2 - \dfrac{k_i}{\xi_1-\xi_2} \dfrac{k_{1x}k_{2y}-k_{1y}k_{2x}}{k_1k_2} \tan\theta_m
    \right] \tan\theta_i\\
    a_{i3}=& (-1)^i\dfrac{k_m}{k_i}  \tan\theta_i,
\end{align}
\end{subequations}
\end{linenomath}
where $i,m\in\{1, 2\}$ so that $i\neq m$ and $\tan\theta_i={k_{iy}}/{k_{ix}}$. The undetermined coefficients $B_{1\pm}$ is solved by inserting (\ref{eq:w2n}) into the combined boundary condition (\ref{eq:cbc_2nd}). Then, the surface elevation is obtained from \eqref{eq:zeta_snd}. 


\section{Effects of current continuation on skewness}\label{sec:3profile}
\begin{figure}
\includegraphics[width=\linewidth]{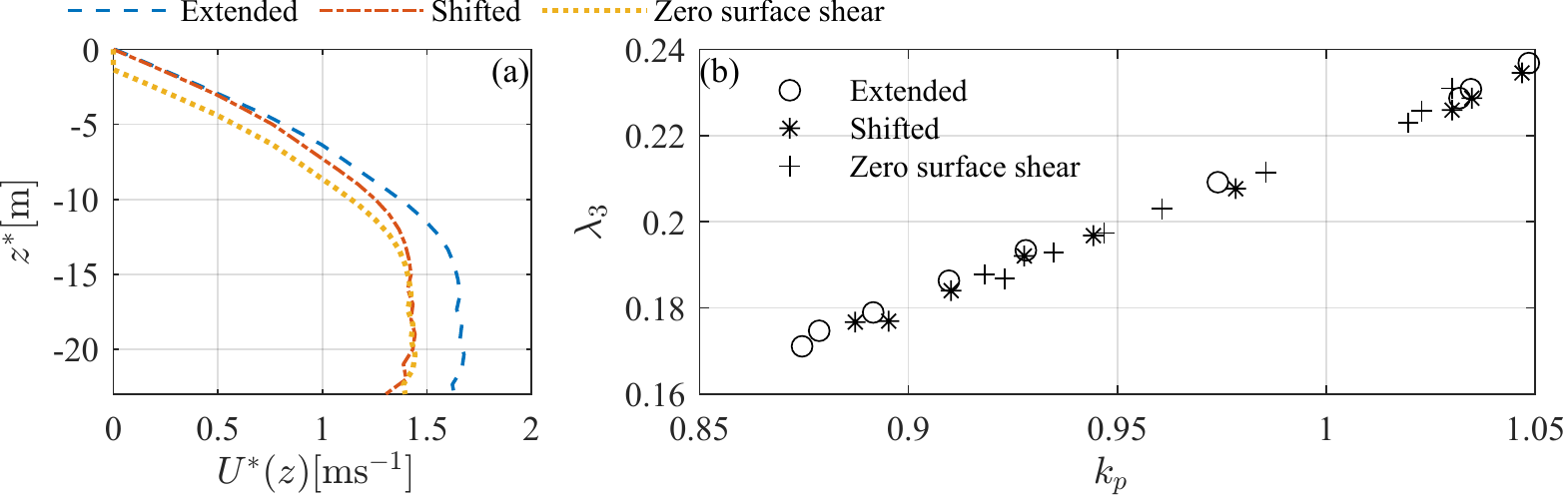}
\caption{
Skewness of wave surface elevation for different profiles. (a) Comparison of shear profiles with three approaches. (b) Numerically obtained skewness, `o': extended profiles, `*': shifted profiles, `+': zero surface shear profiles. Case chosen is same with Fig. \ref{fig:skku_zip} except that two strongest opposing shears are excluded here.}
\label{fig:sk_3shear}
\end{figure}

We here compare three alternative, physically reasonable ways in which profiles measured using ADCP can be extended from the shallowest measurement point --- $z=-1.35$\,m for the Columbia River measurements we use \cite{kilcher10} --- up to the surface. These are: extrapolation using a polynomial fit, shifting the profile upwards so that the shallowest measurement point is set to surface level (used, \emph{inter alia}, in refs.~\cite{smeltzer17,li19b}), and the highly conservative approach of continuing the current profile to the surface with zero shear. These are referred as extended profile, shifted profile and zero surface shear profile, respectively and are shown in figure \ref{fig:sk_3shear}a. 

We compare wave skewness in these three case, the results are given in Fig. \ref{fig:sk_3shear}. Again, the $k_p$ in Fig.\ref{fig:sk_3shear}b is the dimensionless peak wavenumber as in Fig. \ref{fig:skku_zip}, where $k_p=1$ corresponds to the case without shear current whereas the modifications to the dispersion relation due to shear shifts the value. Values $k_p > 1$ correspond to adverse shear and \emph{vice versa}. A plot of the calculated skewness for the different cases shows that the difference in skewness is hardly discernable. 

\section{Dimensionless weak--shear parameter for given $\omega$}
 \label{app:SJ}

Let the depth-averaged shear be small, of order a small parameter $\delta\ll 1$. Assuming the wave number $k$ given, \citet{Stewart74} derived the approximate dispersion relation $\omega(k)$ which may be written \cite{Ellingsen17}
\begin{equation}\label{eq:sj}
    \omega^*(k^*)\approx \sqrt{gk^*}[1-\delta(k^*)] + \order{\delta^2},
\end{equation}
with the small-shear parameter $\delta(k^*)$ defined in \eqref{eq:delta}.
It was shown \cite{Ellingsen17} that a sufficient criterion for the Stewart \& Joy approximation to be good is that $\delta_\omega\ll 1$.

Conversely (i.e., for given $\omega^*$) the presence of shear modifies $k$ slightly, and we write
\begin{equation}\label{eq:sjk}
    k^* = k^*_0[1 + \delta_\omega(\omega^*)]+\order{\delta_\omega^2}
\end{equation}
with $k^*_0=(\omega^*)^2/g$, and clearly $\delta_\omega \sim \delta$.
We seek to find $\delta_\omega$. Inserting \eqref{eq:sjk} into \eqref{eq:sj} via \eqref{eq:delta} and noting that $\sqrt{gk^*_0}=\omega^*$,
\begin{align}
    \omega^* =& \omega^*\sqrt{1 +\delta_\omega}[1-\delta(k_0^*)]+\order{\delta^2}\notag\\
    =& \omega^*[1+ \half\delta_\omega -\delta(k_0^*)] + \order{\delta^2}.
\end{align}
Internal consistency thus demands 
\begin{equation}
    \delta_\omega(\omega^*) = 2\delta(k_0^*).
\end{equation}

\bibliography{references}

\begin{thebibliography}{95}%
\makeatletter
\providecommand \@ifxundefined [1]{%
 \@ifx{#1\undefined}
}%
\providecommand \@ifnum [1]{%
 \ifnum #1\expandafter \@firstoftwo
 \else \expandafter \@secondoftwo
 \fi
}%
\providecommand \@ifx [1]{%
 \ifx #1\expandafter \@firstoftwo
 \else \expandafter \@secondoftwo
 \fi
}%
\providecommand \natexlab [1]{#1}%
\providecommand \enquote  [1]{``#1''}%
\providecommand \bibnamefont  [1]{#1}%
\providecommand \bibfnamefont [1]{#1}%
\providecommand \citenamefont [1]{#1}%
\providecommand \href@noop [0]{\@secondoftwo}%
\providecommand \href [0]{\begingroup \@sanitize@url \@href}%
\providecommand \@href[1]{\@@startlink{#1}\@@href}%
\providecommand \@@href[1]{\endgroup#1\@@endlink}%
\providecommand \@sanitize@url [0]{\catcode `\\12\catcode `\$12\catcode
  `\&12\catcode `\#12\catcode `\^12\catcode `\_12\catcode `\%12\relax}%
\providecommand \@@startlink[1]{}%
\providecommand \@@endlink[0]{}%
\providecommand \url  [0]{\begingroup\@sanitize@url \@url }%
\providecommand \@url [1]{\endgroup\@href {#1}{\urlprefix }}%
\providecommand \urlprefix  [0]{URL }%
\providecommand \Eprint [0]{\href }%
\providecommand \doibase [0]{https://doi.org/}%
\providecommand \selectlanguage [0]{\@gobble}%
\providecommand \bibinfo  [0]{\@secondoftwo}%
\providecommand \bibfield  [0]{\@secondoftwo}%
\providecommand \translation [1]{[#1]}%
\providecommand \BibitemOpen [0]{}%
\providecommand \bibitemStop [0]{}%
\providecommand \bibitemNoStop [0]{.\EOS\space}%
\providecommand \EOS [0]{\spacefactor3000\relax}%
\providecommand \BibitemShut  [1]{\csname bibitem#1\endcsname}%
\let\auto@bib@innerbib\@empty
\bibitem [{\citenamefont {Kharif}\ \emph {et~al.}(2008)\citenamefont {Kharif},
  \citenamefont {Pelinovsky},\ and\ \citenamefont {Slunyaev}}]{kharif08}%
  \BibitemOpen
  \bibfield  {author} {\bibinfo {author} {\bibfnamefont {C.}~\bibnamefont
  {Kharif}}, \bibinfo {author} {\bibfnamefont {E.}~\bibnamefont {Pelinovsky}},\
  and\ \bibinfo {author} {\bibfnamefont {A.}~\bibnamefont {Slunyaev}},\
  }\href@noop {} {\emph {\bibinfo {title} {Rogue waves in the ocean}}}\
  (\bibinfo  {publisher} {Springer Science \& Business Media},\ \bibinfo {year}
  {2008})\BibitemShut {NoStop}%
\bibitem [{\citenamefont {Cavaleri}\ \emph {et~al.}(2018)\citenamefont
  {Cavaleri}, \citenamefont {Abdalla}, \citenamefont {Benetazzo}, \citenamefont
  {Bertotti}, \citenamefont {Bidlot}, \citenamefont {Breivik}, \citenamefont
  {Carniel}, \citenamefont {Jensen}, \citenamefont {Portilla-Yandun},
  \citenamefont {Rogers}, \citenamefont {Roland}, \citenamefont
  {Sanchez-Arcilla}, \citenamefont {Smith}, \citenamefont {Staneva},
  \citenamefont {Toledo}, \citenamefont {van Vledder},\ and\ \citenamefont
  {van~der Westhuysen}}]{Cavaleri18}%
  \BibitemOpen
  \bibfield  {author} {\bibinfo {author} {\bibfnamefont {L.}~\bibnamefont
  {Cavaleri}}, \bibinfo {author} {\bibfnamefont {S.}~\bibnamefont {Abdalla}},
  \bibinfo {author} {\bibfnamefont {A.}~\bibnamefont {Benetazzo}}, \bibinfo
  {author} {\bibfnamefont {L.}~\bibnamefont {Bertotti}}, \bibinfo {author}
  {\bibfnamefont {J.~R.}\ \bibnamefont {Bidlot}}, \bibinfo {author}
  {\bibnamefont {Breivik}}, \bibinfo {author} {\bibfnamefont {S.}~\bibnamefont
  {Carniel}}, \bibinfo {author} {\bibfnamefont {R.~E.}\ \bibnamefont {Jensen}},
  \bibinfo {author} {\bibfnamefont {J.}~\bibnamefont {Portilla-Yandun}},
  \bibinfo {author} {\bibfnamefont {W.~E.}\ \bibnamefont {Rogers}}, \bibinfo
  {author} {\bibfnamefont {A.}~\bibnamefont {Roland}}, \bibinfo {author}
  {\bibfnamefont {A.}~\bibnamefont {Sanchez-Arcilla}}, \bibinfo {author}
  {\bibfnamefont {J.~M.}\ \bibnamefont {Smith}}, \bibinfo {author}
  {\bibfnamefont {J.}~\bibnamefont {Staneva}}, \bibinfo {author} {\bibfnamefont
  {Y.}~\bibnamefont {Toledo}}, \bibinfo {author} {\bibfnamefont {G.~P.}\
  \bibnamefont {van Vledder}},\ and\ \bibinfo {author} {\bibfnamefont {A.~J.}\
  \bibnamefont {van~der Westhuysen}},\ }\bibfield  {title} {\bibinfo {title}
  {Wave modelling in coastal and inner seas},\ }\href@noop {} {\bibfield
  {journal} {\bibinfo  {journal} {Prog. Oceanogr.}\ }\textbf {\bibinfo {volume}
  {167}},\ \bibinfo {pages} {164} (\bibinfo {year} {2018})}\BibitemShut
  {NoStop}%
\bibitem [{\citenamefont {Dudley}\ \emph {et~al.}(2019)\citenamefont {Dudley},
  \citenamefont {Genty}, \citenamefont {Mussot}, \citenamefont {Chabchoub},\
  and\ \citenamefont {Dias}}]{dudley19}%
  \BibitemOpen
  \bibfield  {author} {\bibinfo {author} {\bibfnamefont {J.~M.}\ \bibnamefont
  {Dudley}}, \bibinfo {author} {\bibfnamefont {G.}~\bibnamefont {Genty}},
  \bibinfo {author} {\bibfnamefont {A.}~\bibnamefont {Mussot}}, \bibinfo
  {author} {\bibfnamefont {A.}~\bibnamefont {Chabchoub}},\ and\ \bibinfo
  {author} {\bibfnamefont {F.}~\bibnamefont {Dias}},\ }\bibfield  {title}
  {\bibinfo {title} {Rogue waves and analogies in optics and oceanography},\
  }\href@noop {} {\bibfield  {journal} {\bibinfo  {journal} {Nat. Rev. Phys}\
  }\textbf {\bibinfo {volume} {1}},\ \bibinfo {pages} {675} (\bibinfo {year}
  {2019})}\BibitemShut {NoStop}%
\bibitem [{\citenamefont {Benjamin}\ and\ \citenamefont
  {Feir}(1967)}]{benjamin67}%
  \BibitemOpen
  \bibfield  {author} {\bibinfo {author} {\bibfnamefont {T.~B.}\ \bibnamefont
  {Benjamin}}\ and\ \bibinfo {author} {\bibfnamefont {J.~E.}\ \bibnamefont
  {Feir}},\ }\bibfield  {title} {\bibinfo {title} {The disintegration of wave
  trains on deep water {P}art 1. theory},\ }\href@noop {} {\bibfield  {journal}
  {\bibinfo  {journal} {J. Fluid Mech.}\ }\textbf {\bibinfo {volume} {27}},\
  \bibinfo {pages} {417} (\bibinfo {year} {1967})}\BibitemShut {NoStop}%
\bibitem [{\citenamefont {Janssen}(2003)}]{janssen03}%
  \BibitemOpen
  \bibfield  {author} {\bibinfo {author} {\bibfnamefont {P.~A. E.~M.}\
  \bibnamefont {Janssen}},\ }\bibfield  {title} {\bibinfo {title} {Nonlinear
  four-wave interactions and freak waves},\ }\href@noop {} {\bibfield
  {journal} {\bibinfo  {journal} {J. Phys. Oceanogr.}\ }\textbf {\bibinfo
  {volume} {33}},\ \bibinfo {pages} {863} (\bibinfo {year} {2003})}\BibitemShut
  {NoStop}%
\bibitem [{\citenamefont {White}\ and\ \citenamefont
  {Fornberg}(1998)}]{white98}%
  \BibitemOpen
  \bibfield  {author} {\bibinfo {author} {\bibfnamefont {B.}~\bibnamefont
  {White}}\ and\ \bibinfo {author} {\bibfnamefont {B.}~\bibnamefont
  {Fornberg}},\ }\bibfield  {title} {\bibinfo {title} {On the chance of freak
  waves at sea},\ }\href@noop {} {\bibfield  {journal} {\bibinfo  {journal} {J.
  Fluid Mech.}\ }\textbf {\bibinfo {volume} {335}},\ \bibinfo {pages} {113}
  (\bibinfo {year} {1998})}\BibitemShut {NoStop}%
\bibitem [{\citenamefont {Janssen}\ and\ \citenamefont
  {Herbers}(2009)}]{janssen09}%
  \BibitemOpen
  \bibfield  {author} {\bibinfo {author} {\bibfnamefont {T.}~\bibnamefont
  {Janssen}}\ and\ \bibinfo {author} {\bibfnamefont {T.}~\bibnamefont
  {Herbers}},\ }\bibfield  {title} {\bibinfo {title} {Nonlinear wave statistics
  in a focal zone},\ }\href@noop {} {\bibfield  {journal} {\bibinfo  {journal}
  {J. Phys. Oceanogr.}\ }\textbf {\bibinfo {volume} {39}},\ \bibinfo {pages}
  {1948} (\bibinfo {year} {2009})}\BibitemShut {NoStop}%
\bibitem [{\citenamefont {Gao}\ \emph {et~al.}(2021)\citenamefont {Gao},
  \citenamefont {Ma}, \citenamefont {Dong}, \citenamefont {Chen}, \citenamefont
  {Liu},\ and\ \citenamefont {Zang}}]{gao21}%
  \BibitemOpen
  \bibfield  {author} {\bibinfo {author} {\bibfnamefont {J.}~\bibnamefont
  {Gao}}, \bibinfo {author} {\bibfnamefont {X.}~\bibnamefont {Ma}}, \bibinfo
  {author} {\bibfnamefont {G.}~\bibnamefont {Dong}}, \bibinfo {author}
  {\bibfnamefont {H.}~\bibnamefont {Chen}}, \bibinfo {author} {\bibfnamefont
  {Q.}~\bibnamefont {Liu}},\ and\ \bibinfo {author} {\bibfnamefont
  {J.}~\bibnamefont {Zang}},\ }\bibfield  {title} {\bibinfo {title}
  {Investigation on the effects of {Bragg} reflection on harbor oscillations},\
  }\href@noop {} {\bibfield  {journal} {\bibinfo  {journal} {Coast. Eng.}\
  }\textbf {\bibinfo {volume} {170}},\ \bibinfo {pages} {103977} (\bibinfo
  {year} {2021})}\BibitemShut {NoStop}%
\bibitem [{\citenamefont {Trulsen}\ \emph {et~al.}(2020)\citenamefont
  {Trulsen}, \citenamefont {Raust{\o}l}, \citenamefont {Jorde},\ and\
  \citenamefont {Rye}}]{trulsen20}%
  \BibitemOpen
  \bibfield  {author} {\bibinfo {author} {\bibfnamefont {K.}~\bibnamefont
  {Trulsen}}, \bibinfo {author} {\bibfnamefont {A.}~\bibnamefont {Raust{\o}l}},
  \bibinfo {author} {\bibfnamefont {S.}~\bibnamefont {Jorde}},\ and\ \bibinfo
  {author} {\bibfnamefont {L.}~\bibnamefont {Rye}},\ }\bibfield  {title}
  {\bibinfo {title} {Extreme wave statistics of long-crested irregular waves
  over a shoal},\ }\href@noop {} {\bibfield  {journal} {\bibinfo  {journal} {J.
  Fluid Mech.}\ }\textbf {\bibinfo {volume} {882}} (\bibinfo {year}
  {2020})}\BibitemShut {NoStop}%
\bibitem [{\citenamefont {Li}\ \emph {et~al.}(2021)\citenamefont {Li},
  \citenamefont {Draycott}, \citenamefont {Zheng}, \citenamefont {Lin},
  \citenamefont {Adcock},\ and\ \citenamefont {van~den Bremer}}]{li21}%
  \BibitemOpen
  \bibfield  {author} {\bibinfo {author} {\bibfnamefont {Y.}~\bibnamefont
  {Li}}, \bibinfo {author} {\bibfnamefont {S.}~\bibnamefont {Draycott}},
  \bibinfo {author} {\bibfnamefont {Y.}~\bibnamefont {Zheng}}, \bibinfo
  {author} {\bibfnamefont {Z.}~\bibnamefont {Lin}}, \bibinfo {author}
  {\bibfnamefont {T.~A.~A.}\ \bibnamefont {Adcock}},\ and\ \bibinfo {author}
  {\bibfnamefont {T.~S.}\ \bibnamefont {van~den Bremer}},\ }\bibfield  {title}
  {\bibinfo {title} {Why rogue waves occur atop abrupt depth transitions},\
  }\href@noop {} {\bibfield  {journal} {\bibinfo  {journal} {J. Fluid Mech.}\
  }\textbf {\bibinfo {volume} {919}},\ \bibinfo {pages} {R5} (\bibinfo {year}
  {2021})}\BibitemShut {NoStop}%
\bibitem [{\citenamefont {Longuet-Higgins}(1962)}]{longuet62}%
  \BibitemOpen
  \bibfield  {author} {\bibinfo {author} {\bibfnamefont {M.~S.}\ \bibnamefont
  {Longuet-Higgins}},\ }\bibfield  {title} {\bibinfo {title} {Resonant
  interactions between two trains of gravity waves},\ }\href@noop {} {\bibfield
   {journal} {\bibinfo  {journal} {J. Fluid Mech.}\ }\textbf {\bibinfo {volume}
  {12}},\ \bibinfo {pages} {321} (\bibinfo {year} {1962})}\BibitemShut
  {NoStop}%
\bibitem [{\citenamefont {Longuet-Higgins}(1963)}]{Longuet-Higgins63}%
  \BibitemOpen
  \bibfield  {author} {\bibinfo {author} {\bibfnamefont {M.~S.}\ \bibnamefont
  {Longuet-Higgins}},\ }\bibfield  {title} {\bibinfo {title} {The effect of
  non-linearities on statistical distributions in the theory of sea waves},\
  }\href@noop {} {\bibfield  {journal} {\bibinfo  {journal} {J. Fluid Mech.}\
  }\textbf {\bibinfo {volume} {17}},\ \bibinfo {pages} {459} (\bibinfo {year}
  {1963})}\BibitemShut {NoStop}%
\bibitem [{\citenamefont {Tayfun}(1980)}]{Tayfun80}%
  \BibitemOpen
  \bibfield  {author} {\bibinfo {author} {\bibfnamefont {M.~A.}\ \bibnamefont
  {Tayfun}},\ }\bibfield  {title} {\bibinfo {title} {Narrow-band nonlinear sea
  waves},\ }\href@noop {} {\bibfield  {journal} {\bibinfo  {journal} {J.
  Geophys. Res.}\ }\textbf {\bibinfo {volume} {85}},\ \bibinfo {pages} {1548}
  (\bibinfo {year} {1980})}\BibitemShut {NoStop}%
\bibitem [{\citenamefont {Tayfun}(1983)}]{Tayfun83}%
  \BibitemOpen
  \bibfield  {author} {\bibinfo {author} {\bibfnamefont {M.~A.}\ \bibnamefont
  {Tayfun}},\ }\bibfield  {title} {\bibinfo {title} {Effects of spectrum band
  width on the distribution of wave heights and periods},\ }\href@noop {}
  {\bibfield  {journal} {\bibinfo  {journal} {Ocean Eng.}\ }\textbf {\bibinfo
  {volume} {10}},\ \bibinfo {pages} {107} (\bibinfo {year} {1983})}\BibitemShut
  {NoStop}%
\bibitem [{\citenamefont {Tayfun}(1986)}]{Tayfun86}%
  \BibitemOpen
  \bibfield  {author} {\bibinfo {author} {\bibfnamefont {M.~A.}\ \bibnamefont
  {Tayfun}},\ }\bibfield  {title} {\bibinfo {title} {On narrow‐band
  representation of ocean waves: 1. theory},\ }\href@noop {} {\bibfield
  {journal} {\bibinfo  {journal} {J. Geophys. Res.: Oceans}\ }\textbf {\bibinfo
  {volume} {91}},\ \bibinfo {pages} {7743} (\bibinfo {year}
  {1986})}\BibitemShut {NoStop}%
\bibitem [{\citenamefont {Dalzell}(1999)}]{dalzell99}%
  \BibitemOpen
  \bibfield  {author} {\bibinfo {author} {\bibfnamefont {J.}~\bibnamefont
  {Dalzell}},\ }\bibfield  {title} {\bibinfo {title} {A note on finite depth
  second-order wave–wave interactions},\ }\href@noop {} {\bibfield  {journal}
  {\bibinfo  {journal} {Appl. Ocean Res.}\ }\textbf {\bibinfo {volume} {21}},\
  \bibinfo {pages} {105} (\bibinfo {year} {1999})}\BibitemShut {NoStop}%
\bibitem [{\citenamefont {Forristall}(2000)}]{Forristall00}%
  \BibitemOpen
  \bibfield  {author} {\bibinfo {author} {\bibfnamefont {G.~Z.}\ \bibnamefont
  {Forristall}},\ }\bibfield  {title} {\bibinfo {title} {Wave crest
  distributions: Observations and second-order theory},\ }\href@noop {}
  {\bibfield  {journal} {\bibinfo  {journal} {J. Phys. Oceanogr.}\ }\textbf
  {\bibinfo {volume} {30}},\ \bibinfo {pages} {1931} (\bibinfo {year}
  {2000})}\BibitemShut {NoStop}%
\bibitem [{\citenamefont {Arena}\ and\ \citenamefont {Fedele}(2002)}]{Arena02}%
  \BibitemOpen
  \bibfield  {author} {\bibinfo {author} {\bibfnamefont {F.}~\bibnamefont
  {Arena}}\ and\ \bibinfo {author} {\bibfnamefont {F.}~\bibnamefont {Fedele}},\
  }\bibfield  {title} {\bibinfo {title} {A family of narrow-band non-linear
  stochastic processes for the mechanics of sea waves},\ }\href@noop {}
  {\bibfield  {journal} {\bibinfo  {journal} {Eur. J. Mech. B Fluids}\ }\textbf
  {\bibinfo {volume} {21}},\ \bibinfo {pages} {125} (\bibinfo {year}
  {2002})}\BibitemShut {NoStop}%
\bibitem [{\citenamefont {Toffoli}\ \emph {et~al.}(2007)\citenamefont
  {Toffoli}, \citenamefont {Onorato}, \citenamefont {Babanin}, \citenamefont
  {Bitner-Gregersen}, \citenamefont {Osborne},\ and\ \citenamefont
  {Monbaliu}}]{Toffoli07}%
  \BibitemOpen
  \bibfield  {author} {\bibinfo {author} {\bibfnamefont {A.~A.}\ \bibnamefont
  {Toffoli}}, \bibinfo {author} {\bibfnamefont {M.}~\bibnamefont {Onorato}},
  \bibinfo {author} {\bibfnamefont {A.~V.}\ \bibnamefont {Babanin}}, \bibinfo
  {author} {\bibfnamefont {E.}~\bibnamefont {Bitner-Gregersen}}, \bibinfo
  {author} {\bibfnamefont {A.~R.}\ \bibnamefont {Osborne}},\ and\ \bibinfo
  {author} {\bibfnamefont {J.}~\bibnamefont {Monbaliu}},\ }\bibfield  {title}
  {\bibinfo {title} {Second-order theory and setup in surface gravity waves: A
  comparison with experimental data},\ }\href@noop {} {\bibfield  {journal}
  {\bibinfo  {journal} {J. Phys. Oceanogr.}\ }\textbf {\bibinfo {volume}
  {37}},\ \bibinfo {pages} {2726} (\bibinfo {year} {2007})}\BibitemShut
  {NoStop}%
\bibitem [{\citenamefont {Toffoli}\ \emph {et~al.}(2008)\citenamefont
  {Toffoli}, \citenamefont {Onorato}, \citenamefont {Bitner-Gregersen},
  \citenamefont {Osborne},\ and\ \citenamefont {Babanin}}]{Toffoli08}%
  \BibitemOpen
  \bibfield  {author} {\bibinfo {author} {\bibfnamefont {A.}~\bibnamefont
  {Toffoli}}, \bibinfo {author} {\bibfnamefont {M.}~\bibnamefont {Onorato}},
  \bibinfo {author} {\bibfnamefont {E.}~\bibnamefont {Bitner-Gregersen}},
  \bibinfo {author} {\bibfnamefont {A.~R.}\ \bibnamefont {Osborne}},\ and\
  \bibinfo {author} {\bibfnamefont {A.~V.}\ \bibnamefont {Babanin}},\
  }\bibfield  {title} {\bibinfo {title} {Surface gravity waves from direct
  numerical simulations of the {E}uler equations: A comparison with
  second-order theory},\ }\href@noop {} {\bibfield  {journal} {\bibinfo
  {journal} {Ocean Eng.}\ }\textbf {\bibinfo {volume} {35}},\ \bibinfo {pages}
  {367} (\bibinfo {year} {2008})}\BibitemShut {NoStop}%
\bibitem [{\citenamefont {Longuet-Higgins}(1952)}]{longuet52}%
  \BibitemOpen
  \bibfield  {author} {\bibinfo {author} {\bibfnamefont {M.~S.}\ \bibnamefont
  {Longuet-Higgins}},\ }\bibfield  {title} {\bibinfo {title} {On the
  statistical distribution of the height of sea waves},\ }\href@noop {}
  {\bibfield  {journal} {\bibinfo  {journal} {J. Mar. Res.}\ }\textbf {\bibinfo
  {volume} {11}},\ \bibinfo {pages} {245} (\bibinfo {year} {1952})}\BibitemShut
  {NoStop}%
\bibitem [{\citenamefont {Petrova}\ \emph {et~al.}(2006)\citenamefont
  {Petrova}, \citenamefont {Cherneva},\ and\ \citenamefont
  {Soares}}]{petrova06}%
  \BibitemOpen
  \bibfield  {author} {\bibinfo {author} {\bibfnamefont {P.}~\bibnamefont
  {Petrova}}, \bibinfo {author} {\bibfnamefont {Z.}~\bibnamefont {Cherneva}},\
  and\ \bibinfo {author} {\bibfnamefont {C.~G.}\ \bibnamefont {Soares}},\
  }\bibfield  {title} {\bibinfo {title} {Distribution of crest heights in sea
  states with abnormal waves},\ }\href@noop {} {\bibfield  {journal} {\bibinfo
  {journal} {Appl. Ocean Res.}\ }\textbf {\bibinfo {volume} {28}},\ \bibinfo
  {pages} {235} (\bibinfo {year} {2006})}\BibitemShut {NoStop}%
\bibitem [{\citenamefont {Fedele}\ and\ \citenamefont
  {Tayfun}(2009)}]{fedele09}%
  \BibitemOpen
  \bibfield  {author} {\bibinfo {author} {\bibfnamefont {F.}~\bibnamefont
  {Fedele}}\ and\ \bibinfo {author} {\bibfnamefont {M.~A.}\ \bibnamefont
  {Tayfun}},\ }\bibfield  {title} {\bibinfo {title} {On nonlinear wave groups
  and crest statistics},\ }\href@noop {} {\bibfield  {journal} {\bibinfo
  {journal} {J. Fluid Mech.}\ }\textbf {\bibinfo {volume} {620}},\ \bibinfo
  {pages} {221} (\bibinfo {year} {2009})}\BibitemShut {NoStop}%
\bibitem [{\citenamefont {Fedele}\ \emph {et~al.}(2019)\citenamefont {Fedele},
  \citenamefont {Herterich}, \citenamefont {Tayfun},\ and\ \citenamefont
  {Dias}}]{Fedele19}%
  \BibitemOpen
  \bibfield  {author} {\bibinfo {author} {\bibfnamefont {F.}~\bibnamefont
  {Fedele}}, \bibinfo {author} {\bibfnamefont {J.}~\bibnamefont {Herterich}},
  \bibinfo {author} {\bibfnamefont {A.}~\bibnamefont {Tayfun}},\ and\ \bibinfo
  {author} {\bibfnamefont {F.}~\bibnamefont {Dias}},\ }\bibfield  {title}
  {\bibinfo {title} {Large nearshore storm waves off the {Irish} coast},\
  }\href@noop {} {\bibfield  {journal} {\bibinfo  {journal} {Sci. Rep.}\
  }\textbf {\bibinfo {volume} {9}},\ \bibinfo {pages} {15406} (\bibinfo {year}
  {2019})}\BibitemShut {NoStop}%
\bibitem [{\citenamefont {Shrira}\ and\ \citenamefont
  {Slunyaev}(2014{\natexlab{a}})}]{shrira14}%
  \BibitemOpen
  \bibfield  {author} {\bibinfo {author} {\bibfnamefont {V.~I.}\ \bibnamefont
  {Shrira}}\ and\ \bibinfo {author} {\bibfnamefont {A.~V.}\ \bibnamefont
  {Slunyaev}},\ }\bibfield  {title} {\bibinfo {title} {Trapped waves on jet
  currents: asymptotic modal approach},\ }\href@noop {} {\bibfield  {journal}
  {\bibinfo  {journal} {J. Fluid Mech.}\ }\textbf {\bibinfo {volume} {738}},\
  \bibinfo {pages} {65} (\bibinfo {year} {2014}{\natexlab{a}})}\BibitemShut
  {NoStop}%
\bibitem [{\citenamefont {Shrira}\ and\ \citenamefont
  {Slunyaev}(2014{\natexlab{b}})}]{shrira14b}%
  \BibitemOpen
  \bibfield  {author} {\bibinfo {author} {\bibfnamefont {V.~I.}\ \bibnamefont
  {Shrira}}\ and\ \bibinfo {author} {\bibfnamefont {A.~V.}\ \bibnamefont
  {Slunyaev}},\ }\bibfield  {title} {\bibinfo {title} {Nonlinear dynamics of
  trapped waves on jet currents and rogue waves},\ }\href@noop {} {\bibfield
  {journal} {\bibinfo  {journal} {Phys. Rev. E}\ }\textbf {\bibinfo {volume}
  {89}},\ \bibinfo {pages} {041002(R)} (\bibinfo {year}
  {2014}{\natexlab{b}})}\BibitemShut {NoStop}%
\bibitem [{\citenamefont {Peregrine}(1976)}]{peregrine76}%
  \BibitemOpen
  \bibfield  {author} {\bibinfo {author} {\bibfnamefont {D.~H.}\ \bibnamefont
  {Peregrine}},\ }\bibfield  {title} {\bibinfo {title} {Interaction of water
  waves and currents},\ }\href@noop {} {\bibfield  {journal} {\bibinfo
  {journal} {Adv. Appl. Mech.}\ }\textbf {\bibinfo {volume} {16}},\ \bibinfo
  {pages} {9} (\bibinfo {year} {1976})}\BibitemShut {NoStop}%
\bibitem [{\citenamefont {Stocker}\ and\ \citenamefont
  {Peregrine}(1999)}]{stocker99}%
  \BibitemOpen
  \bibfield  {author} {\bibinfo {author} {\bibfnamefont {J.~R.}\ \bibnamefont
  {Stocker}}\ and\ \bibinfo {author} {\bibfnamefont {D.~H.}\ \bibnamefont
  {Peregrine}},\ }\bibfield  {title} {\bibinfo {title} {The current-modified
  nonlinear {S}chr{\"o}dinger equation},\ }\href@noop {} {\bibfield  {journal}
  {\bibinfo  {journal} {J. Fluid Mech.}\ }\textbf {\bibinfo {volume} {399}},\
  \bibinfo {pages} {335} (\bibinfo {year} {1999})}\BibitemShut {NoStop}%
\bibitem [{\citenamefont {Curtis}\ \emph {et~al.}(2018)\citenamefont {Curtis},
  \citenamefont {Carter},\ and\ \citenamefont {Kalisch}}]{Curtis18}%
  \BibitemOpen
  \bibfield  {author} {\bibinfo {author} {\bibfnamefont {C.~W.}\ \bibnamefont
  {Curtis}}, \bibinfo {author} {\bibfnamefont {J.~D.}\ \bibnamefont {Carter}},\
  and\ \bibinfo {author} {\bibfnamefont {H.}~\bibnamefont {Kalisch}},\
  }\bibfield  {title} {\bibinfo {title} {Particle paths in nonlinear
  {Schr{\"o}dinger} models in the presence of linear shear currents},\
  }\href@noop {} {\bibfield  {journal} {\bibinfo  {journal} {J. Fluid Mech.}\
  }\textbf {\bibinfo {volume} {855}},\ \bibinfo {pages} {322} (\bibinfo {year}
  {2018})}\BibitemShut {NoStop}%
\bibitem [{\citenamefont {Hjelmervik}\ and\ \citenamefont
  {Trulsen}(2009)}]{Hjelmervik09}%
  \BibitemOpen
  \bibfield  {author} {\bibinfo {author} {\bibfnamefont {K.~B.}\ \bibnamefont
  {Hjelmervik}}\ and\ \bibinfo {author} {\bibfnamefont {K.}~\bibnamefont
  {Trulsen}},\ }\bibfield  {title} {\bibinfo {title} {Freak wave statistics on
  collinear currents},\ }\href@noop {} {\bibfield  {journal} {\bibinfo
  {journal} {J. Fluid Mech.}\ }\textbf {\bibinfo {volume} {637}},\ \bibinfo
  {pages} {267} (\bibinfo {year} {2009})}\BibitemShut {NoStop}%
\bibitem [{\citenamefont {Onorato}\ \emph {et~al.}(2011)\citenamefont
  {Onorato}, \citenamefont {Proment},\ and\ \citenamefont
  {Toffoli}}]{onorato11}%
  \BibitemOpen
  \bibfield  {author} {\bibinfo {author} {\bibfnamefont {M.}~\bibnamefont
  {Onorato}}, \bibinfo {author} {\bibfnamefont {D.}~\bibnamefont {Proment}},\
  and\ \bibinfo {author} {\bibfnamefont {A.}~\bibnamefont {Toffoli}},\
  }\bibfield  {title} {\bibinfo {title} {Triggering rogue waves in opposing
  currents},\ }\href@noop {} {\bibfield  {journal} {\bibinfo  {journal} {Phys.
  Rev. Lett.}\ }\textbf {\bibinfo {volume} {107}},\ \bibinfo {pages} {184502}
  (\bibinfo {year} {2011})}\BibitemShut {NoStop}%
\bibitem [{\citenamefont {Ellingsen}(2016)}]{Ellingsen16}%
  \BibitemOpen
  \bibfield  {author} {\bibinfo {author} {\bibfnamefont {S.~{\AA}.}\
  \bibnamefont {Ellingsen}},\ }\bibfield  {title} {\bibinfo {title} {Oblique
  waves on a vertically sheared current are rotational},\ }\href@noop {}
  {\bibfield  {journal} {\bibinfo  {journal} {Eur. J. Mech. B. Fluids}\
  }\textbf {\bibinfo {volume} {56}},\ \bibinfo {pages} {156} (\bibinfo {year}
  {2016})}\BibitemShut {NoStop}%
\bibitem [{\citenamefont {Dalrymple}(1974)}]{Dalrymple74}%
  \BibitemOpen
  \bibfield  {author} {\bibinfo {author} {\bibfnamefont {R.~A.}\ \bibnamefont
  {Dalrymple}},\ }\bibfield  {title} {\bibinfo {title} {A finite amplitude wave
  on a linear shear current},\ }\href@noop {} {\bibfield  {journal} {\bibinfo
  {journal} {J. Geophys. Res.}\ }\textbf {\bibinfo {volume} {79}},\ \bibinfo
  {pages} {4498} (\bibinfo {year} {1974})}\BibitemShut {NoStop}%
\bibitem [{\citenamefont {Thomas}\ \emph {et~al.}(2012)\citenamefont {Thomas},
  \citenamefont {Kharif},\ and\ \citenamefont {Manna}}]{thomas12}%
  \BibitemOpen
  \bibfield  {author} {\bibinfo {author} {\bibfnamefont {R.}~\bibnamefont
  {Thomas}}, \bibinfo {author} {\bibfnamefont {C.}~\bibnamefont {Kharif}},\
  and\ \bibinfo {author} {\bibfnamefont {M.}~\bibnamefont {Manna}},\ }\bibfield
   {title} {\bibinfo {title} {A nonlinear {Schr{\"o}dinger} equation for water
  waves on finite depth with constant vorticity},\ }\href@noop {} {\bibfield
  {journal} {\bibinfo  {journal} {Phys. Fluids}\ }\textbf {\bibinfo {volume}
  {24}},\ \bibinfo {pages} {127102} (\bibinfo {year} {2012})}\BibitemShut
  {NoStop}%
\bibitem [{\citenamefont {Touboul}\ and\ \citenamefont
  {Kharif}(2016)}]{touboul16}%
  \BibitemOpen
  \bibfield  {author} {\bibinfo {author} {\bibfnamefont {J.}~\bibnamefont
  {Touboul}}\ and\ \bibinfo {author} {\bibfnamefont {C.}~\bibnamefont
  {Kharif}},\ }\bibfield  {title} {\bibinfo {title} {Effect of vorticity on the
  generation of rogue waves due to dispersive focusing},\ }\href@noop {}
  {\bibfield  {journal} {\bibinfo  {journal} {Nat. Hazards}\ }\textbf {\bibinfo
  {volume} {84}},\ \bibinfo {pages} {585} (\bibinfo {year} {2016})}\BibitemShut
  {NoStop}%
\bibitem [{\citenamefont {Liao}\ \emph {et~al.}(2017)\citenamefont {Liao},
  \citenamefont {Dong}, \citenamefont {Ma},\ and\ \citenamefont
  {Gao}}]{liao17}%
  \BibitemOpen
  \bibfield  {author} {\bibinfo {author} {\bibfnamefont {B.}~\bibnamefont
  {Liao}}, \bibinfo {author} {\bibfnamefont {G.}~\bibnamefont {Dong}}, \bibinfo
  {author} {\bibfnamefont {Y.}~\bibnamefont {Ma}},\ and\ \bibinfo {author}
  {\bibfnamefont {J.~L.}\ \bibnamefont {Gao}},\ }\bibfield  {title} {\bibinfo
  {title} {Linear-shear-current modified {Schr{\"o}dinger} equation for gravity
  waves in finite water depth},\ }\href@noop {} {\bibfield  {journal} {\bibinfo
   {journal} {Phys. Rev. E}\ }\textbf {\bibinfo {volume} {96}},\ \bibinfo
  {pages} {043111} (\bibinfo {year} {2017})}\BibitemShut {NoStop}%
\bibitem [{\citenamefont {Hsu}\ \emph {et~al.}(2018)\citenamefont {Hsu},
  \citenamefont {Kharif}, \citenamefont {Abid},\ and\ \citenamefont
  {Chen}}]{hsu18}%
  \BibitemOpen
  \bibfield  {author} {\bibinfo {author} {\bibfnamefont {H.~C.}\ \bibnamefont
  {Hsu}}, \bibinfo {author} {\bibfnamefont {C.}~\bibnamefont {Kharif}},
  \bibinfo {author} {\bibfnamefont {M.}~\bibnamefont {Abid}},\ and\ \bibinfo
  {author} {\bibfnamefont {Y.~Y.}\ \bibnamefont {Chen}},\ }\bibfield  {title}
  {\bibinfo {title} {A nonlinear {Schr{\"o}dinger} equation for
  gravity--capillary water waves on arbitrary depth with constant vorticity.
  part 1},\ }\href@noop {} {\bibfield  {journal} {\bibinfo  {journal} {J. Fluid
  Mech.}\ }\textbf {\bibinfo {volume} {854}},\ \bibinfo {pages} {146} (\bibinfo
  {year} {2018})}\BibitemShut {NoStop}%
\bibitem [{\citenamefont {Akselsen}\ and\ \citenamefont
  {Ellingsen}(2019)}]{Akselsen19}%
  \BibitemOpen
  \bibfield  {author} {\bibinfo {author} {\bibfnamefont {A.~H.}\ \bibnamefont
  {Akselsen}}\ and\ \bibinfo {author} {\bibfnamefont {S.}~\bibnamefont
  {Ellingsen}},\ }\bibfield  {title} {\bibinfo {title} {Weakly nonlinear
  transient waves on a shear current: Ring waves and skewed langmuir rolls},\
  }\href@noop {} {\bibfield  {journal} {\bibinfo  {journal} {J. Fluid Mech.}\
  }\textbf {\bibinfo {volume} {863}},\ \bibinfo {pages} {114} (\bibinfo {year}
  {2019})}\BibitemShut {NoStop}%
\bibitem [{\citenamefont {Baumstein}(1998)}]{baumstein98}%
  \BibitemOpen
  \bibfield  {author} {\bibinfo {author} {\bibfnamefont {A.~I.}\ \bibnamefont
  {Baumstein}},\ }\bibfield  {title} {\bibinfo {title} {Modulation of gravity
  waves with shear in water},\ }\href@noop {} {\bibfield  {journal} {\bibinfo
  {journal} {Stud. Appl. Math.}\ }\textbf {\bibinfo {volume} {100}},\ \bibinfo
  {pages} {365} (\bibinfo {year} {1998})}\BibitemShut {NoStop}%
\bibitem [{\citenamefont {Steer}\ \emph {et~al.}(2020)\citenamefont {Steer},
  \citenamefont {Borthwick}, \citenamefont {Stagonas}, \citenamefont
  {Buldakov},\ and\ \citenamefont {van~den Bremer}}]{steer20}%
  \BibitemOpen
  \bibfield  {author} {\bibinfo {author} {\bibfnamefont {J.~N.}\ \bibnamefont
  {Steer}}, \bibinfo {author} {\bibfnamefont {A.~G.}\ \bibnamefont
  {Borthwick}}, \bibinfo {author} {\bibfnamefont {D.}~\bibnamefont {Stagonas}},
  \bibinfo {author} {\bibfnamefont {E.}~\bibnamefont {Buldakov}},\ and\
  \bibinfo {author} {\bibfnamefont {T.~S.}\ \bibnamefont {van~den Bremer}},\
  }\bibfield  {title} {\bibinfo {title} {Experimental study of dispersion and
  modulational instability of surface gravity waves on constant vorticity
  currents},\ }\href@noop {} {\bibfield  {journal} {\bibinfo  {journal} {J.
  Fluid Mech.}\ }\textbf {\bibinfo {volume} {884}} (\bibinfo {year}
  {2020})}\BibitemShut {NoStop}%
\bibitem [{\citenamefont {Pizzo}\ \emph {et~al.}(2023)\citenamefont {Pizzo},
  \citenamefont {Lenain}, \citenamefont {R{\o}mcke}, \citenamefont
  {Ellingsen},\ and\ \citenamefont {Smeltzer}}]{pizzo23}%
  \BibitemOpen
  \bibfield  {author} {\bibinfo {author} {\bibfnamefont {N.}~\bibnamefont
  {Pizzo}}, \bibinfo {author} {\bibfnamefont {L.}~\bibnamefont {Lenain}},
  \bibinfo {author} {\bibfnamefont {O.}~\bibnamefont {R{\o}mcke}}, \bibinfo
  {author} {\bibfnamefont {S.~{\AA}.}\ \bibnamefont {Ellingsen}},\ and\
  \bibinfo {author} {\bibfnamefont {B.~K.}\ \bibnamefont {Smeltzer}},\
  }\bibfield  {title} {\bibinfo {title} {The role of {Lagrangian} drift in the
  geometry, kinematics and dynamics of surface waves},\ }\href@noop {}
  {\bibfield  {journal} {\bibinfo  {journal} {J. Fluid Mech.}\ }\textbf
  {\bibinfo {volume} {954}},\ \bibinfo {pages} {R4} (\bibinfo {year}
  {2023})}\BibitemShut {NoStop}%
\bibitem [{\citenamefont {Francius}\ and\ \citenamefont
  {Kharif}(2017)}]{francius17}%
  \BibitemOpen
  \bibfield  {author} {\bibinfo {author} {\bibfnamefont {M.}~\bibnamefont
  {Francius}}\ and\ \bibinfo {author} {\bibfnamefont {C.}~\bibnamefont
  {Kharif}},\ }\bibfield  {title} {\bibinfo {title} {Two-dimensional stability
  of finite-amplitude gravity waves on water of finite depth with constant
  vorticity},\ }\href@noop {} {\bibfield  {journal} {\bibinfo  {journal} {J.
  Fluid Mech.}\ }\textbf {\bibinfo {volume} {830}},\ \bibinfo {pages} {631}
  (\bibinfo {year} {2017})}\BibitemShut {NoStop}%
\bibitem [{\citenamefont {Abrashkin}\ and\ \citenamefont
  {Pelinovsky}(2017)}]{abrashkin17}%
  \BibitemOpen
  \bibfield  {author} {\bibinfo {author} {\bibfnamefont {A.}~\bibnamefont
  {Abrashkin}}\ and\ \bibinfo {author} {\bibfnamefont {E.}~\bibnamefont
  {Pelinovsky}},\ }\bibfield  {title} {\bibinfo {title} {Lagrange form of the
  nonlinear {Schr{\"o}dinger} equation for low-vorticity waves in deep water},\
  }\href@noop {} {\bibfield  {journal} {\bibinfo  {journal} {Nonlinear Process
  Geophys.}\ }\textbf {\bibinfo {volume} {24}},\ \bibinfo {pages} {255}
  (\bibinfo {year} {2017})}\BibitemShut {NoStop}%
\bibitem [{\citenamefont {Voronovich}(1976)}]{voronovich76}%
  \BibitemOpen
  \bibfield  {author} {\bibinfo {author} {\bibfnamefont {A.~G.}\ \bibnamefont
  {Voronovich}},\ }\bibfield  {title} {\bibinfo {title} {Propagation of
  internal and surface gravity waves in the approximation of geometrical
  optics},\ }\href@noop {} {\bibfield  {journal} {\bibinfo  {journal} {Izv. -
  Atmos. Ocean. Phys.}\ }\textbf {\bibinfo {volume} {12}},\ \bibinfo {pages}
  {850} (\bibinfo {year} {1976})}\BibitemShut {NoStop}%
\bibitem [{\citenamefont {Quinn}\ \emph {et~al.}(2017)\citenamefont {Quinn},
  \citenamefont {Toledo},\ and\ \citenamefont {Shrira}}]{quinn17}%
  \BibitemOpen
  \bibfield  {author} {\bibinfo {author} {\bibfnamefont {B.}~\bibnamefont
  {Quinn}}, \bibinfo {author} {\bibfnamefont {Y.}~\bibnamefont {Toledo}},\ and\
  \bibinfo {author} {\bibfnamefont {V.}~\bibnamefont {Shrira}},\ }\bibfield
  {title} {\bibinfo {title} {Explicit wave action conservation for water waves
  on vertically sheared flows},\ }\href@noop {} {\bibfield  {journal} {\bibinfo
   {journal} {Ocean Model.}\ }\textbf {\bibinfo {volume} {112}},\ \bibinfo
  {pages} {33} (\bibinfo {year} {2017})}\BibitemShut {NoStop}%
\bibitem [{\citenamefont {Banihashemi}\ \emph {et~al.}(2017)\citenamefont
  {Banihashemi}, \citenamefont {Kirby},\ and\ \citenamefont
  {Dong}}]{Banihashemi17}%
  \BibitemOpen
  \bibfield  {author} {\bibinfo {author} {\bibfnamefont {S.}~\bibnamefont
  {Banihashemi}}, \bibinfo {author} {\bibfnamefont {J.~T.}\ \bibnamefont
  {Kirby}},\ and\ \bibinfo {author} {\bibfnamefont {Z.}~\bibnamefont {Dong}},\
  }\bibfield  {title} {\bibinfo {title} {Approximation of wave action flux
  velocity in strongly sheared mean flows},\ }\href@noop {} {\bibfield
  {journal} {\bibinfo  {journal} {Ocean Model.}\ }\textbf {\bibinfo {volume}
  {116}},\ \bibinfo {pages} {33} (\bibinfo {year} {2017})}\BibitemShut
  {NoStop}%
\bibitem [{\citenamefont {Li}\ and\ \citenamefont {Ellingsen}(2019)}]{li19}%
  \BibitemOpen
  \bibfield  {author} {\bibinfo {author} {\bibfnamefont {Y.}~\bibnamefont
  {Li}}\ and\ \bibinfo {author} {\bibfnamefont {S.}~\bibnamefont {Ellingsen}},\
  }\bibfield  {title} {\bibinfo {title} {A framework for modeling linear
  surface waves on shear currents in slowly varying waters},\ }\href@noop {}
  {\bibfield  {journal} {\bibinfo  {journal} {J. Geophys. Res.: Oceans}\
  }\textbf {\bibinfo {volume} {124}},\ \bibinfo {pages} {2527} (\bibinfo {year}
  {2019})}\BibitemShut {NoStop}%
\bibitem [{\citenamefont {Banihashemi}\ and\ \citenamefont
  {Kirby}(2019)}]{banihashemi19}%
  \BibitemOpen
  \bibfield  {author} {\bibinfo {author} {\bibfnamefont {S.}~\bibnamefont
  {Banihashemi}}\ and\ \bibinfo {author} {\bibfnamefont {J.~T.}\ \bibnamefont
  {Kirby}},\ }\bibfield  {title} {\bibinfo {title} {Approximation of wave
  action conservation in vertically sheared mean flows},\ }\href@noop {}
  {\bibfield  {journal} {\bibinfo  {journal} {Ocean Model.}\ }\textbf {\bibinfo
  {volume} {143}},\ \bibinfo {pages} {101460} (\bibinfo {year}
  {2019})}\BibitemShut {NoStop}%
\bibitem [{\citenamefont {Cummins}\ and\ \citenamefont
  {Swan}(1995)}]{Cummins94}%
  \BibitemOpen
  \bibfield  {author} {\bibinfo {author} {\bibfnamefont {I.}~\bibnamefont
  {Cummins}}\ and\ \bibinfo {author} {\bibfnamefont {C.}~\bibnamefont {Swan}},\
  }\bibfield  {title} {\bibinfo {title} {Vorticity effects in combined waves
  and currents},\ }in\ \href@noop {} {\emph {\bibinfo {booktitle} {Coastal
  Engineering 1994}}}\ (\bibinfo {year} {1995})\ pp.\ \bibinfo {pages}
  {113--127}\BibitemShut {NoStop}%
\bibitem [{\citenamefont {Waseda}\ \emph {et~al.}(2015)\citenamefont {Waseda},
  \citenamefont {Kinoshita}, \citenamefont {Cavaleri},\ and\ \citenamefont
  {Toffoli}}]{Waseda15}%
  \BibitemOpen
  \bibfield  {author} {\bibinfo {author} {\bibfnamefont {T.}~\bibnamefont
  {Waseda}}, \bibinfo {author} {\bibfnamefont {T.}~\bibnamefont {Kinoshita}},
  \bibinfo {author} {\bibfnamefont {L.}~\bibnamefont {Cavaleri}},\ and\
  \bibinfo {author} {\bibfnamefont {A.}~\bibnamefont {Toffoli}},\ }\bibfield
  {title} {\bibinfo {title} {Third-order resonant wave interactions under the
  influence of background current fields},\ }\href@noop {} {\bibfield
  {journal} {\bibinfo  {journal} {J. Fluid Mech.}\ }\textbf {\bibinfo {volume}
  {784}},\ \bibinfo {pages} {51} (\bibinfo {year} {2015})}\BibitemShut
  {NoStop}%
\bibitem [{\citenamefont {Smeltzer}\ \emph {et~al.}(2019)\citenamefont
  {Smeltzer}, \citenamefont {Æsøy},\ and\ \citenamefont
  {Ellingsen}}]{Smeltzer19}%
  \BibitemOpen
  \bibfield  {author} {\bibinfo {author} {\bibfnamefont {B.~K.}\ \bibnamefont
  {Smeltzer}}, \bibinfo {author} {\bibfnamefont {E.}~\bibnamefont {Æsøy}},\
  and\ \bibinfo {author} {\bibfnamefont {S.~A.}\ \bibnamefont {Ellingsen}},\
  }\bibfield  {title} {\bibinfo {title} {Observation of surface wave patterns
  modified by sub-surface shear currents},\ }\href@noop {} {\bibfield
  {journal} {\bibinfo  {journal} {J. Fluid Mech.}\ }\textbf {\bibinfo {volume}
  {873}},\ \bibinfo {pages} {508} (\bibinfo {year} {2019})}\BibitemShut
  {NoStop}%
\bibitem [{\citenamefont {Ardhuin}(2017)}]{Ardhuin17}%
  \BibitemOpen
  \bibfield  {author} {\bibinfo {author} {\bibfnamefont {F.}~\bibnamefont
  {Ardhuin}},\ }\bibfield  {title} {\bibinfo {title} {Small-scale open ocean
  currents have large effects on wind wave heights},\ }\href@noop {} {\bibfield
   {journal} {\bibinfo  {journal} {J. Geophys. Res.: Oceans}\ }\textbf
  {\bibinfo {volume} {122}},\ \bibinfo {pages} {1} (\bibinfo {year}
  {2017})}\BibitemShut {NoStop}%
\bibitem [{\citenamefont {Ardhuin}\ \emph {et~al.}(2009)\citenamefont
  {Ardhuin}, \citenamefont {Marié}, \citenamefont {Rascle}, \citenamefont
  {Forget},\ and\ \citenamefont {Roland}}]{Ardhuin09}%
  \BibitemOpen
  \bibfield  {author} {\bibinfo {author} {\bibfnamefont {F.}~\bibnamefont
  {Ardhuin}}, \bibinfo {author} {\bibfnamefont {L.}~\bibnamefont {Marié}},
  \bibinfo {author} {\bibfnamefont {N.}~\bibnamefont {Rascle}}, \bibinfo
  {author} {\bibfnamefont {P.}~\bibnamefont {Forget}},\ and\ \bibinfo {author}
  {\bibfnamefont {A.}~\bibnamefont {Roland}},\ }\bibfield  {title} {\bibinfo
  {title} {Observation and estimation of {L}agrangian, {S}tokes, and {E}ulerian
  currents induced by wind and waves at the sea surface},\ }\href@noop {}
  {\bibfield  {journal} {\bibinfo  {journal} {J. Phys. Oceanogr.}\ }\textbf
  {\bibinfo {volume} {39}},\ \bibinfo {pages} {2820} (\bibinfo {year}
  {2009})}\BibitemShut {NoStop}%
\bibitem [{\citenamefont {Zippel}\ and\ \citenamefont
  {Thomson}(2017)}]{Zippel17}%
  \BibitemOpen
  \bibfield  {author} {\bibinfo {author} {\bibfnamefont {S.}~\bibnamefont
  {Zippel}}\ and\ \bibinfo {author} {\bibfnamefont {J.}~\bibnamefont
  {Thomson}},\ }\bibfield  {title} {\bibinfo {title} {Surface wave breaking
  over sheared currents: Observations from the mouth of the {C}olumbia
  {R}iver},\ }\href@noop {} {\bibfield  {journal} {\bibinfo  {journal} {J.
  Geophys. Res.: Oceans}\ }\textbf {\bibinfo {volume} {122}},\ \bibinfo {pages}
  {3311} (\bibinfo {year} {2017})}\BibitemShut {NoStop}%
\bibitem [{\citenamefont {Stewart}\ and\ \citenamefont
  {Joy}(1974)}]{Stewart74}%
  \BibitemOpen
  \bibfield  {author} {\bibinfo {author} {\bibfnamefont {R.~H.}\ \bibnamefont
  {Stewart}}\ and\ \bibinfo {author} {\bibfnamefont {J.~W.}\ \bibnamefont
  {Joy}},\ }\bibfield  {title} {\bibinfo {title} {{HF} radio measurements of
  surface currents},\ }\href@noop {} {\bibfield  {journal} {\bibinfo  {journal}
  {Deep-Sea Res. Oceanogr. Abstracts}\ }\textbf {\bibinfo {volume} {21}},\
  \bibinfo {pages} {1039} (\bibinfo {year} {1974})}\BibitemShut {NoStop}%
\bibitem [{\citenamefont {Skop}(1987)}]{skop87}%
  \BibitemOpen
  \bibfield  {author} {\bibinfo {author} {\bibfnamefont {R.~A.}\ \bibnamefont
  {Skop}},\ }\bibfield  {title} {\bibinfo {title} {Approximate dispersion
  relation for wave-current interactions},\ }\href@noop {} {\bibfield
  {journal} {\bibinfo  {journal} {J. Waterw. Port, Coast. Ocean Eng.}\ }\textbf
  {\bibinfo {volume} {113}},\ \bibinfo {pages} {187} (\bibinfo {year}
  {1987})}\BibitemShut {NoStop}%
\bibitem [{\citenamefont {Kirby}\ and\ \citenamefont {Chen}(1989)}]{kirby89}%
  \BibitemOpen
  \bibfield  {author} {\bibinfo {author} {\bibfnamefont {J.~T.}\ \bibnamefont
  {Kirby}}\ and\ \bibinfo {author} {\bibfnamefont {T.}~\bibnamefont {Chen}},\
  }\bibfield  {title} {\bibinfo {title} {Surface waves on vertically sheared
  flows: approximate dispersion relations},\ }\href@noop {} {\bibfield
  {journal} {\bibinfo  {journal} {J. Geophys. Res. Oceans}\ }\textbf {\bibinfo
  {volume} {94}},\ \bibinfo {pages} {1013} (\bibinfo {year}
  {1989})}\BibitemShut {NoStop}%
\bibitem [{\citenamefont {Zakharov}\ and\ \citenamefont
  {Shrira}(1990)}]{zakharov90}%
  \BibitemOpen
  \bibfield  {author} {\bibinfo {author} {\bibfnamefont {V.~E.}\ \bibnamefont
  {Zakharov}}\ and\ \bibinfo {author} {\bibfnamefont {V.~I.}\ \bibnamefont
  {Shrira}},\ }\bibfield  {title} {\bibinfo {title} {Formation of the angular
  spectrum of wind waves},\ }\href@noop {} {\bibfield  {journal} {\bibinfo
  {journal} {Sov. phys. JETP}\ }\textbf {\bibinfo {volume} {71}},\ \bibinfo
  {pages} {1091} (\bibinfo {year} {1990})}\BibitemShut {NoStop}%
\bibitem [{\citenamefont {Shrira}(1993)}]{shrira93}%
  \BibitemOpen
  \bibfield  {author} {\bibinfo {author} {\bibfnamefont {V.~I.}\ \bibnamefont
  {Shrira}},\ }\bibfield  {title} {\bibinfo {title} {Surface waves on shear
  currents: solution of the boundary-value problem},\ }\href@noop {} {\bibfield
   {journal} {\bibinfo  {journal} {J. Fluid Mech.}\ }\textbf {\bibinfo {volume}
  {252}},\ \bibinfo {pages} {565} (\bibinfo {year} {1993})}\BibitemShut
  {NoStop}%
\bibitem [{\citenamefont {Ellingsen}\ and\ \citenamefont
  {Li}(2017)}]{Ellingsen17}%
  \BibitemOpen
  \bibfield  {author} {\bibinfo {author} {\bibfnamefont {S.~{\AA}.}\
  \bibnamefont {Ellingsen}}\ and\ \bibinfo {author} {\bibfnamefont
  {Y.}~\bibnamefont {Li}},\ }\bibfield  {title} {\bibinfo {title} {Approximate
  dispersion relations for waves on arbitrary shear flows},\ }\href@noop {}
  {\bibfield  {journal} {\bibinfo  {journal} {J. Geophys. Res.: Oceans}\
  }\textbf {\bibinfo {volume} {122}},\ \bibinfo {pages} {9889} (\bibinfo {year}
  {2017})}\BibitemShut {NoStop}%
\bibitem [{\citenamefont {Laxague}\ \emph {et~al.}(2017)\citenamefont
  {Laxague}, \citenamefont {Haus}, \citenamefont {Ortiz-Suslow}, \citenamefont
  {Smith}, \citenamefont {Novelli}, \citenamefont {Dai}, \citenamefont
  {{\"O}zg{\"o}kmen},\ and\ \citenamefont {Graber}}]{laxague17}%
  \BibitemOpen
  \bibfield  {author} {\bibinfo {author} {\bibfnamefont {N.~J.}\ \bibnamefont
  {Laxague}}, \bibinfo {author} {\bibfnamefont {B.~K.}\ \bibnamefont {Haus}},
  \bibinfo {author} {\bibfnamefont {D.~G.}\ \bibnamefont {Ortiz-Suslow}},
  \bibinfo {author} {\bibfnamefont {C.~J.}\ \bibnamefont {Smith}}, \bibinfo
  {author} {\bibfnamefont {G.}~\bibnamefont {Novelli}}, \bibinfo {author}
  {\bibfnamefont {H.}~\bibnamefont {Dai}}, \bibinfo {author} {\bibfnamefont
  {T.}~\bibnamefont {{\"O}zg{\"o}kmen}},\ and\ \bibinfo {author} {\bibfnamefont
  {H.~C.}\ \bibnamefont {Graber}},\ }\bibfield  {title} {\bibinfo {title}
  {Passive optical sensing of the near-surface wind-driven current profile},\
  }\href@noop {} {\bibfield  {journal} {\bibinfo  {journal} {J. Atmos. Ocean.
  Technol.}\ }\textbf {\bibinfo {volume} {34}},\ \bibinfo {pages} {1097}
  (\bibinfo {year} {2017})}\BibitemShut {NoStop}%
\bibitem [{\citenamefont {Laxague}\ \emph {et~al.}(2018)\citenamefont
  {Laxague}, \citenamefont {{\"O}zg{\"o}kmen}, \citenamefont {Haus},
  \citenamefont {Novelli}, \citenamefont {Shcherbina}, \citenamefont
  {Sutherland}, \citenamefont {Guigand}, \citenamefont {Lund}, \citenamefont
  {Mehta}, \citenamefont {Alday} \emph {et~al.}}]{laxague18}%
  \BibitemOpen
  \bibfield  {author} {\bibinfo {author} {\bibfnamefont {N.~J.}\ \bibnamefont
  {Laxague}}, \bibinfo {author} {\bibfnamefont {T.~M.}\ \bibnamefont
  {{\"O}zg{\"o}kmen}}, \bibinfo {author} {\bibfnamefont {B.~K.}\ \bibnamefont
  {Haus}}, \bibinfo {author} {\bibfnamefont {G.}~\bibnamefont {Novelli}},
  \bibinfo {author} {\bibfnamefont {A.}~\bibnamefont {Shcherbina}}, \bibinfo
  {author} {\bibfnamefont {P.}~\bibnamefont {Sutherland}}, \bibinfo {author}
  {\bibfnamefont {C.~M.}\ \bibnamefont {Guigand}}, \bibinfo {author}
  {\bibfnamefont {B.}~\bibnamefont {Lund}}, \bibinfo {author} {\bibfnamefont
  {S.}~\bibnamefont {Mehta}}, \bibinfo {author} {\bibfnamefont
  {M.}~\bibnamefont {Alday}}, \emph {et~al.},\ }\bibfield  {title} {\bibinfo
  {title} {Observations of near-surface current shear help describe oceanic oil
  and plastic transport},\ }\href@noop {} {\bibfield  {journal} {\bibinfo
  {journal} {Geophys. Res. Lett.}\ }\textbf {\bibinfo {volume} {45}},\ \bibinfo
  {pages} {245} (\bibinfo {year} {2018})}\BibitemShut {NoStop}%
\bibitem [{\citenamefont {Wu}(1983)}]{wu83}%
  \BibitemOpen
  \bibfield  {author} {\bibinfo {author} {\bibfnamefont {J.}~\bibnamefont
  {Wu}},\ }\bibfield  {title} {\bibinfo {title} {Sea-surface drift currents
  induced by wind and waves},\ }\href@noop {} {\bibfield  {journal} {\bibinfo
  {journal} {J. Phys. Oceanogr.}\ }\textbf {\bibinfo {volume} {13}},\ \bibinfo
  {pages} {1441} (\bibinfo {year} {1983})}\BibitemShut {NoStop}%
\bibitem [{\citenamefont {Kilcher}\ and\ \citenamefont
  {Nash}(2010)}]{kilcher10}%
  \BibitemOpen
  \bibfield  {author} {\bibinfo {author} {\bibfnamefont {L.~F.}\ \bibnamefont
  {Kilcher}}\ and\ \bibinfo {author} {\bibfnamefont {J.~D.}\ \bibnamefont
  {Nash}},\ }\bibfield  {title} {\bibinfo {title} {Structure and dynamics of
  the {Columbia} {River} tidal plume front},\ }\href@noop {} {\bibfield
  {journal} {\bibinfo  {journal} {J. Geophys. Res.: Oceans}\ }\textbf {\bibinfo
  {volume} {115}} (\bibinfo {year} {2010})}\BibitemShut {NoStop}%
\bibitem [{\citenamefont {Tucker}\ \emph {et~al.}(1984)\citenamefont {Tucker},
  \citenamefont {Challenor},\ and\ \citenamefont {Carter}}]{Tucker84}%
  \BibitemOpen
  \bibfield  {author} {\bibinfo {author} {\bibfnamefont {M.~J.}\ \bibnamefont
  {Tucker}}, \bibinfo {author} {\bibfnamefont {P.~G.}\ \bibnamefont
  {Challenor}},\ and\ \bibinfo {author} {\bibfnamefont {D.~J.~T.}\ \bibnamefont
  {Carter}},\ }\bibfield  {title} {\bibinfo {title} {Numerical simulation of a
  random sea: a common error and its effect upon wave group statistics},\
  }\href@noop {} {\bibfield  {journal} {\bibinfo  {journal} {Appl. Ocean Res.}\
  }\textbf {\bibinfo {volume} {6}},\ \bibinfo {pages} {118} (\bibinfo {year}
  {1984})}\BibitemShut {NoStop}%
\bibitem [{\citenamefont {Hasselmann}(1962)}]{hasselmann62}%
  \BibitemOpen
  \bibfield  {author} {\bibinfo {author} {\bibfnamefont {K.}~\bibnamefont
  {Hasselmann}},\ }\bibfield  {title} {\bibinfo {title} {On the non-linear
  energy transfer in a gravity-wave spectrum part 1. general theory},\
  }\href@noop {} {\bibfield  {journal} {\bibinfo  {journal} {J. Fluid Mech.}\
  }\textbf {\bibinfo {volume} {12}},\ \bibinfo {pages} {481} (\bibinfo {year}
  {1962})}\BibitemShut {NoStop}%
\bibitem [{\citenamefont {Dommermuth}\ and\ \citenamefont
  {Yue}(1987)}]{dommermuth87}%
  \BibitemOpen
  \bibfield  {author} {\bibinfo {author} {\bibfnamefont {D.~G.}\ \bibnamefont
  {Dommermuth}}\ and\ \bibinfo {author} {\bibfnamefont {D.~K.~P.}\ \bibnamefont
  {Yue}},\ }\bibfield  {title} {\bibinfo {title} {A high-order spectral method
  for the study of nonlinear gravity waves},\ }\href@noop {} {\bibfield
  {journal} {\bibinfo  {journal} {J. Fluid Mech.}\ }\textbf {\bibinfo {volume}
  {184}},\ \bibinfo {pages} {267} (\bibinfo {year} {1987})}\BibitemShut
  {NoStop}%
\bibitem [{\citenamefont {West}\ \emph {et~al.}(1987)\citenamefont {West},
  \citenamefont {Brueckner}, \citenamefont {Janda}, \citenamefont {Milder},\
  and\ \citenamefont {Milton}}]{west87}%
  \BibitemOpen
  \bibfield  {author} {\bibinfo {author} {\bibfnamefont {B.~J.}\ \bibnamefont
  {West}}, \bibinfo {author} {\bibfnamefont {K.~A.}\ \bibnamefont {Brueckner}},
  \bibinfo {author} {\bibfnamefont {R.~S.}\ \bibnamefont {Janda}}, \bibinfo
  {author} {\bibfnamefont {D.~M.}\ \bibnamefont {Milder}},\ and\ \bibinfo
  {author} {\bibfnamefont {R.~L.}\ \bibnamefont {Milton}},\ }\bibfield  {title}
  {\bibinfo {title} {A new numerical method for surface hydrodynamics},\
  }\href@noop {} {\bibfield  {journal} {\bibinfo  {journal} {J. Geophys. Res.:
  Oceans}\ }\textbf {\bibinfo {volume} {92}},\ \bibinfo {pages} {11803}
  (\bibinfo {year} {1987})}\BibitemShut {NoStop}%
\bibitem [{\citenamefont {Li}\ and\ \citenamefont {Li}(2021)}]{li21b}%
  \BibitemOpen
  \bibfield  {author} {\bibinfo {author} {\bibfnamefont {Y.}~\bibnamefont
  {Li}}\ and\ \bibinfo {author} {\bibfnamefont {X.}~\bibnamefont {Li}},\
  }\bibfield  {title} {\bibinfo {title} {Weakly nonlinear broadband and
  multi-directional surface waves on an arbitrary depth: a framework, stokes
  drift, and particle trajectories},\ }\href@noop {} {\bibfield  {journal}
  {\bibinfo  {journal} {Phys. Fluids}\ }\textbf {\bibinfo {volume} {33}},\
  \bibinfo {pages} {076609} (\bibinfo {year} {2021})}\BibitemShut {NoStop}%
\bibitem [{\citenamefont {Srokosz}\ and\ \citenamefont
  {Longuet-Higgins}(1986)}]{Srokosz86}%
  \BibitemOpen
  \bibfield  {author} {\bibinfo {author} {\bibfnamefont {M.~A.}\ \bibnamefont
  {Srokosz}}\ and\ \bibinfo {author} {\bibfnamefont {M.~S.}\ \bibnamefont
  {Longuet-Higgins}},\ }\bibfield  {title} {\bibinfo {title} {On the skewness
  of sea-surface elevation},\ }\href@noop {} {\bibfield  {journal} {\bibinfo
  {journal} {J. Fluid Mech.}\ }\textbf {\bibinfo {volume} {164}},\ \bibinfo
  {pages} {487} (\bibinfo {year} {1986})}\BibitemShut {NoStop}%
\bibitem [{\citenamefont {Craik}(1968)}]{Craik68}%
  \BibitemOpen
  \bibfield  {author} {\bibinfo {author} {\bibfnamefont {A.~D.~D.}\
  \bibnamefont {Craik}},\ }\bibfield  {title} {\bibinfo {title} {Resonant
  gravity-wave interactions in a shear flow},\ }\href@noop {} {\bibfield
  {journal} {\bibinfo  {journal} {J. Fluid Mech.}\ }\textbf {\bibinfo {volume}
  {34}},\ \bibinfo {pages} {531} (\bibinfo {year} {1968})}\BibitemShut
  {NoStop}%
\bibitem [{\citenamefont {Hasselmann}\ \emph {et~al.}(1973)\citenamefont
  {Hasselmann}, \citenamefont {Barnett}, \citenamefont {Bouws}, \citenamefont
  {Carlson}, \citenamefont {Cartwright}, \citenamefont {Eake}, \citenamefont
  {Euring}, \citenamefont {Gicnapp}, \citenamefont {Hasselmann},\ and\
  \citenamefont {Kruseman}}]{hasselmann73}%
  \BibitemOpen
  \bibfield  {author} {\bibinfo {author} {\bibfnamefont {K.~F.}\ \bibnamefont
  {Hasselmann}}, \bibinfo {author} {\bibfnamefont {T.~P.}\ \bibnamefont
  {Barnett}}, \bibinfo {author} {\bibfnamefont {E.}~\bibnamefont {Bouws}},
  \bibinfo {author} {\bibfnamefont {H.}~\bibnamefont {Carlson}}, \bibinfo
  {author} {\bibfnamefont {D.~E.}\ \bibnamefont {Cartwright}}, \bibinfo
  {author} {\bibfnamefont {K.}~\bibnamefont {Eake}}, \bibinfo {author}
  {\bibfnamefont {J.}~\bibnamefont {Euring}}, \bibinfo {author} {\bibfnamefont
  {A.}~\bibnamefont {Gicnapp}}, \bibinfo {author} {\bibfnamefont
  {D.}~\bibnamefont {Hasselmann}},\ and\ \bibinfo {author} {\bibfnamefont
  {P.}~\bibnamefont {Kruseman}},\ }\bibfield  {title} {\bibinfo {title}
  {Measurements of wind wave growth and swell decay during the {J}oint {N}orth
  {S}ea {W}ave {P}roject ({JONSWAP})},\ }\href@noop {} {\bibfield  {journal}
  {\bibinfo  {journal} {Deut. Hydrogr. Z.}\ }\textbf {\bibinfo {volume} {8}},\
  \bibinfo {pages} {1} (\bibinfo {year} {1973})}\BibitemShut {NoStop}%
\bibitem [{\citenamefont {Dysthe}\ \emph {et~al.}(2005)\citenamefont {Dysthe},
  \citenamefont {Socquet-Juglard}, \citenamefont {Trulsen}, \citenamefont
  {Krogstad},\ and\ \citenamefont {Liu}}]{Dysthe05}%
  \BibitemOpen
  \bibfield  {author} {\bibinfo {author} {\bibfnamefont {K.}~\bibnamefont
  {Dysthe}}, \bibinfo {author} {\bibfnamefont {H.}~\bibnamefont
  {Socquet-Juglard}}, \bibinfo {author} {\bibfnamefont {K.}~\bibnamefont
  {Trulsen}}, \bibinfo {author} {\bibfnamefont {H.~E.}\ \bibnamefont
  {Krogstad}},\ and\ \bibinfo {author} {\bibfnamefont {J.}~\bibnamefont
  {Liu}},\ }\bibfield  {title} {\bibinfo {title} {"freak" waves and large-scale
  simulations of surface gravity waves},\ }in\ \href@noop {} {\emph {\bibinfo
  {booktitle} {Proc. 14th ‘Aha Huliko’ a Hawaiian Winter Workshop}}}\
  (\bibinfo  {publisher} {Citeseer},\ \bibinfo {address} {University of Hawaii,
  U.S.A.},\ \bibinfo {year} {2005})\BibitemShut {NoStop}%
\bibitem [{\citenamefont {Socquet-Juglard}\ \emph {et~al.}(2005)\citenamefont
  {Socquet-Juglard}, \citenamefont {Dysthe}, \citenamefont {Trulsen},
  \citenamefont {Krogstad},\ and\ \citenamefont {Liu}}]{Socquet-Juglard05}%
  \BibitemOpen
  \bibfield  {author} {\bibinfo {author} {\bibfnamefont {H.}~\bibnamefont
  {Socquet-Juglard}}, \bibinfo {author} {\bibfnamefont {K.}~\bibnamefont
  {Dysthe}}, \bibinfo {author} {\bibfnamefont {K.}~\bibnamefont {Trulsen}},
  \bibinfo {author} {\bibfnamefont {H.~E.}\ \bibnamefont {Krogstad}},\ and\
  \bibinfo {author} {\bibfnamefont {J.}~\bibnamefont {Liu}},\ }\bibfield
  {title} {\bibinfo {title} {Probability distributions of surface gravity waves
  during spectral changes},\ }\href@noop {} {\bibfield  {journal} {\bibinfo
  {journal} {J. Fluid Mech.}\ }\textbf {\bibinfo {volume} {542}},\ \bibinfo
  {pages} {195} (\bibinfo {year} {2005})}\BibitemShut {NoStop}%
\bibitem [{\citenamefont {Longuet-Higgins}(1975)}]{Longuet75}%
  \BibitemOpen
  \bibfield  {author} {\bibinfo {author} {\bibfnamefont {M.~S.}\ \bibnamefont
  {Longuet-Higgins}},\ }\bibfield  {title} {\bibinfo {title} {On the joint
  distribution of the periods and amplitudes of sea waves},\ }\href@noop {}
  {\bibfield  {journal} {\bibinfo  {journal} {J. Geophys. Res.}\ }\textbf
  {\bibinfo {volume} {80}},\ \bibinfo {pages} {2688} (\bibinfo {year}
  {1975})}\BibitemShut {NoStop}%
\bibitem [{\citenamefont {Dong}\ and\ \citenamefont {Kirby}(2012)}]{Dong12}%
  \BibitemOpen
  \bibfield  {author} {\bibinfo {author} {\bibfnamefont {Z.}~\bibnamefont
  {Dong}}\ and\ \bibinfo {author} {\bibfnamefont {J.~T.}\ \bibnamefont
  {Kirby}},\ }\bibfield  {title} {\bibinfo {title} {Theoretical and numerical
  study of wave-current interaction in strongly-sheared flows},\ }\href@noop {}
  {\bibfield  {journal} {\bibinfo  {journal} {Coast. Eng. Proc.}\ }\textbf
  {\bibinfo {volume} {1}},\ \bibinfo {pages} {2} (\bibinfo {year}
  {2012})}\BibitemShut {NoStop}%
\bibitem [{\citenamefont {Elias}\ \emph {et~al.}(2012)\citenamefont {Elias},
  \citenamefont {Gelfenbaum},\ and\ \citenamefont {Van~der
  Westhuysen}}]{elias12}%
  \BibitemOpen
  \bibfield  {author} {\bibinfo {author} {\bibfnamefont {E.~P.}\ \bibnamefont
  {Elias}}, \bibinfo {author} {\bibfnamefont {G.}~\bibnamefont {Gelfenbaum}},\
  and\ \bibinfo {author} {\bibfnamefont {A.~J.}\ \bibnamefont {Van~der
  Westhuysen}},\ }\bibfield  {title} {\bibinfo {title} {Validation of a coupled
  wave‐flow model in a high‐energy setting: The mouth of the {C}olumbia
  {R}iver},\ }\href@noop {} {\bibfield  {journal} {\bibinfo  {journal} {J.
  Geophys. Res. Oceans}\ }\textbf {\bibinfo {volume} {117}} (\bibinfo {year}
  {2012})}\BibitemShut {NoStop}%
\bibitem [{\citenamefont {Maxwell}\ \emph {et~al.}(2020)\citenamefont
  {Maxwell}, \citenamefont {Smeltzer},\ and\ \citenamefont
  {Ellingsen}}]{maxwell19}%
  \BibitemOpen
  \bibfield  {author} {\bibinfo {author} {\bibfnamefont {P.}~\bibnamefont
  {Maxwell}}, \bibinfo {author} {\bibfnamefont {B.~K.}\ \bibnamefont
  {Smeltzer}},\ and\ \bibinfo {author} {\bibfnamefont {S.~{\AA}.}\ \bibnamefont
  {Ellingsen}},\ }\bibfield  {title} {\bibinfo {title} {The error in predicted
  phase velocity of surface waves atop a shear current with uncertainty},\
  }\href@noop {} {\bibfield  {journal} {\bibinfo  {journal} {Water Waves}\
  }\textbf {\bibinfo {volume} {2}},\ \bibinfo {pages} {79} (\bibinfo {year}
  {2020})}\BibitemShut {NoStop}%
\bibitem [{\citenamefont {Campana}\ \emph {et~al.}(2015)\citenamefont
  {Campana}, \citenamefont {Terrill},\ and\ \citenamefont
  {De~Paolo}}]{campana15}%
  \BibitemOpen
  \bibfield  {author} {\bibinfo {author} {\bibfnamefont {J.}~\bibnamefont
  {Campana}}, \bibinfo {author} {\bibfnamefont {E.}~\bibnamefont {Terrill}},\
  and\ \bibinfo {author} {\bibfnamefont {T.}~\bibnamefont {De~Paolo}},\
  }\bibfield  {title} {\bibinfo {title} {Observations of surface current and
  current shear using {X}-band radar},\ }in\ \href@noop {} {\emph {\bibinfo
  {booktitle} {Current, Waves and Turbulence Measurement (CWTM), 2015 IEEE/OES
  Eleventh}}}\ (\bibinfo  {publisher} {IEEE},\ \bibinfo {year} {2015})\ pp.\
  \bibinfo {pages} {1--5}\BibitemShut {NoStop}%
\bibitem [{\citenamefont {Lund}\ \emph {et~al.}(2018)\citenamefont {Lund},
  \citenamefont {Haus}, \citenamefont {Horstmann}, \citenamefont {Graber},
  \citenamefont {Carrasco}, \citenamefont {Laxague}, \citenamefont {Novelli},
  \citenamefont {Guigand},\ and\ \citenamefont {\"{O}zg\"{o}kmen}}]{lund18}%
  \BibitemOpen
  \bibfield  {author} {\bibinfo {author} {\bibfnamefont {B.}~\bibnamefont
  {Lund}}, \bibinfo {author} {\bibfnamefont {B.~K.}\ \bibnamefont {Haus}},
  \bibinfo {author} {\bibfnamefont {J.}~\bibnamefont {Horstmann}}, \bibinfo
  {author} {\bibfnamefont {H.~C.}\ \bibnamefont {Graber}}, \bibinfo {author}
  {\bibfnamefont {R.}~\bibnamefont {Carrasco}}, \bibinfo {author}
  {\bibfnamefont {N.~J.}\ \bibnamefont {Laxague}}, \bibinfo {author}
  {\bibfnamefont {G.}~\bibnamefont {Novelli}}, \bibinfo {author} {\bibfnamefont
  {C.~M.}\ \bibnamefont {Guigand}},\ and\ \bibinfo {author} {\bibfnamefont
  {T.~M.}\ \bibnamefont {\"{O}zg\"{o}kmen}},\ }\bibfield  {title} {\bibinfo
  {title} {Near-surface current mapping by shipboard marine {X}-band radar: A
  validation},\ }\href@noop {} {\bibfield  {journal} {\bibinfo  {journal} {J.
  Atmos. Ocean. Technol.}\ }\textbf {\bibinfo {volume} {35}},\ \bibinfo {pages}
  {1077} (\bibinfo {year} {2018})}\BibitemShut {NoStop}%
\bibitem [{\citenamefont {Kudryavtsev}\ \emph {et~al.}(2008)\citenamefont
  {Kudryavtsev}, \citenamefont {Shrira}, \citenamefont {Dulov},\ and\
  \citenamefont {Malinovsky}}]{Kudryavtsev08}%
  \BibitemOpen
  \bibfield  {author} {\bibinfo {author} {\bibfnamefont {V.}~\bibnamefont
  {Kudryavtsev}}, \bibinfo {author} {\bibfnamefont {V.}~\bibnamefont {Shrira}},
  \bibinfo {author} {\bibfnamefont {V.}~\bibnamefont {Dulov}},\ and\ \bibinfo
  {author} {\bibfnamefont {V.}~\bibnamefont {Malinovsky}},\ }\bibfield  {title}
  {\bibinfo {title} {On the vertical structure of wind-driven sea currents},\
  }\href@noop {} {\bibfield  {journal} {\bibinfo  {journal} {J. Phys.
  Oceanogr.}\ }\textbf {\bibinfo {volume} {38}},\ \bibinfo {pages} {2121}
  (\bibinfo {year} {2008})}\BibitemShut {NoStop}%
\bibitem [{\citenamefont {Li}\ \emph {et~al.}(2019)\citenamefont {Li},
  \citenamefont {Smeltzer},\ and\ \citenamefont {Ellingsen}}]{li19b}%
  \BibitemOpen
  \bibfield  {author} {\bibinfo {author} {\bibfnamefont {Y.}~\bibnamefont
  {Li}}, \bibinfo {author} {\bibfnamefont {B.~K.}\ \bibnamefont {Smeltzer}},\
  and\ \bibinfo {author} {\bibfnamefont {S.~{\AA}.}\ \bibnamefont
  {Ellingsen}},\ }\bibfield  {title} {\bibinfo {title} {Transient wave
  resistance upon a real shear current},\ }\href@noop {} {\bibfield  {journal}
  {\bibinfo  {journal} {Eur. J. Mech. B/Fluids.}\ }\textbf {\bibinfo {volume}
  {73}},\ \bibinfo {pages} {180} (\bibinfo {year} {2019})}\BibitemShut
  {NoStop}%
\bibitem [{\citenamefont {Goda}(2010)}]{goda10}%
  \BibitemOpen
  \bibfield  {author} {\bibinfo {author} {\bibfnamefont {Y.}~\bibnamefont
  {Goda}},\ }\href@noop {} {\emph {\bibinfo {title} {Random seas and design of
  maritime structures}}},\ \bibinfo {edition} {3rd}\ ed.,\ Vol.~\bibinfo
  {volume} {33}\ (\bibinfo  {publisher} {World Scientific Publishing Company},\
  \bibinfo {year} {2010})\BibitemShut {NoStop}%
\bibitem [{\citenamefont {Janssen}(2014)}]{janssen14}%
  \BibitemOpen
  \bibfield  {author} {\bibinfo {author} {\bibfnamefont {P.~A. E.~M.}\
  \bibnamefont {Janssen}},\ }\bibfield  {title} {\bibinfo {title} {On a random
  time series analysis valid for arbitrary spectral shape},\ }\href
  {https://doi.org/10.1017/jfm.2014.565} {\bibfield  {journal} {\bibinfo
  {journal} {J. Fluid Mech.}\ }\textbf {\bibinfo {volume} {759}},\ \bibinfo
  {pages} {236} (\bibinfo {year} {2014})}\BibitemShut {NoStop}%
\bibitem [{\citenamefont {Barbariol}\ \emph {et~al.}(2019)\citenamefont
  {Barbariol}, \citenamefont {Bidlot}, \citenamefont {Cavaleri}, \citenamefont
  {Sclavo}, \citenamefont {Thomson},\ and\ \citenamefont
  {Benetazzo}}]{barbiarol19}%
  \BibitemOpen
  \bibfield  {author} {\bibinfo {author} {\bibfnamefont {F.}~\bibnamefont
  {Barbariol}}, \bibinfo {author} {\bibfnamefont {J.-R.}\ \bibnamefont
  {Bidlot}}, \bibinfo {author} {\bibfnamefont {L.}~\bibnamefont {Cavaleri}},
  \bibinfo {author} {\bibfnamefont {M.}~\bibnamefont {Sclavo}}, \bibinfo
  {author} {\bibfnamefont {J.}~\bibnamefont {Thomson}},\ and\ \bibinfo {author}
  {\bibfnamefont {A.}~\bibnamefont {Benetazzo}},\ }\bibfield  {title} {\bibinfo
  {title} {Maximum wave heights from global model reanalysis},\ }\href@noop {}
  {\bibfield  {journal} {\bibinfo  {journal} {Prog. Oceanogr.}\ }\textbf
  {\bibinfo {volume} {175}},\ \bibinfo {pages} {139} (\bibinfo {year}
  {2019})}\BibitemShut {NoStop}%
\bibitem [{\citenamefont {Cartwright}\ \emph {et~al.}(1956)\citenamefont
  {Cartwright}, \citenamefont {Longuet-Higgins},\ and\ \citenamefont
  {Deacon}}]{Cartwright56}%
  \BibitemOpen
  \bibfield  {author} {\bibinfo {author} {\bibfnamefont {D.~E.}\ \bibnamefont
  {Cartwright}}, \bibinfo {author} {\bibfnamefont {M.~S.}\ \bibnamefont
  {Longuet-Higgins}},\ and\ \bibinfo {author} {\bibfnamefont {G.~E.~R.}\
  \bibnamefont {Deacon}},\ }\bibfield  {title} {\bibinfo {title} {The
  statistical distribution of the maxima of a random function},\ }\href@noop {}
  {\bibfield  {journal} {\bibinfo  {journal} {Proc. R. Soc. Lond.}\ }\textbf
  {\bibinfo {volume} {237}},\ \bibinfo {pages} {212} (\bibinfo {year}
  {1956})}\BibitemShut {NoStop}%
\bibitem [{\citenamefont {Rice}(1944)}]{Rice45}%
  \BibitemOpen
  \bibfield  {author} {\bibinfo {author} {\bibfnamefont {S.~O.}\ \bibnamefont
  {Rice}},\ }\bibfield  {title} {\bibinfo {title} {Mathematical analysis of
  random noise},\ }\href@noop {} {\bibfield  {journal} {\bibinfo  {journal}
  {Bell Syst. Tech. J.}\ }\textbf {\bibinfo {volume} {23}},\ \bibinfo {pages}
  {282} (\bibinfo {year} {1944})}\BibitemShut {NoStop}%
\bibitem [{\citenamefont {Haring}\ \emph {et~al.}(1976)\citenamefont {Haring},
  \citenamefont {Osborne},\ and\ \citenamefont {Spencer}}]{Haring77}%
  \BibitemOpen
  \bibfield  {author} {\bibinfo {author} {\bibfnamefont {R.~E.}\ \bibnamefont
  {Haring}}, \bibinfo {author} {\bibfnamefont {A.~R.}\ \bibnamefont
  {Osborne}},\ and\ \bibinfo {author} {\bibfnamefont {L.~P.}\ \bibnamefont
  {Spencer}},\ }\bibfield  {title} {\bibinfo {title} {Extreme wave parameters
  based on continental shelf storm wave records},\ }in\ \href@noop {} {\emph
  {\bibinfo {booktitle} {Proceedings of the fifteenth Coastal Engineering
  Conference}}}\ (\bibinfo {year} {1976})\ pp.\ \bibinfo {pages}
  {151--170}\BibitemShut {NoStop}%
\bibitem [{\citenamefont {Kriebel}\ and\ \citenamefont
  {Dawson}(1993)}]{Kriebel94}%
  \BibitemOpen
  \bibfield  {author} {\bibinfo {author} {\bibfnamefont {D.~L.}\ \bibnamefont
  {Kriebel}}\ and\ \bibinfo {author} {\bibfnamefont {T.~H.}\ \bibnamefont
  {Dawson}},\ }\bibfield  {title} {\bibinfo {title} {Nonlinearity in wave crest
  statistics},\ }in\ \href@noop {} {\emph {\bibinfo {booktitle} {Proceedings of
  the Second International Conference on Wave Measurement and Analysis}}}\
  (\bibinfo {year} {1993})\ pp.\ \bibinfo {pages} {61--75}\BibitemShut
  {NoStop}%
\bibitem [{\citenamefont {Huang}\ \emph {et~al.}(1986)\citenamefont {Huang},
  \citenamefont {Bliven}, \citenamefont {Long},\ and\ \citenamefont
  {Tung}}]{Huang86}%
  \BibitemOpen
  \bibfield  {author} {\bibinfo {author} {\bibfnamefont {N.~E.}\ \bibnamefont
  {Huang}}, \bibinfo {author} {\bibfnamefont {L.~F.}\ \bibnamefont {Bliven}},
  \bibinfo {author} {\bibfnamefont {S.~R.}\ \bibnamefont {Long}},\ and\
  \bibinfo {author} {\bibfnamefont {C.-C.}\ \bibnamefont {Tung}},\ }\bibfield
  {title} {\bibinfo {title} {An analytical model for oceanic whitecap
  coverage},\ }\href@noop {} {\bibfield  {journal} {\bibinfo  {journal} {J.
  Phys. Oceanogr.}\ }\textbf {\bibinfo {volume} {16}},\ \bibinfo {pages} {1597}
  (\bibinfo {year} {1986})}\BibitemShut {NoStop}%
\bibitem [{\citenamefont {Kriebel}\ and\ \citenamefont
  {Dawson}(1991)}]{Kriebel91}%
  \BibitemOpen
  \bibfield  {author} {\bibinfo {author} {\bibfnamefont {D.~L.}\ \bibnamefont
  {Kriebel}}\ and\ \bibinfo {author} {\bibfnamefont {T.~H.}\ \bibnamefont
  {Dawson}},\ }\bibfield  {title} {\bibinfo {title} {Nonlinear effects on wave
  groups in random seas},\ }\href@noop {} {\bibfield  {journal} {\bibinfo
  {journal} {J. Offshore Mech. Arct. Eng.}\ }\textbf {\bibinfo {volume}
  {113}},\ \bibinfo {pages} {142} (\bibinfo {year} {1991})}\BibitemShut
  {NoStop}%
\bibitem [{\citenamefont {Prevosto}\ \emph {et~al.}(2000)\citenamefont
  {Prevosto}, \citenamefont {Krogstad},\ and\ \citenamefont
  {Robin}}]{Prevosto00}%
  \BibitemOpen
  \bibfield  {author} {\bibinfo {author} {\bibfnamefont {M.}~\bibnamefont
  {Prevosto}}, \bibinfo {author} {\bibfnamefont {H.}~\bibnamefont {Krogstad}},\
  and\ \bibinfo {author} {\bibfnamefont {A.}~\bibnamefont {Robin}},\ }\bibfield
   {title} {\bibinfo {title} {Probability distributions for maximum wave and
  crest heights},\ }\href@noop {} {\bibfield  {journal} {\bibinfo  {journal}
  {Coast. Eng.}\ }\textbf {\bibinfo {volume} {40}},\ \bibinfo {pages} {329}
  (\bibinfo {year} {2000})}\BibitemShut {NoStop}%
\bibitem [{\citenamefont {Dysthe}\ \emph {et~al.}(2008)\citenamefont {Dysthe},
  \citenamefont {Krogstad},\ and\ \citenamefont {Muller}}]{Dysthe08}%
  \BibitemOpen
  \bibfield  {author} {\bibinfo {author} {\bibfnamefont {K.}~\bibnamefont
  {Dysthe}}, \bibinfo {author} {\bibfnamefont {H.~E.}\ \bibnamefont
  {Krogstad}},\ and\ \bibinfo {author} {\bibfnamefont {P.}~\bibnamefont
  {Muller}},\ }\bibfield  {title} {\bibinfo {title} {Oceanic rogue waves},\
  }\href@noop {} {\bibfield  {journal} {\bibinfo  {journal} {Annu. Rev. Fluid
  Mech.}\ }\textbf {\bibinfo {volume} {40}},\ \bibinfo {pages} {287} (\bibinfo
  {year} {2008})}\BibitemShut {NoStop}%
\bibitem [{\citenamefont {Krogstad}\ \emph {et~al.}(2004)\citenamefont
  {Krogstad}, \citenamefont {Liu}, \citenamefont {Socquet-Juglard},
  \citenamefont {Dysthe},\ and\ \citenamefont {Trulsen}}]{Krogstad04}%
  \BibitemOpen
  \bibfield  {author} {\bibinfo {author} {\bibfnamefont {H.~E.}\ \bibnamefont
  {Krogstad}}, \bibinfo {author} {\bibfnamefont {J.}~\bibnamefont {Liu}},
  \bibinfo {author} {\bibfnamefont {H.}~\bibnamefont {Socquet-Juglard}},
  \bibinfo {author} {\bibfnamefont {K.~B.}\ \bibnamefont {Dysthe}},\ and\
  \bibinfo {author} {\bibfnamefont {K.}~\bibnamefont {Trulsen}},\ }\bibfield
  {title} {\bibinfo {title} {Spatial extreme value analysis of nonlinear
  simulations of random surface waves},\ }in\ \href@noop {} {\emph {\bibinfo
  {booktitle} {International Conference on Offshore Mechanics and Arctic
  Engineering}}},\ Vol.\ \bibinfo {volume} {37440}\ (\bibinfo {year} {2004})\
  pp.\ \bibinfo {pages} {285--295}\BibitemShut {NoStop}%
\bibitem [{\citenamefont {Smeltzer}\ and\ \citenamefont
  {Ellingsen}(2017)}]{smeltzer17}%
  \BibitemOpen
  \bibfield  {author} {\bibinfo {author} {\bibfnamefont {B.~K.}\ \bibnamefont
  {Smeltzer}}\ and\ \bibinfo {author} {\bibfnamefont {S.~{\AA}.}\ \bibnamefont
  {Ellingsen}},\ }\bibfield  {title} {\bibinfo {title} {Surface waves on
  currents with arbitrary vertical shear},\ }\href@noop {} {\bibfield
  {journal} {\bibinfo  {journal} {Phys. Fluids}\ }\textbf {\bibinfo {volume}
  {29}},\ \bibinfo {pages} {047102} (\bibinfo {year} {2017})}\BibitemShut
  {NoStop}%
\end{thebibliography}%
\end{document}